\documentclass[a4paper,longauth]{aa}  

\usepackage{graphicx}
\usepackage{txfonts}
\usepackage{ae,aecompl}
\usepackage[caption=false]{subfig}
\usepackage{graphicx}   % Including figure files
\usepackage{amsmath}    % Advanced maths commands
\usepackage{mathtools}
\usepackage{amssymb}    % Extra maths symbols
\usepackage{amsbsy}
\usepackage{color}
\usepackage{booktabs}
\usepackage{lipsum}
\usepackage{rotating}
\usepackage{tabularx}
\usepackage[dvipsnames]{xcolor}

\usepackage{CJKutf8}

\usepackage[hidelinks]{hyperref}
\hypersetup{
    colorlinks,
    linkcolor={blue!90!black},
    citecolor={blue!90!black},
    urlcolor={blue}
}
\usepackage{xspace}
\usepackage{catchfile}
\usepackage{multirow} 
\usepackage{multicol} 
\usepackage{tabularx}

\usepackage[switch]{lineno}
\definecolor{ur}{HTML}{e59db5}
\definecolor{trueor}{HTML}{df4853}
\definecolor{noselection}{HTML}{d7bf5e}
\definecolor{recor}{HTML}{5675a5}

\newcommand{\dr}{{\mathrm{d}}}

\newcommand{\ziangtxt}[1]{{\color{black}{{#1}}}}

\newcommand{\wtheta}{$w(\theta)\,$}
\newcommand{\omm}{$\Omega_{\mathrm{m}}\,$}

\begin{document}

   \title{KiDS-Legacy: Angular galaxy clustering from deep surveys with complex selection effects}

    \authorrunning{Yan et al.}

   \author{Ziang Yan \inst{1}\thanks{E-mail:yanza21@astro.rub.de}
          %Ziang Yan ({\cjkfont 颜子昂})
          \and Angus H. Wright \inst{1}   
          \and Nora Elisa Chisari \inst{2, 14}%g2
          \and Christos Georgiou \inst{2}
          \and Shahab Joudaki \inst{3,4}
          \and Arthur Loureiro \inst{5,6}
          \and Robert Reischke \inst{7}
          \and Marika Asgari \inst{8} %g3
          \and Maciej Bilicki \inst{9}
          \and Andrej Dvornik \inst{1}
          \and Catherine Heymans \inst{10,1}
          \and Hendrik Hildebrandt \inst{1} 
          \and Priyanka Jalan \inst{9}
          \and Benjamin Joachimi \inst{11}
          \and Giorgio Francesco Lesci \inst{12,13}
          %\and Shun-Sheng Li ({\cjkfont 李顺生}) \inst{14,15}
          \and Shun-Sheng Li \inst{14,15}          
          \and Laila Linke \inst{16}
          \and Constance Mahony \inst{17,18,1}
          \and Lauro Moscardini \inst{12,13,19}
          \and Nicola R. Napolitano\inst{20}
          \and Benjamin Stölzner\inst{1}
          \and Maximilian Von Wietersheim-Kramsta \inst{21,22}
          \and Mijin Yoon \inst{14}}
            %group 2 starts from here:
            
            %group 3 starts from here:

   \institute{Ruhr University Bochum, Faculty of Physics and Astronomy, Astronomical Institute (AIRUB), German Centre for Cosmological Lensing, 44780 Bochum, Germany %1
   \and 
   Institute for Theoretical Physics, Utrecht University, Princetonplein 5, 3584CC Utrecht, The Netherlands %2
   \and 
   Centro de Investigaciones Energéticas, Medioambientales y Tecnológicas (CIEMAT), Av. Complutense 40, E-28040 Madrid, Spain %3 
   \and 
   Institute of Cosmology \& Gravitation, Dennis Sciama Building, University of Portsmouth, Portsmouth, PO1 3FX, United Kingdom %4
   \and
   The Oskar Klein Centre, Department of Physics, Stockholm University, AlbaNova University Centre, SE-106 91 Stockholm, Sweden %5
   \and
   Imperial Centre for Inference and Cosmology (ICIC) Blackett Laboratory, Imperial College London, Prince Consort Road, London SW7 2AZ, UK %6
   \and
    Argelander-Institut für Astronomie, Universität Bonn, Auf dem Hügel 71, D-53121 Bonn, Germany %7
   \and
   School of Mathematics, Statistics and Physics, Newcastle University, Herschel Building, NE1 7RU, Newcastle-upon-Tyne, UK %8
   \and
   Center for Theoretical Physics, Polish Academy of Sciences, al. Lotników 32/46, 02-668 Warsaw, Poland %9
   \and
   Institute for Astronomy, University of Edinburgh, Royal Observatory, Blackford Hill, Edinburgh, EH9 3HJ, UK. %10
   \and
   Department of Physics and Astronomy, University College London, Gower Street, London WC1E 6BT, UK %11
   \and
   Dipartimento di Fisica e Astronomia "Augusto Righi" - Alma Mater Studiorum Università di Bologna, via Piero Gobetti 93/2, 40129 Bologna, Italy %12
   \and
   INAF-Osservatorio di Astrofisica e Scienza dello Spazio di Bologna, Via Piero Gobetti 93/3, 40129 Bologna, Italy %13
   \and
   Leiden Observatory, Leiden University, Einsteinweg 55, 2333 CC Leiden, The Netherlands %14
   \and
   Aix-Marseille Université, CNRS, CNES, LAM, Marseille, France %15
   \and Universität Innsbruck, Institut für Astro- und Teilchenphysik, Technikerstr. 25/8, 6020 Innsbruck, Austria %16
   \and
   Department of Physics, University of Oxford, Denys Wilkinson Building, Keble Road, Oxford OX1 3RH, United Kingdom %17
   \and
   Donostia International Physics Center, Manuel Lardizabal Ibilbidea, 4, 20018 Donostia, Gipuzkoa, Spain %18
   \and
   Istituto Nazionale di Fisica Nucleare (INFN) - Sezione di Bologna, viale Berti Pichat 6/2, I-40127 Bologna, Italy %19
   \and 
   Department of Physics “E. Pancini” University of Naples Federico II C.U. di Monte Sant’Angelo Via Cintia, 21 ed. 6, 80126 Naples, Italy %20
   \and
   Institute for Computational Cosmology, Ogden Centre for Fundament Physics - West, Department of Physics, Durham University, South Road, Durham DH1 3LE, UK %21
   \and 
   Centre for Extragalactic Astronomy, Ogden Centre for Fundament Physics - West, Department of Physics, Durham University, South Road, Durham DH1 3LE, UK %22
   }

   \date{}

 %\abstract{}{}{}{}{} 
% 5 {} token are mandatory
 
\abstract{Photometric galaxy surveys, despite their limited resolution along the line of sight, encode rich information about the large-scale structure (LSS) of the Universe thanks to the high number density and extensive depth of the data. However, the complicated selection effects in wide and deep surveys can potentially cause significant bias in the angular two-point correlation function (2PCF) measured from those surveys. In this paper, we measure the 2PCF from the newly published KiDS-Legacy sample. Given an $r$-band $5\sigma$ magnitude limit of $24.8$ and survey footprint of $1347$ deg$^2$, it achieves an excellent combination of sky coverage and depth for such a measurement. We find that complex selection effects, primarily induced by varying seeing, introduce over-estimation of the 2PCF by approximately an order of magnitude. To correct for such effects, we apply a machine learning-based method to recover an organised random (OR) that presents the same selection pattern as the galaxy sample. The basic idea is to find the selection-induced clustering of galaxies using a combination of self-organising maps (SOMs) and hierarchical clustering (HC). This unsupervised machine learning method is able to recover complicated selection effects without specifying their functional forms. We validate this SOM+HC method on mock deep galaxy samples with realistic systematics and selections derived from the KiDS-Legacy catalogue. Using mock data, we demonstrate that the OR delivers unbiased 2PCF cosmological parameter constraints, removing the $27\sigma$ offset in the galaxy bias parameter that is recovered when adopting uniform randoms.
Blinded measurements on the real KiDS-Legacy data show that the corrected 2PCF is robust to the SOM+HC configuration near the optimal set-up suggested by the mock tests. }

     \titlerunning{KiDS-Legacy: Angular galaxy clustering}

\begin{CJK}{UTF8}{gbsn}
\maketitle
\end{CJK}

%
%-------------------------------------------------------------------
%Comment out for A&A submission

\tableofcontents
\clearpage

\section{Introduction}
\label{sect:intro}

The fluctuation patterns of the mass distribution in the Universe (i.e. the large-scale structure, LSS) is one of the central topics of modern cosmology \citep[see, for example, ][ for a comprehensive introduction to the LSS]{dodelson2020modern}. Because the matter in the Universe is mainly made up of invisible dark matter, one needs to observe visible LSS tracers to infer the matter distribution. Among observations of various LSS tracers, large-scale galaxy surveys have provided robust constraints on cosmological parameters. Spectroscopic surveys such as the Sloan Digital Sky Survey \citep[SDSS-IV;][]{2021PhRvD.103h3533A} and the Dark Energy Spectroscopic Instrument \citep[DESI;][]{2024arXiv241112022D} map  3D galaxy distributions 
%owing to their extensive measurement of galaxy spectra 
from which they measure baryon acoustic oscillations (BAOs) and redshift-space distortions (RSDs) to constrain the Hubble constant, matter density, dark energy equation of state, and neutrino mass \citep{2019JCAP...10..044C, 2019PhRvL.123h1301L, 2020MNRAS.499..210N}. Recent photometric surveys, including the Kilo-Degree Survey \citep[KiDS;][]{asgari2020kids1000}{}{}, the Dark Energy Survey \citep[DES;][]{amon_dark_2022, 2022PhRvD.105b3515S}{}{}, and the Hyper Suprime-Cam Subaru Strategic Program \citep[HSC-SSP;][]{2023PhRvD.108l3518L, 2023PhRvD.108l3519D}{}{} have obtained consistent joint constraints on the matter density, $\Omega_{\mathrm{m}}$, and fluctuation amplitude, $\sigma_8$, of the density field with a precision of $\sim 5\%$. As the precision improves, tensions between the probes have begun to emerge, an important one being the tension in $S_8$, the matter clustering amplitude parameter, between the CMB and the late-time Universe probes. \citep[for a review, see][]{Abdalla_2022}. Future surveys, including the Legacy Survey of Space and Time (LSST) at the \textit{Vera Rubin} Observatory \citep[][]{lsstsciencecollaboration2009lsst}{}{}, the {\textit{Euclid}} Survey \citep[][]{2024EuclidOverview}{}{}, the {{Xuntian}} \citep[][also known as the Chinese Space Station Telescope, CSST]{2019ApJ...883..203G}{}{} are promising to tackle these problems with better data quality and systematics estimations. 

The galaxy distribution encodes rich information about the LSS of the Universe and galaxy formation, so it has long been used as a tracer to study the LSS \citep{1973ApJ...185..413P,1983ApJ...274..529S,1990MNRAS.242P..43M,1990Natur.348..705E,1996MNRAS.283.1227M, Nichol_2007}. On large scales, it can be estimated to be linearly biased relative to the matter distribution\citep[see][for a review on galaxy bias]{2018PhR...733....1D}. Subtler bias terms can be modelled via an effective field theory \citep[][]{Baumann_2012, Carrasco_2012}. On small scales, the galaxy distribution can be described by a halo occupation distribution model \citep[HOD;][]{Zheng_2005}{}{} in the context of the halo model \citep[][]{2000MNRAS.318.1144P, Seljak_2000, COORAY_2002, 2023OJAp....6E..39A}{}{}. An important summary statistic of the galaxy distribution is the two-point correlation function (2PCF), which describes the galaxy clustering between two points as a function of their separations. The measurements of 2PCF have been used to constrain the matter density and clustering amplitude of the LSS since the 1980s \citep[][]{1983ApJ...267..465D}{}{}. Recently, a combination of galaxy clustering, galaxy-galaxy lensing, and cosmic shear, called the $3\times2$pt analysis, has greatly improved the constraining power on $\Omega_{\mathrm{m}}$ and $S_8$ \citep{heymans2020kids1000, 2022PhRvD.105b3520A, Sugiyama_2023, 2023A&A...675A.189D}. 

Measurements of three-dimensional galaxy clustering depend on reliable galaxy redshift measurements, typically carried out by spectroscopic observations that are generally time-consuming and limited in depth. While photometric surveys can reach deeper, they cannot obtain precise redshifts for each galaxy. In this case, given the angular coordinates of galaxy positions, one can measure the angular galaxy clustering with the 2PCF with respect to angular separation. With the calibrated redshift distribution of a photometric sample, the angular 2PCF can be  modelled as projected 3-D 2PCF weighted by the redshift distribution. Angular 2PCF also encodes a significant amount of cosmological information \citep[see, for example;][]{2012cfht}{}{} and is relatively easy to measure. However, several systematic uncertainties need to be considered to obtain accurate constraints. In terms of modelling, galaxy distributions are affected by redshift space distortions \citep[RSDs;][]{1987MNRAS.227....1K}{}{} and cosmic magnification \citep[][]{menard2002cosmic}{}{}. On the observational side, galaxies are subject to selection effects by various observational systematics, including seeing, sky background, instrument response, survey strategy, galactic extinction, and so on. In practice, the selection effects are anisotropic given the various observational conditions and survey strategies, such that the galaxy sample will show a variable depth pattern that will introduce additional non-cosmological correlations that bias the angular 2PCF measurement and subsequent cosmological constraints\footnote{This work primarily deals with angular 2PCF, so in the following text, if not explicitly stated otherwise, 2PCF means angular two-point correlation function.}.

The selection effects of bright samples tend to be mild \citep{harryOR}, but they are expected to be more significant for deeper samples since fainter galaxies are more sensitive to the selection. However, we can benefit from the higher number density and higher redshift of deep samples with their lower shot noise and richer information about the higher redshifts. In this work we measure the angular galaxy clustering with the Legacy catalogue selected from the fifth data release of the KiDS survey \citep[][]{wright2024fifth}. It is a deep sample that reaches an $r$-band magnitude limit of $24.8$ which is deeper than the KiDS-Bright catalogue \citep{Bilicki_2021} used in \citet{harryOR} by five magnitudes. Notably, the KiDS-Legacy catalogue has a 5$\sigma$ magnitude limit in $i$-band of 23.5, which is deeper than the MagLim sample selected from DES data \citep{PhysRevD.103.043503} with an $i$-band magnitude limit of 22.2. Although \citet{Morrison_2015} and \citet{Nicola_2020} have measured 2PCF using deeper galaxy samples from Canada-France-Hawaii Telescope Lensing Survey \citep[CFHTLenS;][]{heymans2012cfhtlens} and HSC, the KiDS-Legacy sample has the advantage of larger sky coverage and enhanced accuracy on photo-$z$ at $z>1$ thanks to additional near-infrared bands. The KiDS-Legacy 2PCF will be included in the KiDS-Legacy $6\times2$pt analysis \citep[see][for forecasts of the $6\times2$pt analysis]{johnston20246x2ptforecastinggainsjoint}, which is  a combination of the six two-point statistics among KiDS-Legacy galaxy shapes, galaxy positions, and the spectroscopic samples of 2dFLenS \citep{2016MNRAS.462.4240B} and BOSS DR12 \citep{Alam_2015}. The advantages of such a measurement include low shot noise due to high number densities and self-calibration of the redshift distribution, given the same $n(z)$ as the shear sample. As we show here,  the strong and complicated selection effects in the KiDS-Legacy catalogue  causes order-of-magnitude bias in the 2PCF. Therefore, it is crucial to correct such a bias to ensure the accuracy of subsequent cosmological analyses.

Various methods have been proposed and applied to correct for selection effects. Following \citet{harryOR}, we classify them into three types:

\begin{itemize}
    \item \ziangtxt{Mitigating selection bias in statistics. Correcting the selection effects with the following two methods \citep[][]{Weaverdyck_2021}{}{}:  1) template subtraction method \citep[][]{Ho_2012, Ross_2011}{}{},  which models contamination terms as templates multiplied by an amplitude factor. After determining the amplitude factor, these terms are subtracted from the statistics; 2) mode projection \citep[][]{Leistedt_2013, Leistedt_2014, Nicola_2020, berlfein2024multiplicative}{}{}, which mitigates systematics contamination by marginalising the functional contribution of systematic templates.}
    
    \item Forward modelling with synthetic objects. This approach calibrates the observational selections by injecting artificial galaxies into realistic images \citep[][]{berge2012ultra, 2016MNRAS.457..786S}{}{}. The injected objects are passed through a realistic measurement process, resulting in a random catalogue mimicking realistic selection functions. This method is computationally demanding, but it is under active development for ongoing and future surveys \citep[][]{Everett_2022, 2024arXiv240516299K}{}{}.

    \item \ziangtxt{Reconstructing the selection function by regression. This method aims to probe the relationship between observed galaxy number densities and systematics values to construct weight maps or random catalogues that reflect the reconstructed selection functions. For example, \citet{2018PhRvD..98d2006E}, \citet{Rezaie_2020}, and \citet{Rodr_guez_Monroy_2022} regress galaxy number and systematics with the help of systematics templates, while \citet{Morrison_2015} accounts for selection effects by finding selection-induced galaxy clustering in high-dimensional systematics space.}
\end{itemize}

\ziangtxt{A subclass of regression methods uses machine learning to learn the relationship between galaxy number densities and systematics values. This class of methods is more capable of finding complex,  non-linear, and potentially correlated selection effects. For example, \citet{Rezaie_2020} uses deep neural networks to derive weights of galaxies selected from the Dark Energy Camera Legacy Survey \citep[DECaLS;][]{2019AJ....157..168D}; \citet{Morrison_2015} applied a $k$-means clustering method in the high-dimensional density-systematics space to create weight maps. The method used in this work belongs to this category.}

%Correcting bias in statistics depends on the functional assumptions of systematic contamination. For example, \citet{Nicola_2020} assumes additive systematics contamination in the galaxy over-density and tries to fit the linear factor from systematics maps. This method works when the contamination level is low (as in that work), but will potentially fail when the selection functions are complicated and correlated. On the other hand, forward modelling and regression-based methods try to recover the overall selection effects without modelling the selection function of each systematic. Therefore, these methods are capable of capturing complicated selection effects, but they are valid only when the sample is large enough for the algorithm to learn the selection pattern.

We focus on 2PCF in this work. In practice, 2PCF can be estimated with the Landy-Szalay estimator \citep[][]{landeysz}{}{}, which employs a random catalogue to factor out non-cosmological correlations. If the catalogue has significant selection effects, a uniform random catalogue  (UR)  will fail to capture them and will not prevent a bias in the estimated 2PCF. An organised random catalogue (OR)  is a tailored random catalogue reflecting the same selection effects that can be used to correct such biases. \citet{harryOR} proposed a machine-learning-based method to recover the OR. It uses a combination of self-organising maps and hierarchical clustering (SOM+HC) to identify galaxy clusters\footnote{Here the galaxy clusters are in the systematics space, and are not related to gravitationally bounded galaxy clusters in the Universe. In the following text we use the term cluster to refer to the cluster of galaxies in the systematics space.} in the systematics space. Galaxies from the same cluster are assumed to be selected uniformly, so a UR with the same galaxy number is generated in the sky region occupied by those galaxies. The OR is constructed by combining the URs corresponding to all the clusters. This method has been validated in \citet{harryOR} for the KiDS-1000 bright sample.

In this work, we apply the SOM+HC method to correct for complex selection effects in the full KiDS-Legacy sample. We first present the SOM+HC methodology, then test it on mock galaxy samples generated from the Generator for Large-Scale Structure \citep[GLASS;][]{glass}{}{}. The mock systematics and selections are applied based on the real KiDS-Legacy sample. We evaluate the method by comparing the 2PCF measured with the recovered OR with an unbiased 2PCF with no selection effect. Then we perform a preliminary, blinded 2PCF measurement from the real KiDS-Legacy catalogue. This work focuses on comparing and modelling the 2PCFs on linear scales, as most photometric surveys use such scales due to limitations in the theoretical modelling \citep{Crocce_2015, Rodr_guez_Monroy_2022}. In forthcoming work, the OR recovered from the KiDS-Legacy catalogue will be used to measure the 2PCF for the KiDS-Legacy $6\times2$-pt analysis, and will also be useful for future LSS surveys.

The paper is structured as follows: Sect. \ref{sect:model} describes the definition and measurements of galaxy 2PCF; Sect. \ref{sect:data} presents the data and mock that we use to run and test the method; Sect. \ref{sect:method} introduces the SOM+HC method to correct the selection effects; Sect. \ref{sect:validation} deals with validations of the SOM+HC methods; Sect. \ref{sect:kidslegacy} presents a blinded measurement with the KiDS-legacy catalogue; and Sect. \ref{sect:discussions} summarises the conclusion of this work and discusses the advantages of our method and future prospects. Throughout this study we assume a flat $\Lambda$CDM cosmology with the fixed cosmological parameters from \cite{planckcosmo18} as our background: $ (h,\Omega_\mathrm{c} h^2,\Omega_\mathrm{b} h^2, \sigma_8,n_\mathrm{s}) = (0.676, 0.119, 0.022, 0.81, 0.967)$. The code used in this work is published as the \textsc{tiaogeng}\footnote{\begin{CJK}{UTF8}{gbsn}
Tiaogeng (调羹) is the Chinese word for `spoon', more commonly used in southern China. It contains two characters: tiao (调) meaning to reconcile and geng (羹) referring to a Chinese-style thick soup. The code reconciles the unevenly observed sky, just as a tiaogeng stirs soup to make it taste more balanced and delicious.
\end{CJK}} package\footnote{\href{https://github.com/yanzastro/tiaogeng/}{https://github.com/yanzastro/tiaogeng/}} for future use.

\section{The galaxy clustering correlation function}
\label{sect:model}

\subsection{Definition and connection with matter distribution}

Galaxy 3-D 2PCFs describe the correlation of galaxy number over-density between points separated by a certain distance. For two galaxy samples $a$ and $b$, their 2PCF is defined as \citep[][]{1973ApJ...185..413P}
\begin{equation}
    w^{ab}(r) := \left\langle \delta^{a}(\boldsymbol{x})\delta^{b}(\boldsymbol{x}-\boldsymbol{r}) \right\rangle\,,
\end{equation}
where $\left\langle \cdot \right\rangle$ denotes the ensemble average. Due to the large-scale ergodicity of the Universe, the ensemble average can be approximated as a spatial average. In addition, due to the large-scale isotropy and homogeneity of the Universe, the 2PCF only depends on $r$, the length of spatial separation between two points. $\delta(\boldsymbol{x})$ denotes the galaxy number over-density at point $\boldsymbol{x}$:
\begin{equation}
    \delta_{\mathrm{g}}(\boldsymbol{x}) := \frac{n(\boldsymbol{x})-\bar{n}}{\bar{n}}\,.
\end{equation}
Here $n(\boldsymbol{x})$ denotes the galaxy number density at $\boldsymbol{x}$ and $\Bar{n}$ is the mean galaxy number density. Since we mainly focus on the galaxy 2PCF, we omit the subscript $g$ in the following formulas. We note that the 2PCF can also be defined for fluctuations of any LSS fields.

According to Parseval's theorem, the 2PCF is the Fourier transform of the power spectrum, the correlation function in Fourier space. They are both widely used summary statistics of the spatial distribution of cosmological fields, and they both depend on the evolution and ingredients of the Universe. For detailed mathematical descriptions of the 2PCF and power spectra, we refer to \citet{dodelson2020modern}.

For photometric surveys, we usually cannot measure the distance of an individual galaxy. Thus we can only project galaxies onto the celestial sphere and study their angular distribution. We can define the angular two-point correlation function by replacing $\boldsymbol{x}$ with a 2-dimensional angular vector $\boldsymbol{\theta}$. Since we only care about the angular galaxy distribution in this paper, we use 2PCF to denote the angular two-point correlation function unless otherwise stated. Under the `flat-sky approximation', the angular 2PCF is related to the angular power spectrum via \citep{1973ApJ...185..413P}
\begin{equation}
    w^{ab}(\theta) = \int \frac{\dr \ell}{2\pi} \ell J_0(\ell \theta) C^{ab}_{\ell}  \, ,
\end{equation}
where $J_0(x)$ is the zeroth order Bessel function of the first kind; $C^{ab}_{\ell}$ is the angular galaxy cross-correlation power spectrum of samples $a$ and $b$. We note that this formula works for 2PCF and angular power spectra between any cosmological fields. At sufficiently large $\ell$, the angular power spectra can be calculated via the Limber approximation \citep{limber1953analysis}:
\begin{equation}
    C^{ab}_{\ell} = \int_0^{\chi_{\mathrm{H}}} \frac{\dr \chi}{\chi^2}W^{a}(\chi)W^{b}(\chi)P^{ab}_{\mathrm{gg}}\left( \frac{\ell+1/2}{\chi}, z(\chi)\right)\,.
\end{equation}
Here $\chi$ is the comoving distance and the subscript $\rm H$ is the horizon; $P_{\mathrm{gg}}(k, z)$ is the galaxy power spectra at redshift $z$; $W^{a}(\chi)$ and $W^{b}(\chi)$ are the radial kernels of the two galaxy samples. We note that this formulation works for flat cosmology. For sample $a$, given its normalised galaxy redshift distributions $n^a(z)$, its radial kernel is
\begin{equation}
    W^{a}(\chi) = \frac{H\left(z(\chi)\right)}{c}n^{a}\left(z(\chi)\right)\,,
\end{equation}
where $H(z)$ is the Hubble parameter at redshift $z$. The radial kernel for field $b$ is defined likewise.

The galaxy power spectra $P_{\mathrm{gg}}(k, z)$ encode the 3-D fluctuations of galaxy distributions, which follow the fluctuations of underlying mass distribution. On small scales, the non-linear clustering of galaxies can be modelled as a halo model-based halo occupation distribution \citep[HOD;][]{Seljak_2000, COORAY_2002, Zheng_2005}. On large scales, galaxy fluctuations can be approximated as linearly biased matter fluctuations with $\delta_{\mathrm{g}}(\boldsymbol{x}, z) = b_{\mathrm{g}}(z)\delta_{\mathrm{m}}(\boldsymbol{x}, z)$ where $b_{\mathrm{g}}(z)$ is the redshift-dependent galaxy bias. Thus $P^{ab}_{\mathrm{gg}}(k, z)=b^a_{\mathrm{g}}(z)b^b_{\mathrm{g}}(z)P_{\mathrm{m}}(k, z)$. We can take the galaxy biases outside of the Limber integral by approximating them at the mean redshift $\bar{z}^{a}\equiv\int n^a(z)z\dr z / \int n^a(z)\dr z$ of the galaxy samples (denoted as $b^a\equiv b_{\mathrm{g}}(\bar{z}^a)$ and $b^b\equiv b_{\mathrm{g}}(\bar{z}^b)$). Therefore,  on linear scales we have
\begin{equation}
    w^{ab}(\theta) = b^a b^b w_{\mathrm{m}}(\theta)\,,
\end{equation}
where $w_{\mathrm{m}}(\theta)$ is the 2PCF of the matter field which can be calculated similarly to galaxies, but with galaxy power spectra replaced by matter power spectra.

\subsection{The measurement of two-point correlation function}

In practice, the galaxy 2PCF is measured in angular bins by counting the number of galaxy pairs (one galaxy from catalogue $a$ and the other from $b$) with separation angles within each angular bin. Given a pair of galaxy catalogues (denoted as $D$ hereafter) and corresponding random catalogues (denoted as $R$ hereafter), the 2PCF can be measured using a standard \cite{landeysz} estimator,
\begin{equation}
    \hat{w}^{ab}(\theta_i) = \frac{(D^aD^b)_i-(D^aR^b)_i-(R^aD^b)_i+(R^aR^b)_i}{(R^aR^b)_i}\,,
    \label{eq:wtheta_def}
\end{equation}
where $\theta_i$ is the mean angle in the $i$-th angular bin. Here $(D^aD^b)_i$ is the normalised number of galaxy pairs with separation angles within the $i$-th bin, which can be formally expressed as
\begin{equation}
    (D^aD^b)_i \equiv \frac{\sum_{p=1}^{N^a_D}\sum_{q=1}^{N^b_D}\Theta_i(\theta_{pq})}{N^a_D(N^b_D-\delta_{\mathrm{K}}^{ab})}\,,
    \label{eq:dd_def}
\end{equation}
where $N_D$ is the number of galaxies, $\theta_{pq}$ is the separation angle between the $p$-th and $q-$th galaxies, $\Theta_i(\theta)$ is the rect function that equals 1 when $\theta$ falls in the $i-$th bin and 0 otherwise. $\delta_{\mathrm{K}}^{ab}$ is the Kronecker $\delta$ symbol. The denominator in Eq.~\eqref{eq:dd_def} is the total number of galaxy pairs and the $\delta^{ab}$ term concerns removing galaxies paired with themselves for auto-correlation (i.e. $a=b$). The $RR$ term is likewise defined while $DR$ is defined without $\delta_{\mathrm{K}}^{ab}$.

For this work, instead of counting galaxy pairs in angular bins, we pixelised the galaxy distribution into the Healpix scheme \citep{Gorski_2005} and counted pixel pairs weighted by galaxy numbers in each pixel. This loses sub-pixel information, but we had three reasons for doing this: 1) pixelisation can speed up the calculation: the original KiDS catalogue contains $\sim 200 $ million galaxies, while a pixelised galaxy map with \texttt{Nside}=2048 which we use throughout this work only has about one million non-zero pixels, so counting pixel pairs instead of galaxy pairs will significantly speed up the measurement; 2) as we introduce later, our method to recover ORs is pixel-based, so any sub-pixel variable depth cannot be corrected; 3) the pixel size for an \texttt{Nside}=2048 HEALPix map is 1.7 arcmin, corresponding to a physical size of 0.86 Mpc at $z=1.5$, which is much below the linear scales that we probe in this work.

With pixelised galaxy fields, the terms in the Landy-Szalay estimator are modified as (take $DD$ as an example)
\begin{equation}
    (D^aD^b)_i \equiv \frac{\sum_{p=1}^{P^a_D}\sum_{q=1}^{P^b_D}\Theta_i(\theta_{pq})N^{a}_{D,p}N^{b}_{D,q}}{N^a_D(N^b_D-\delta_{\mathrm{K}}^{ab})}\,,
    \label{eq:ddpix_def}
\end{equation}
where $p$ and $q$ are now pixel indices; $\theta_{pq}$ is the angular separation between these two pixels; $P^{a,b}_D$ is the total number of occupied pixels in samples $a$ and $b$; $N^{a}_{D,p}$ is the number of galaxies from sample $a$ in pixel $p$. The random terms are defined likewise. The pixelised UR map can be taken as the footprint map with pixel value $N^{a}_{R,p}$ representing the coverage fraction of pixel $p$.

The random catalogue or map is used to factor out galaxy distributions induced by cosmology-unrelated systematics, including galactic and atmospheric foregrounds, instrument response, and survey strategy. If all those systematics are uniform across the survey footprint, the random catalogue is usually taken as a uniform Poisson random sample within the survey footprint with several times the mean galaxy number density. However, deep surveys, such as the KiDS survey, have non-uniform systematics that cause variable depth in the galaxy sample. A UR will fail to capture the spatial variation in galaxy number density caused by those selection effects and biases the 2PCF.

To de-bias the 2PCF, we can create a random catalogue with the same variable depth, namely an `organised random' catalogue, or, for pixelised measurement, an OR weight map for the $R$ terms in the estimator. The method to recover the OR will be presented in Sect.. \ref{sect:method}. We note that the OR can be used for summary statistics other than the angular 2PCF. We  present the OR application in angular power spectra $C_{\ell}$ in Appendix~\ref{sect:pcl_or}.

\section{The KiDS-Legacy data and GLASS mock data}
\label{sect:data}

\subsection{The Kilo-degree Survey and the Legacy data}

\begin{figure*}[!htp]
    \centering
    \includegraphics[width=\textwidth]{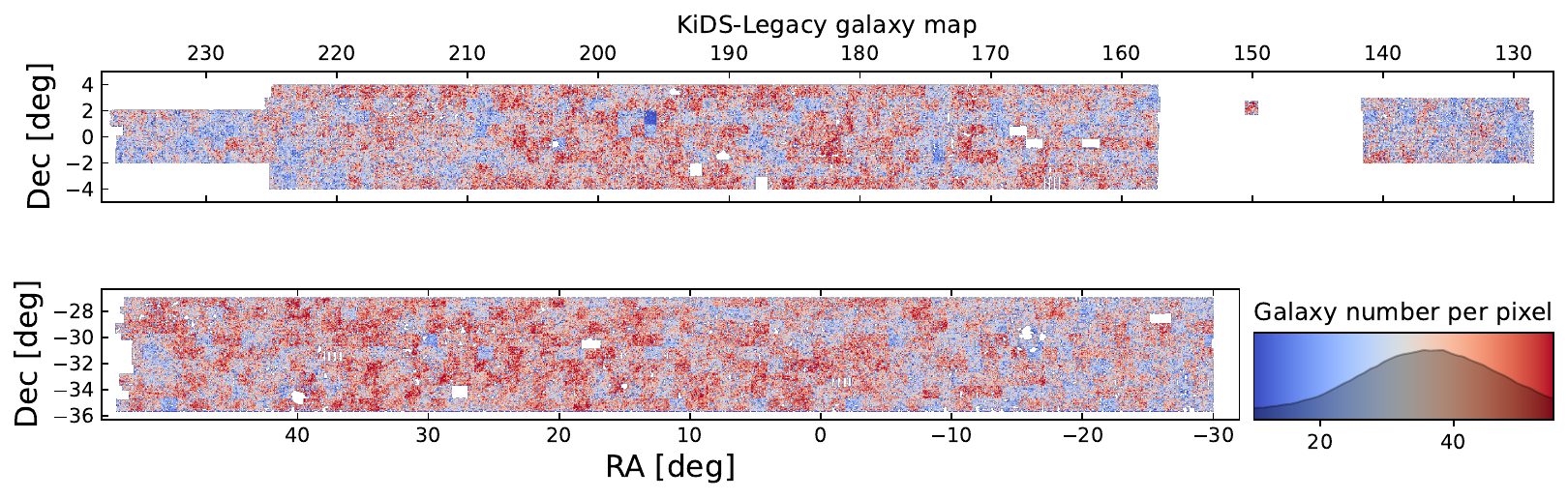}
    \caption{Galaxy distribution of the KiDS-Legacy catalogue. The map is pixelised into \textsc{HEALPix} grids with $\texttt{Nside}=2048$. The map is colour-coded by galaxy number per pixel, which has a size of 1.7 arcmin. A tile-based selection pattern can be seen by eye. \ziangtxt{The shaded region in the colour bar shows the normalised distribution of galaxy number per pixel.}}
    \label{fig:kids_galmap}
\end{figure*}

\begin{table*}
\caption{Summary of the systematics \ziangtxt{of each galaxy} that we used to recover the OR.}
\centering
\begin{tabular}{l|l|p{5cm}|p{3cm}|p{5cm}}
\toprule
Systematics & Type & Definition & Main origin & Selection effect \\
\hline 
    \texttt{Level} & A & Detection threshold (in the $r$-band) above background & background sky luminosity; CCD temperature, etc. & Fainter objects will be lost from an area of observation where \texttt{Level} is higher. \\
\hline 
    \texttt{PSF\_size} & A & Full-width at half-maximum of the $r$-band point-spread function in arcseconds & Seeing & The blurring could cause small or faint objects to drop below detection thresholds. \\
\hline   
    \texttt{PSF\_ell} & B & Ellipticity ($1-q$), where $q$ is the 2D major/minor axis ratio) of the $r$-band point-spread function. & Distortions on the focal plane & Non-isotropic blurring of object isophotes may induce a directional dependence for detections. \\
\hline  
    \texttt{EXTINCTION\_r} & C & Galactic extinction in the $r$-band derived from the \citet{2011ApJ...737..103S} coefficients for the \citet{1998ApJ...500..525S} dust map & The Milky Way & Dust preferentially scatters short-wavelength light from extragalactic objects; the loss of flux could prevent detection. \\
\hline
    \texttt{GAIA\_nstar} & C & The number of Gaia DR3 \citep{2016AA...595A...1G, 2023AA...674A...1G} stars with 14 $<$ G $<$ 17 within 5 arcmin around each KiDS source. & The Milky Way & Light from star-dense regions obscures background objects, and can also result in spurious galaxy detections through the misidentification of PSF-blurred, or blended, stars as galaxies. \\
\bottomrule
\end{tabular}
\tablefoot{The second column indicates the spatial variance type: inter-tile (type A), intra-tile (type B), and tile-independent (type C). The other columns indicate their definitions, physical origins, and how they potentially select the galaxy sample.}
\label{table:systematics}
\end{table*}

In this work, we  use the KiDS-Legacy galaxy catalogue selected from the fifth data release (DR5) of the Kilo-degree Survey \citep[KiDS;][]{wright2024fifth}. KiDS is a wide-field imaging survey that measures the positions and shapes of galaxies using the VLT Survey Telescope (VST) at the European Southern Observatory (ESO). Both the telescope and the survey were primarily designed for weak gravitational lensing applications. High-quality optical images are produced with VST-OmegaCAM, and these data are then combined with imaging from the VISTA Kilo-degree INfrared Galaxy survey (VIKING; \citealt{2013Msngr.154...32E}), allowing all sources in KiDS to be photometrically measured in nine optical and near-infrared bands: $ugriZYJHK_{\mathrm{s}}$ \citep[][]{2019A&A...632A..34W}. Although the sky coverage of KiDS is smaller than some galaxy lensing surveys \citep[such as DES,][]{abbott2016dark}, galaxy photometric redshift estimation and redshift distribution calibration (especially at high redshift) benefit from complementary NIR information from VIKING (which was co-designed with KiDS to reach complementary depths in the NIR bands). Cosmological constraints have already been made available from DR3 and DR4 \citep{Hildebrandt_2016, heymans2020kids1000}. KiDS DR5 covers a survey area of 1347 $\mathrm{deg}^2$. It also includes an $i$-band re-observation of the full footprint, thereby increasing the effective $i$-band depth by 0.4 magnitudes and enabling multi-epoch science. There is a 27 $\mathrm{deg}^2$ overlap with deep spectroscopic surveys, which enables the robust calibration of photometric redshifts across the full survey footprint \citep{Hildebrandt_2021}. 

The ``KiDS-Legacy'' sample used in this work is a subset of the full DR5 sample primarily determined by the availability of reliable shape measurements. It is the lensing sample that will be used for the fiducial KiDS DR5 cosmological analyses. Each galaxy in the KiDS-Legacy sample has ellipticities measured with the \emph{lens}fit algorithm \citep{miller2013bayesian, 2017MNRAS.467.1627F, 2023A&A...670A.100L} for weak lensing analyses. The footprint of the KiDS DR5 is divided into a Northern and a Southern patch. Artefacts around bright, saturated starlight, planets, the Moon, satellite flares, aeroplanes, and higher-order reflections from very bright stars have been masked out, leaving an unmasked $\sim 1000 \mathrm{deg}^2$ footprint. In addition, blended sources, unresolved binaries, transients, point sources flagged by \emph{lens}fit, and sources with failed photometry, badly estimated shapes, low resolution, or zero lensing weight are removed from the full sample, resulting in a Legacy sample of 43,205,156 galaxies, corresponding to 12 galaxies per arcmin$^{-2}$. A Bayesian
photo-$z$ \citep[BPZ,][]{2000ApJ...536..571B} estimation gives an approximated redshift range of $0.1<z_{\mathrm{B}}<2.0$. \ziangtxt{The redshift distribution is calibrated with a combination of SOM and clustering-redshift method\citep[][]{2020A&A...642A.200V}, which will be presented in a companion paper.} For more details about the construction of the KiDS-Legacy catalogue, we refer to Sect. 7 of \citet{wright2024fifth}. The galaxy distribution is shown in Fig. \ref{fig:kids_galmap}. We can tell by eye that the galaxy distribution has a clear tile-like pattern that is not likely the LSS.

The KiDS-Legacy data will be used to constrain cosmology with cosmic shear (Wright, et al., in prep.), the combined analysis of three two-point functions (3$\times$2pt), and the combined analysis of six two-point functions ($6\times2$pt; see \citealt{johnston20246x2ptforecastinggainsjoint} for the forecast). The 3$\times$2pt measurements include cosmic shear, galaxy-galaxy lensing (GGL), and galaxy clustering. The galaxy clustering correlation function in KiDS-1000 3$\times$2pt presented in \cite{heymans2020kids1000} is measured from the Baryon Oscillation Spectroscopic Survey Data Release 12 (BOSS DR12, \citealt{2016MNRAS.455.1553R, 2020A&A...633L..10T}), while for KiDS-Legacy the 3$\times$2pt statistics will be measured with the Bright sample selected from the Legacy data. The KiDS-Legacy 6$\times$2pt analysis measures the 2-point functions among KiDS galaxy shapes, galaxy positions, and the spectroscopic samples of 2dFLenS \citep{2016MNRAS.462.4240B} and BOSS DR12 \citep{Alam_2015}; therefore, six 2-point functions in total. 

The KiDS survey strategy is tile-based. That is, the survey footprint is divided into adjacent $1^{\circ}\times1^{\circ}$ square tiles, each of which gets five exposures and is never re-observed afterwards except for the $i$-band, which was observed twice. Therefore, observational systematics can have three types of spatial variation based on their origin: 

\begin{enumerate}[Type A,]
    \item inter-tile: uniform in each tile but differ across tiles, usually originating from varying observation conditions across tiles;
    \item intra-tile: varying in each tile, usually originating from focal plane distortion;
    \item observation-independent variations: usually generated from the anisotropy of the Milky Way or the Solar System.
\end{enumerate}

A synthesis of systematics results in a complex selection pattern, namely variable depth, in the galaxy distribution. Figure~\ref{fig:kids_galmap} illustrates the galaxy map of the KiDS-Legacy catalogue, which exhibits a tile-based pattern of variable depth. For example, galaxies in the anomalous dark blue tile around ($\alpha=196\deg$, $\delta=1.5\deg$) in the centre of KiDS-North in Fig. \ref{fig:kids_galmap} are significantly depleted. \citet{2020A&A...634A.104H} studied the effect of variable depth on cosmic shear and concluded that the impact is insignificant for KiDS-like surveys. However, the variable depth will introduce significant bias in angular clustering measurement if not corrected. As galaxy clustering measurements form part of the KiDS-Legacy $3\times2$pt and $6\times2$pt analyses, it is essential to correct this bias.

To correct the selection effects, we create an OR which reflects the same variable depth as the galaxy sample from the systematics given by the KiDS-Legacy catalogue. Those systematics generally originate from the Galactic foreground, atmospheric seeing, survey strategy, and instrument set-up, and are ideally independent of cosmology, galaxy properties, and the LSS. In the catalogue, systematics are described by metadata that are calculated or derived from the instrument set-up, observing conditions, image properties, and so on. In the following text, we  use `systematics' to also refer to the metadata we use to obtain the OR. We select five systematics that are most likely to have selection effects: \{\texttt{Level}, \texttt{PSF\_size}, \texttt{PSF\_ell}, \texttt{EXTINCTION\_r}, \texttt{GAIA\_nstar}\}. Their types of distribution, definitions, origins and selection effects are summarised in Table \ref{table:systematics}. These systematics, with the exception of \texttt{GAIA\_nstar}, are measured in the $r$-band because this is the detection band and we do not consider tomographic redshift binning in this analysis.

\begin{figure*}
    \centering
    \includegraphics[width=0.91\textwidth]{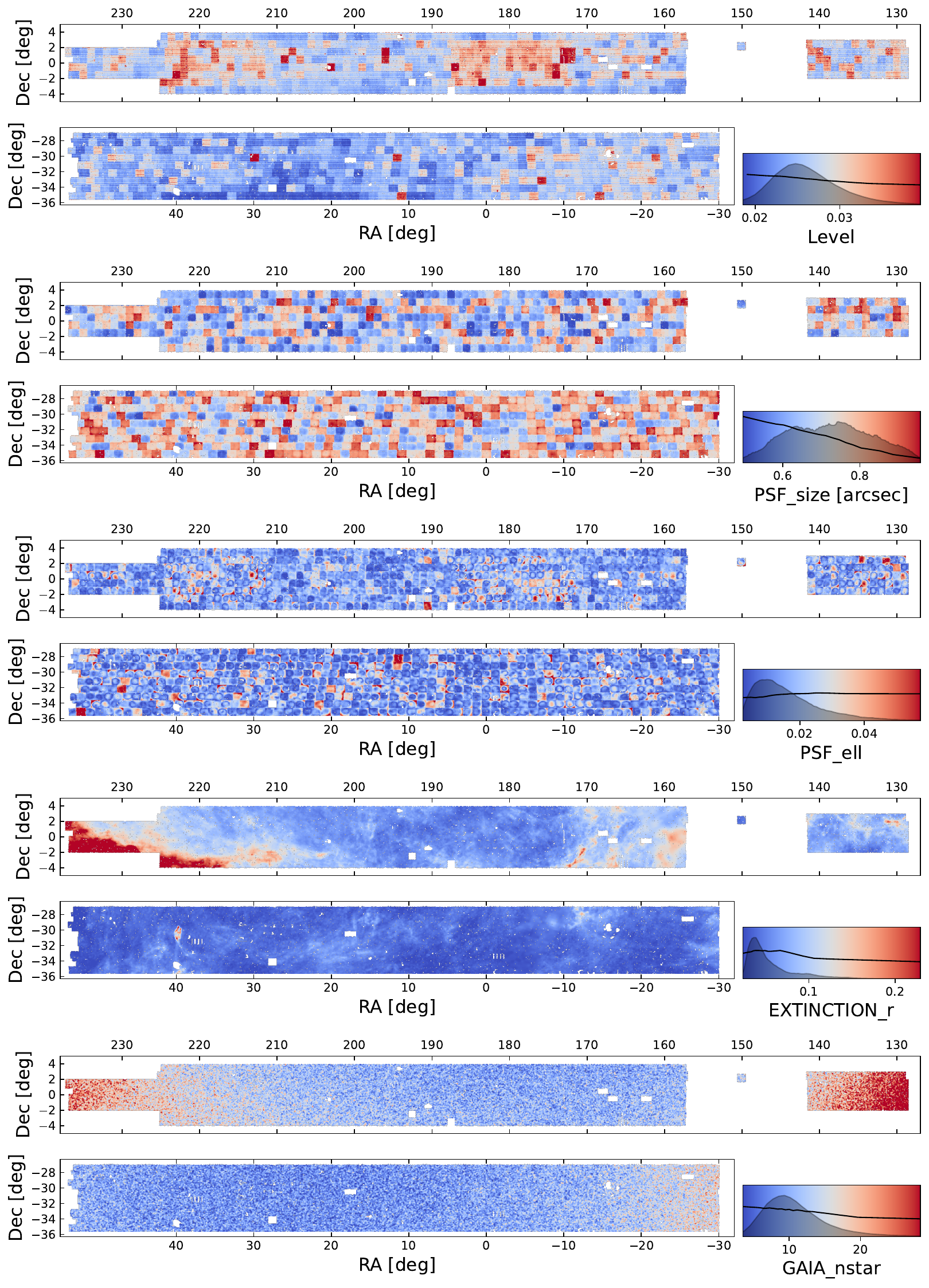}
    \caption{Maps of the KiDS-Legacy systematic considered in this paper. Each map is divided into northern and southern fields, plotted together with their colour bars. The black curves over-plotted in the colour bars show the \ziangtxt{relationship between galaxy contrast and systematics value, and the dynamic ranges are $[-0.25, 0.25]$. The shaded regions show the} probability distributions of each systematics.}
    \label{fig:sys_maps}
\end{figure*}

Figure~\ref{fig:sys_maps} shows the maps of these systematics. Each systematics map is divided into south and north fields, colour-coded by the mean systematics value in each pixel. The \ziangtxt{shaded regions} in the colour bars are the probability distributions of the systematics. From Fig.~\ref{fig:sys_maps}, we can see how \texttt{Level}, \texttt{PSF\_size}, \texttt{PSF\_ell} are distributed based on the $1^{\circ}\times 1^{\circ}$ tiles. Among them, \texttt{Level} and \texttt{PSF\_size} are determined by the seeing of each exposure and are therefore relatively uniform in each tile, but vary from tile to tile; while \texttt{PSF\_ell} depends on the curvature of the focal plane, which changes from the centre to the edge of each tile. On the other hand, \texttt{EXTINCTION\_r}\footnote{Given that the extinction in different bands is commonly just assumed to be the same template scaled by a different factor (thus they are fully correlated), we only use the $r$-band extinction for our model training.} and \texttt{Gaia\_nstar} reflect the spatial distribution of the Milky Way and are therefore more diffuse and independent of the tiles. \ziangtxt{To illustrate the selection effect of each systematics, we plot galaxy number contrast with respect to systematics values as black curves in the colour bars of Fig.~\ref{fig:sys_maps}. We note that the black curves share the same dynamic range between -0.25 and 0.25, so one can see that \texttt{PSF\_size} has the strongest selection effect. In addition, the selection effects of \texttt{EXTINCTION\_r} and \texttt{Gaia\_nstar} are slightly non-linear.}

It should be noted that we could recover the OR with all the systematics such as PSF shape components and background counts, but in practice, it is advantageous to only use a subset of systematics that is most likely to select galaxies to reduce computation time. The rest of the systematics are either not correlated with galaxy numbers or are strongly correlated with the selected systematics. In Appendix \ref{sect:all_sys} we run our method with all the systematics provided in the catalogue and prove that the performance does not improve. In the following sections of this paper, we present and test the method for recovering the organised random from these five systematics.

\subsection{The GLASS mock catalogue}

In this paper, we use mock catalogues generated by the Generator for Large Scale Structure \citep[][GLASS]{glass}\footnote{\href{https://github.com/glass-dev/glass}{https://github.com/glass-dev/glass}} to test and validate our method. GLASS is a public code for generating mock data for LSS surveys. It takes as input pre-calculated matter power spectra that are projected within a sequence of shells through the light cone and generates lognormal matter density fields. Galaxy positions and shears can be generated accordingly with additional input of redshift distributions, galaxy bias and intrinsic alignment model. GLASS can also generate a Poisson random sample according to the galaxy number density in each spherical shell, which can be used as a UR sample. All the samplings are done on \textsc{HEALPix} spheres which specify the minimum spatial resolution of the mock.

In this work, we only need galaxy positions generated by \textsc{GLASS}. We generate realisations of galaxy samples and corresponding random samples to validate our method of recovering organised random samples. The input cosmology is assumed to be the fiducial cosmology. We also note that under the assumption that the selection is independent of cosmology, the cosmology used to generate mock catalogues should not affect the evaluation of our method.

\section{The SOM+HC method to recover organised randoms}
\label{sect:method}
\begin{figure}
    \centering
    \includegraphics[width=\columnwidth]{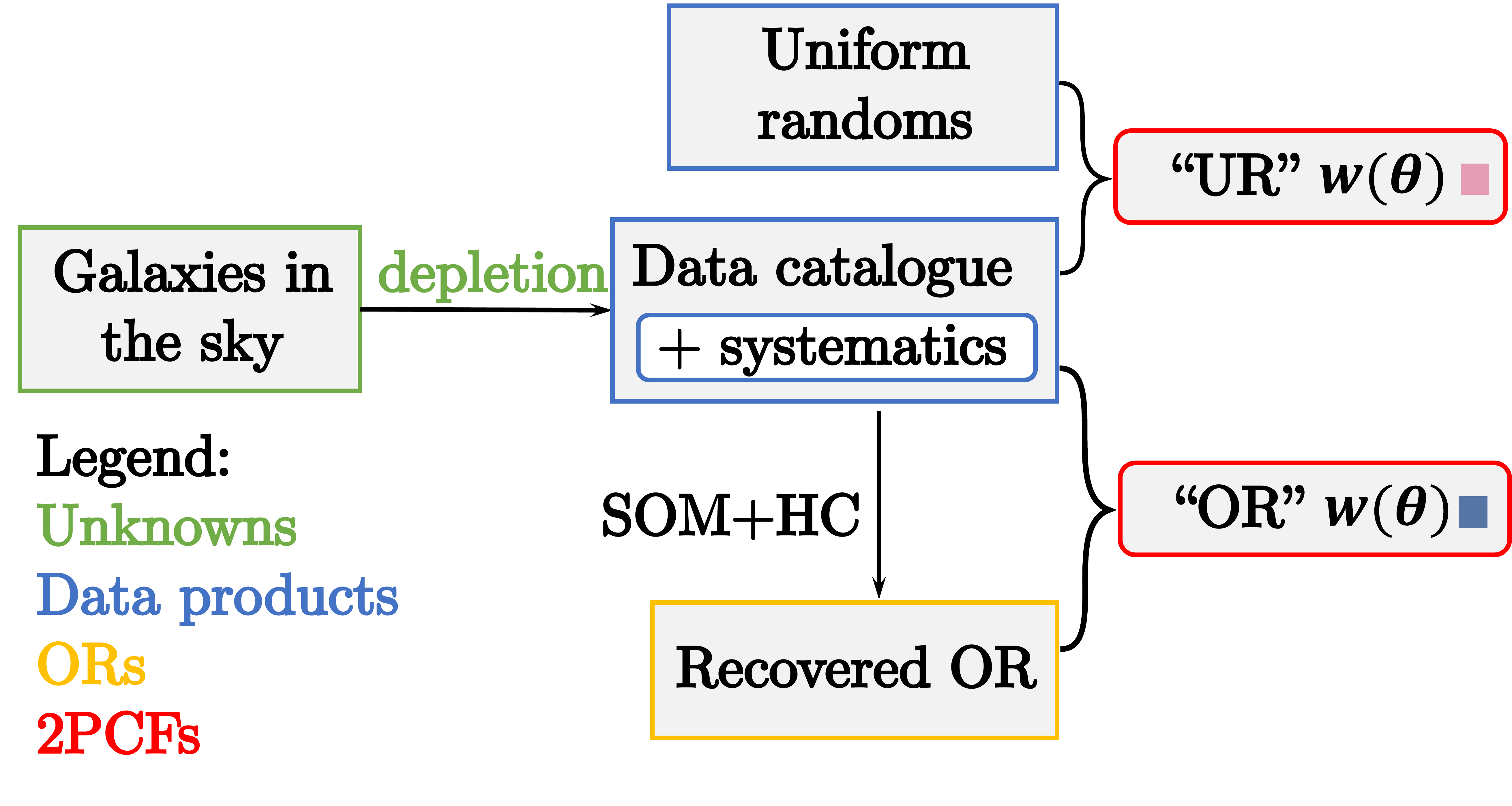}

        \caption{Flowchart of the measurement of the 2PCF with reconstructed organised randoms. The colour block after the 2PCF indicates the colours shown in the following $w(\theta)-\theta$ figures.}
    \label{fig:or_data_flowchart}
\end{figure}

\begin{figure}[!t]
    \centering
    \includegraphics[width=\columnwidth]{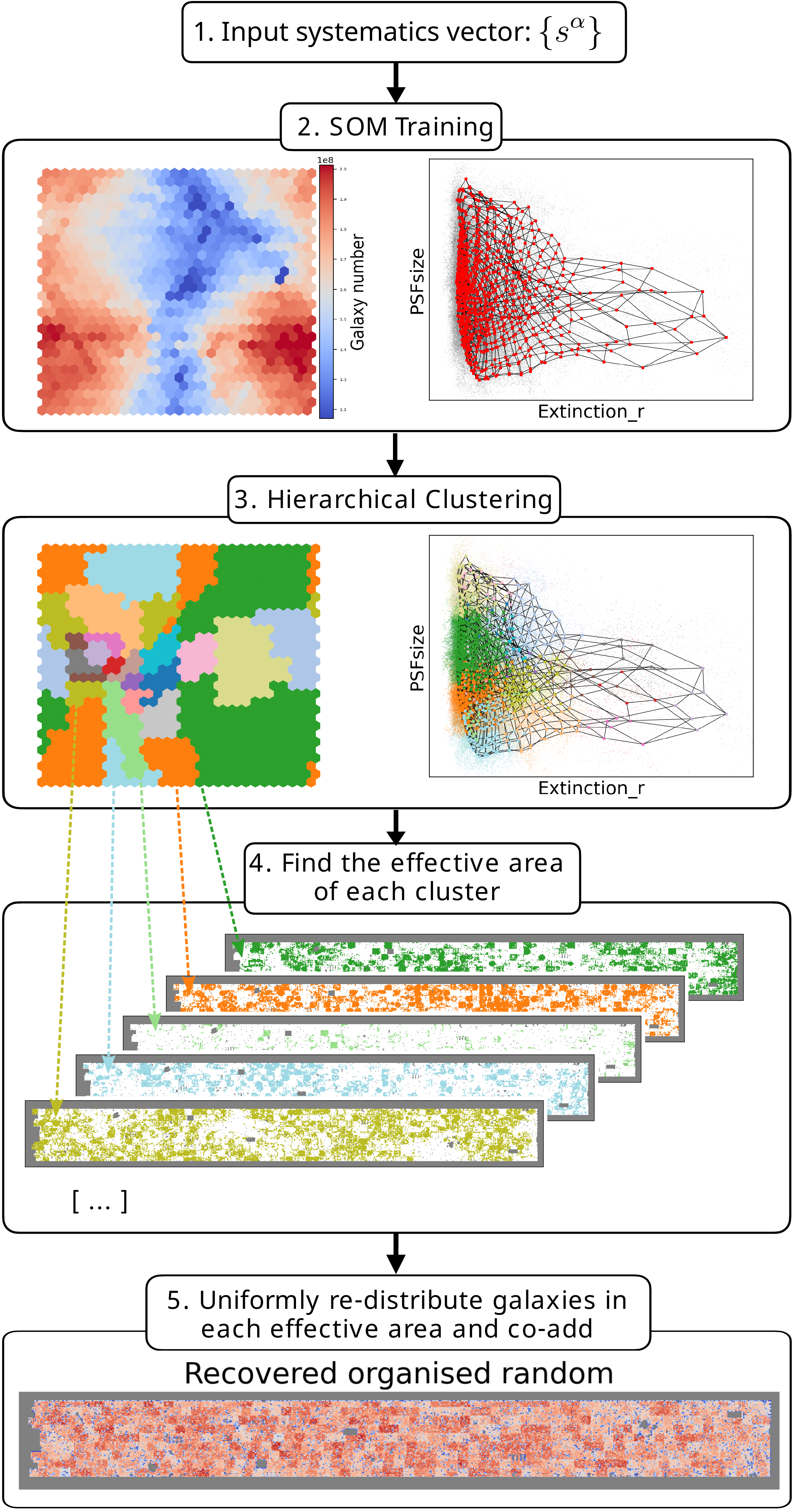}
    \caption{Flowchart illustrating the SOM+HC method to recover OR and correct selection effects in the 2PCF. Starting from the top: 1. Input systematics; 2. SOM training: left panel: SOM grid colour-coded by galaxy number in each cell; right panel: projection of systematics vectors (\ziangtxt{small} grey dots) and SOM cells (red dots connecting) on the plane of two systematics. The projected adjacent SOM cells are connected with black lines; 3. HC output: left panel: SOM cells colour-coded according to hierarchical cluster indices; right panel: systematics vectors and weight vectors colour-coded by corresponding cluster indices; 4. Effective areas corresponding to galaxies from each cluster; 5. Recovered OR weight map which will be used in a subsequent 2PCF measurement.}
    \label{fig:SOMHC_flowchart}
\end{figure}

Figure~\ref{fig:or_data_flowchart} shows a flowchart of the 2PCF measurement of galaxies with variable depth. Through observation, we obtain a depleted catalogue of galaxies with systematics evaluated for each galaxy. If we follow the upper half of the flowchart by using UR to compute the 2PCF, we get a biased ``UR'' 2PCF. Instead, we can recover OR to correct for non-uniform selection effects and obtain the unbiased ``OR'' 2PCF. \ziangtxt{In this section we  introduce the method that we used to generate OR, namely a combination of self-organising map (SOM) and hierarchical clustering (HC). Figure \ref{fig:SOMHC_flowchart} summarises this ``SOM+HC'' method.}

\subsection{The selection effects induced by systematics}

The variable depth of the galaxy catalogue can be treated as non-uniform selection effects due to systematics. In other words, some systematics make certain galaxies more difficult to observe and thus deplete them from the catalogue. For example, poor seeing results in large PSF size, which smooths the galaxy brightness and prevents us from observing low-magnitude galaxies. Spatially variable systematics cause spatially variable selections, hence the variable depth. Our task is to find the spatial distribution of the selection effects from these systematics and create an OR with the same selection distributions.

The selection effect of each systematic can be interpreted as the probability of removing a galaxy from the sample as a result of this systematic. If the unselected spatial distribution of galaxy numbers is $N(\boldsymbol{\theta})$, then the selected sample number $\tilde{N}(\boldsymbol{\theta})$ will distribute as 

\begin{equation}
    \tilde{N}(\boldsymbol{\theta}) =  P_{\mathrm{keep}}\left(\boldsymbol{\theta},\{s^a(\theta)\}\right)N(\boldsymbol{\theta}),
\end{equation}
where $P_{\mathrm{keep}}\left(\boldsymbol{\theta},\{s^a(\theta)\}\right)$ is the selection function describing the probability that a galaxy at $\boldsymbol{\theta}$ is kept in the sample given a set of systematics $\{s^a(\theta)\}$. Thus $P_{\mathrm{keep}}(\boldsymbol{\theta})$ acts as a ``weight'' of the galaxy distribution. When using selected galaxy samples, all the ``galaxy numbers'' in the definition of correlation functions, including both data and random samples for Eqs. \eqref{eq:wtheta_def}, \eqref{eq:dd_def}, \eqref{eq:ddpix_def}, should be depleted galaxy numbers $\tilde{N}$. The OR is just a UR selected by  $P_{\mathrm{keep}}(\boldsymbol{\theta})$.

If the systematics are uncorrelated, the overall selection function is a multiplication of the selection functions of individual systematics. One can reconstruct the total selection function by modelling the selection effect of each systematics \citep{Rodr_guez_Monroy_2022}. In practice, however, it is not always possible to construct reliable quantitative models to describe the effect of each systematics on the galaxy number density. In addition, systematics can be correlated (e.g. extinction in different bands, extinction and GAIA stars), making the overall selection function more complicated.

\ziangtxt{The key to recovering the ORs is to group galaxies into subsamples, each sharing a similar selection effect. Galaxies from each subsample occupy a subregion in the survey footprint that is presumably selected uniformly. Therefore, we can generate a UR in the same sky region for each group, and combine them to obtain the OR. In this work, we use a combination of hierarchical clustering \citep[HC]{murtagh2012algorithms} and SOM, namely the SOM+HC method, to group the galaxies and recover the OR weight to account for the selection effect. This method does not assume formal models for selection functions and their correlations, so it is flexible enough to account for arbitrarily complex selection effects.}

\subsection{Self-organising map and hierarchical clustering}

HC is a widely used clustering algorithm. It has the advantages of flexibility, robustness, and interpretability, but is computationally expensive (the computational time scales with the data volume as $O(N^2)$, where $N$ is the number of galaxies). On the other hand, the SOM has a complexity of $O(N)$. In this work, we combine these two algorithms: first, we group the systematics vectors into SOM cells and then we further group the SOM cells into hierarchical clusters\footnote{Training a small SOM in the first place instead of grouping big SOM into clusters will restrict the accuracy of the manifold’s mapping. Moreover, It's more flexible and faster to change the number of HC clusters from a pre-trained SOM than to change the dimension of a SOM and retrain.}. After the training, the hierarchical clusters are projected back onto the sky, resulting in discrete regions occupied by galaxies with uniform systematics and selection effects. We then randomly redistribute galaxies across these regions and combine galaxies from all the clusters to form the OR. The details of the method are presented below.

SOM is an unsupervised machine-learning algorithm that maps high-dimensional vectors to cells on a two-dimensional map while preserving the topological properties of the high-dimensional vectors by faithfully maintaining the distance between these data vectors. This means that when data vectors are close together in high-dimensional space, they are mapped to the same cell or cells that are close together on the SOM grid. Therefore, SOM can be used to group data, find correlations and visualise data. Specifically, the redshift distribution of the KiDS sample we use is calibrated by the SOM algorithm: mapping data vectors in colour space to SOM, and synthesising the redshift distribution in all SOM cells containing galaxies from spectroscopic samples \citep{Wright_2020, Myles_2021}. In addition, \citet{jalan2024enhancingphotometricredshiftcatalogs} use the SOM technique to quantify the completeness of spectroscopic samples used for photo-$z$ training of the KiDS-Legacy bright sample. 

In this work, instead of training a SOM in the colour space, we train it in the systematics space:  the data vectors are $N_{\mathrm{gal}}$ (number of galaxies) $N_{\mathrm{sys}}$-dimensional vectors, where $N_{\mathrm{sys}}$ is the number of systematics to account for ($N_{\mathrm{sys}}=5$ in our fiducial case). A SOM consists of $N_{\mathrm{dim}}\times N_{\mathrm{dim}}=N_{\mathrm{cell}}/,$ cells to represent the $N_{\mathrm{gal}}$ data vector. The positions of SOM cells in the $N_{\mathrm{sys}}$-dimensional systematics space are called weight vectors.

For the $n$-th galaxy, we use $\boldsymbol{V}_n$ to denote its systematics vector and $V_{n,i}, i=1,2,...,N_{\mathrm{sys}}$ to denote its $i$-th component. We can calculate the Euclidean distance between the weight vector of each SOM cell $\boldsymbol{W}^a, a=1,2,...,N_{\mathrm{cell}}$\footnote{A SOM is two-dimensional, so one could use two cell indices (column and row indices).} and the systematics vector as
\begin{equation}
    d^a_n = \sqrt{\sum_{i=1}^{N_{\mathrm{sys}}}\left( V_{n,i} - W^a_i \right)^2}\,.
\end{equation}
For each data vector, the representing weight vector is chosen as the one with minimum $d^a_n$. The corresponding SOM cell is termed the ``best matching unit'' (BMU). 

By definition, data vectors that share the same BMU are clustered in the systematics space. The training of the SOM is to iteratively update $\boldsymbol{W}^a$ so that the clustered data vectors are mapped to the same or close BMU. In each epoch, all the weight vectors are brought closer to the data vector, with the requirement that distant weight vectors and weight vectors around the BMU do not move much. To achieve this, a ``neighbourhood function'' $\Sigma(a, \sigma)$ is defined as a function of the grid distance between a cell and the BMU (not the distance between weight vectors). $\Sigma$ is close to 1 when the weight vector is close to the BMU, and drops to zero outside a typical width $\sigma$. In each epoch, the optimisation is performed iteratively over the whole galaxy sample.

The training steps are summarised as follows:

\begin{enumerate}
    \item Initialise the SOM by setting up the initial position of $N_{\mathrm{cell}}$ weight vectors as the two highest principal component analysis (PCA) components. This is equivalent to initialising the weight vectors as the $N_{\mathrm{cell}}$ on the 2-dimensional subspace spanned by the first two eigenvectors of the correlation matrix of the data vectors.\footnote{One can also initialise the SOM with random weight vectors.}
    \item In each epoch, perform the following steps iteratively until all the galaxies are iterated over:
    \begin{enumerate}[(i)]
        \item Calculate the distance between the data vector and all the weight vectors, then find the BMU;
        \item Choose a data vector and update the weight vector as
    \begin{equation}
        \boldsymbol{W}^a(t+1) = \boldsymbol{W}^a(t) + L \Sigma(a, \sigma(t)) \left[ \boldsymbol{V}(t)-\boldsymbol{W}^a(t) \right]\,,
        \label{eq:som_update}
    \end{equation}
    where $t$ denotes a time-step (i.e. the presentation of a new data vector to the SOM); $L$ is the learning rate specifying how fast the weight vectors approach $\boldsymbol{V}(t)$ in each step, and $\Sigma$ is the neighbourhood function;
    \item Choose another galaxy and perform Eq.~(\ref{eq:som_update}) until all the galaxies are iterated over. 
    \end{enumerate}
    \item Perform the iteration in Step 2 for several training epochs. During the first few epochs, $\sigma$ is roughly the size of the SOM, meaning that almost all the cells are updated. It decreases through the epochs so that only cells that are close to the BMU are updated. The learning rate also decreases to prevent ``jumping over the minimum''. The training stops when the weight vectors converge.
\end{enumerate}

Here, we use hexagonal SOM cells, which means that each cell has six neighbouring cells. {The SOM will be smoother in the systematics space compared to rectangular cells.} We choose the toroidal topology for the SOM, which means that the top and bottom boundaries and the left and right boundaries are adjacent.  {This can prevent edge effects in the training}. The neighbourhood function is defined as a Gaussian with $\sigma$ equal to half the dimension of the SOM then decreasing linearly to 1 at the last iteration of each epoch. The initial learning rate is 0.1 and decreases linearly to 0.01 in the last iteration of each epoch. We notice that the weight vectors barely update after five epochs, so we train the SOM in 10 epochs to ensure convergence.

After training, each systematic vector is represented by its BMU in the SOM. Since SOMs preserve the topological structure of the systematic vectors, they properly take into account the correlations between systematics. Galaxies that are clustered in the systematics space will be mapped into the same SOM cell, or the cells that are close both in the SOM grids and in the systematics space. 

\begin{figure}
    \centering
    \includegraphics[width=\columnwidth]{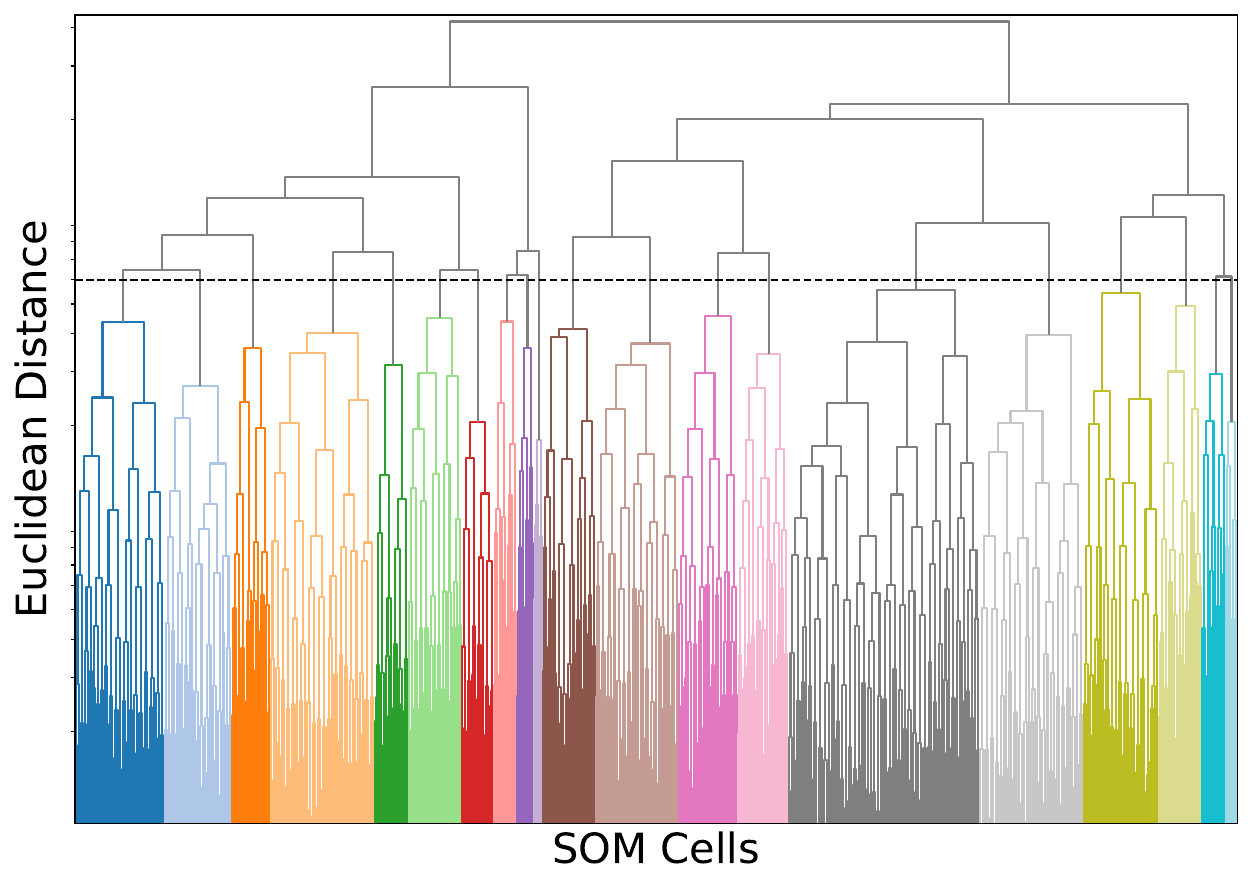}
    \caption{Example of a dendrogram  showing the clustering of 900 SOM cells into 20 HCs. The cells are clustered from the bottom to the top according to their Euclidean distance in the systematics space. The black dashed line shows the distance threshold where the cells are grouped into 20 clusters. SOM cells resulting in the same clusters are colour-coded with the same colour.}
    \label{fig:hc_dgram}
\end{figure}

After training the SOM, we further group galaxies by running HC on weight vectors according to the distance between them. Since the distance between the weight vectors reflects the distance between the systematic vectors by construction, each cluster represents a group of galaxies with similar systematics just like each SOM cell, but with more galaxies to eliminate sample variance. In this work, we use ``agglomerative clustering'', a bottom-up clustering method. The procedure can be briefly described as follows: 
\begin{enumerate}
    \item Treat each SOM cell as a cluster (so we have $N_{\mathrm{cell}}$ clusters in the beginning); 
    \item Iteratively merge two clusters separated by the shortest distance specified by the complete linkage criteria\footnote{ In this work, we use the ``complete linkage criteria'' with Euclidean distance, for which the distance between clusters equals the Euclidean distance between two data vectors (one in each cluster) that are farthest away from each other.}; 
    \item Hierarchically merge clusters until there is only one cluster, and construct the dendrogram of the whole clustering process;
    \item Cut the dendrogram \ziangtxt{ where the cells are merged into the desired number of clusters (we use $NC$ to denote the number of clusters hereafter)}. See Fig. \ref{fig:hc_dgram} for an example of a dendrogram. 
\end{enumerate}

By construction, a galaxy cluster in the systematics space represents a particular combination of systematics. {If the galaxy distribution in systematics space does not depend on the LSS of the Universe (we  validate this assumption in Sect.~\ref{sect:eva_datadriven}), we can assume that the number of galaxies in each cluster is determined by a uniform selection probability given by the synthetic systematics represented by that cluster.} After being grouped into clusters through SOM+HC, the galaxies from one cluster are distributed into discrete regions in the survey footprint and are assumed to be uniformly depleted by the associated combination of systematics. The detailed process of generating OR from clustered galaxies is presented in the next subsection.

The number of hierarchical clusters $NC$ is an important parameter. If $NC$ is too large, there will be higher sampling noise in each cluster. In addition, high $NC$ will cause over-fitting of the OR, resulting in over-correction in the 2PCF. If $NC$ is too small, the resolution in the systematics space will be too low to detect systematics variability. Therefore, we  evaluate the $NC$ value that optimises the trade-off between resolution and sampling noise in Sect.~\ref{sect:validation}.

In this work, we use the \textsc{somoclu} package \citep{Wittek_2017} to train the SOM. Hierarchical clustering is then performed via the \texttt{AgglomerativeClustering} class from \texttt{sklearn.cluster} package \citep[for a more detailed technical explanation of the algorithm, see][]{mullner2011modernhierarchicalagglomerativeclustering}. In the following text, we use ``SOM+HC'' to denote the procedure to create an OR with a combination of SOM and hierarchical clustering.

\subsection{Reconstructing organised randoms}

The basic idea of constructing an OR is to first find the effective sky region containing galaxies from each cluster. Galaxies in this region can be treated as uniformly selected according to the specific synthetic systematics of that cluster. We then generate a uniform random sample in the region according to the mean number density of that cluster and then add up the URs corresponding to all the clusters to get the overall OR. It is desirable to obtain an OR catalogue (i.e. coordinates of points within the observation footprint whose spatial distribution strictly follows the depletion distribution). However, with a finite number of hierarchical clusters, we cannot achieve such a resolution. Instead, we generate pixelised OR maps of the sky. That is, we aim for the OR weight in sky pixels that reflects the selection function for galaxies in that pixel ($P_{\mathrm{keep}}$). When measuring the 2PCF, we use it to calculate the $RR$ term in Eq. \eqref{eq:ddpix_def}. In this work, we use the \textsc{HEALPix} scheme to pixelise the sky. 

The OR is constructed as follows: after identifying the galaxy clusters in the systematics space via SOM+HC, we first find $N^i_p$, the number of galaxies in the $i$-th cluster that are in the $p$-th pixel of the sky; we note that a pixel can contain galaxies from different clusters. For a given cluster index $i$, $N^i_p$ should occupy only discrete sub-regions within the footprint that contain galaxies distributed closely in the high-dimensional space. We calculate the effective area of the $p$-th pixel corresponding to the $i$-th cluster as
\begin{equation}
    A_p^{i} \equiv \frac{N_p^{i}}{N_p} A_{p}\,,
\end{equation}
where $N_p$ is the total number of galaxies in that pixel and $A_{p}$ is the area of the $p$-th pixel in the footprint.\footnote{The pixel area can be calculated as the full pixel area times the fraction between masked and unmasked random sources.} If we assume that the SOM+HC are not affected by the underlying LSS, then we can assume that $N_p^{i}$ and $N_p$ have the same LSS contribution that cancels each other out. So $A_p^{i}$ is the area in the $p$-th pixel occupied by galaxies with uniform selection effect of the $i$-th cluster.

Now we add up $A_p^{i}$ across pixels to get the total effective area for each cluster,
\begin{equation}
    A^{i} \equiv \sum_p A_p^{i}\,,
\end{equation}
and the effective surface number density of the $i$-th cluster,
\begin{equation}
    n^{i} \equiv \frac{N^{i}}{A^{i}}\,.
\end{equation}
The average number of galaxies from the $i$-th cluster in the $p$-th pixel is $n_{i}A^i_p$. We can uniformly sample this number of galaxies within the occupied region of the $i$-th cluster and combine them across clusters to get an OR catalogue. However, as mentioned before, for a pixelised sky, it is more efficient to construct the OR weight by simply adding up $n_{i}A^i_p$ across clusters:
\begin{equation}
    \mathcal{W}_p = \sum_{i} A_p^in^i.
\end{equation}

The OR weight $\mathcal{W}_p$ is an estimate of $P_{\mathrm{keep}}$, so it is used as the $R$ terms in Eq. \eqref{eq:wtheta_def} to measure the unbiased 2PCF. We note that we use the calligraphic font $\mathcal{W}$ for the OR weight to distinguish from the 2PCF $w$. Figure \ref{fig:SOMHC_flowchart} summarises the SOM+HC method in a flowchart.

\begin{figure}
    \centering
    \includegraphics[width=\linewidth]{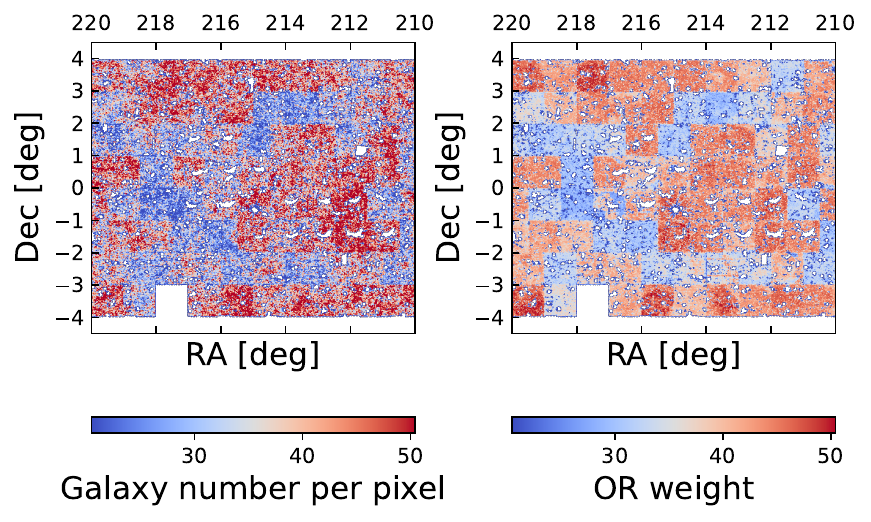}
    \caption{{\textit{Left panel}}: Galaxy number per pixel on \ziangtxt{a subregion} of the KiDS-Legacy footprint; {\textit{right panel}}:  OR weight on the same region.}
    \label{fig:orweight}
\end{figure}

The left panel of Fig.~\ref{fig:orweight} shows the KiDS-Legacy galaxy number fluctuation colour-coded by galaxy number per pixel. The holes in this map are the masked regions. From the left panel, one can notice an obvious tile-based variable depth in the galaxy map by eye. The right panel shows the organised random weight map constructed with a $30\times 30$ SOM, grouped into 400 hierarchical clusters. Both maps are pixelised with \texttt{Nside}=2048. Visually, it is clear that the OR weight map also shows a tile-based pattern that roughly matches the pattern on the galaxy number map.

Pixel size is another important parameter affecting the performance of the OR reconstruction. If the pixel size is too large, we lose the spatial resolution of the variable depth, resulting in an overestimated 2PCF with residual variable depth correlations. If the pixel size is too small, the organised random weight map will be too sparse, with many unintentionally masked pixels in the organised random map, since the organised random weight will only be non-zero in pixels containing galaxies. Thus the organised random will correlate with the LSS. If we correct this, we will remove the LSS from the 2PCF and under-estimate it. Therefore, we need to choose the pixel size carefully to balance the trade-off between resolution and over-correction. For the \textsc{HEALPix} scheme used in this paper, the pixel size is determined by the \texttt{Nside} parameter of the map. \texttt{Nside} can only be a power of 2, and if \texttt{Nside} doubles, the pixel size is halved. Our baseline analysis uses \texttt{Nside}=2048 corresponding to an angular size of 1.7 arcmin. We  validate this choice below.

\section{Validation of the method}
\label{sect:validation}

In this section we validate the SOM+HC method with two case studies. In Sect. \ref{sect:toy_sys}, we recover the OR for mock galaxy samples selected by systematics with simple spatial distributions and selection functions, while in Sect. \ref{sect:eva_datadriven}, we recover the OR for mock galaxy samples selected by realistic systematics distributions and data-driven selection functions. We evaluate the method by calculating the bias in the OR-corrected 2PCF.

\subsection{Testing the organised random with toy systematics model}
\label{sect:toy_sys}

\begin{figure*}[!htb]
    \centering
    \includegraphics[width=\textwidth]{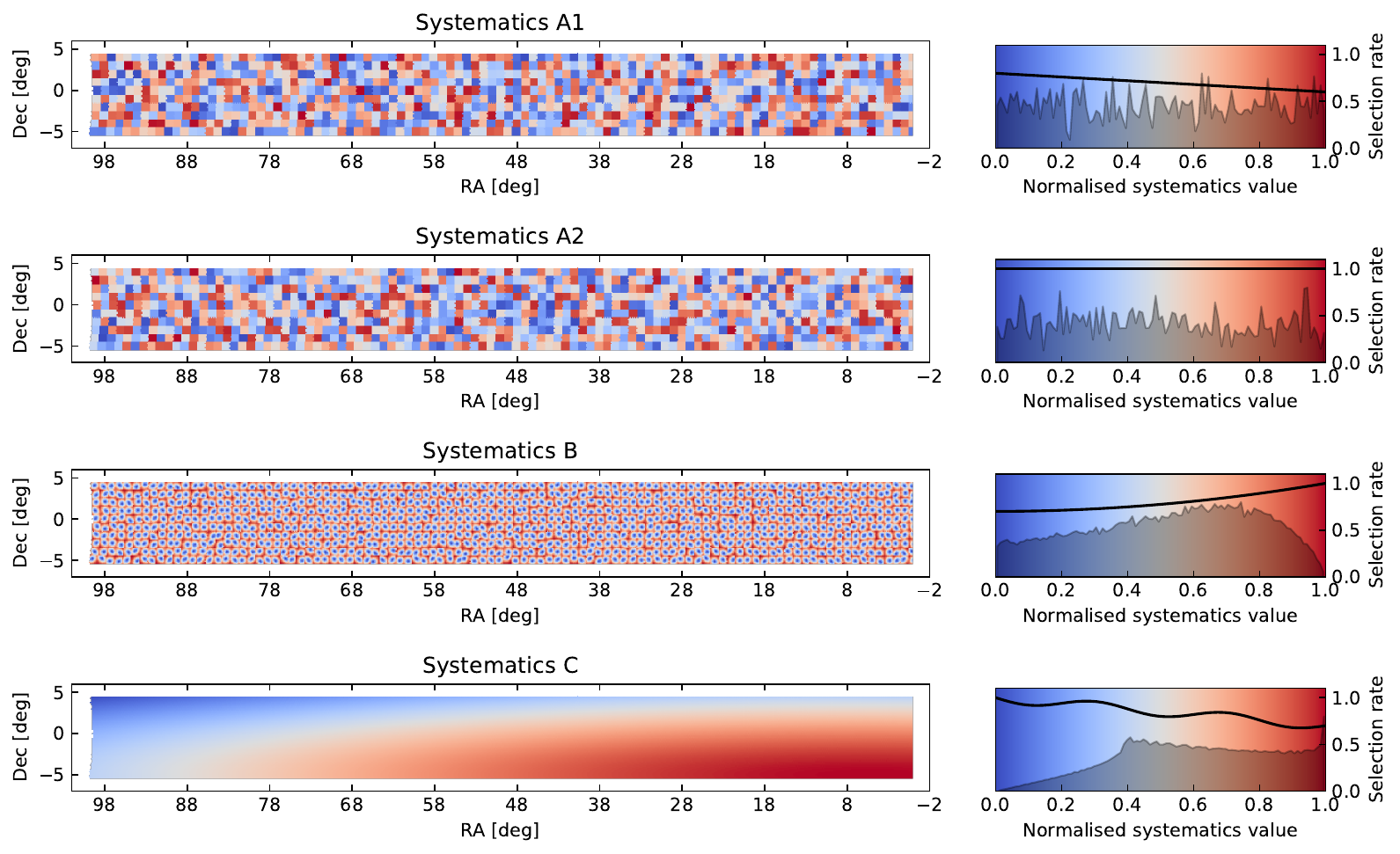}
    \caption{\textit{\ziangtxt{Left panels}}: Spatial distribution of the four toy systematics. \texttt{Systematics A1 and A2} are uniform in each tile, but differ across tiles  (type A); \texttt{Systematics B} varies within each tile (type B); \texttt{Systematics C} is tile-independent (type C);  \textit{\ziangtxt{Right panels}}: Black curves in the colour bar show the selection function of each systematic and the shaded region is the normalised distribution of the systematics. \ziangtxt{The numbers on the right show the selection rate values.}}
    \label{fig:toy_sys}
\end{figure*}

\begin{figure}[!htb]
    \centering
    \includegraphics[width=\columnwidth]{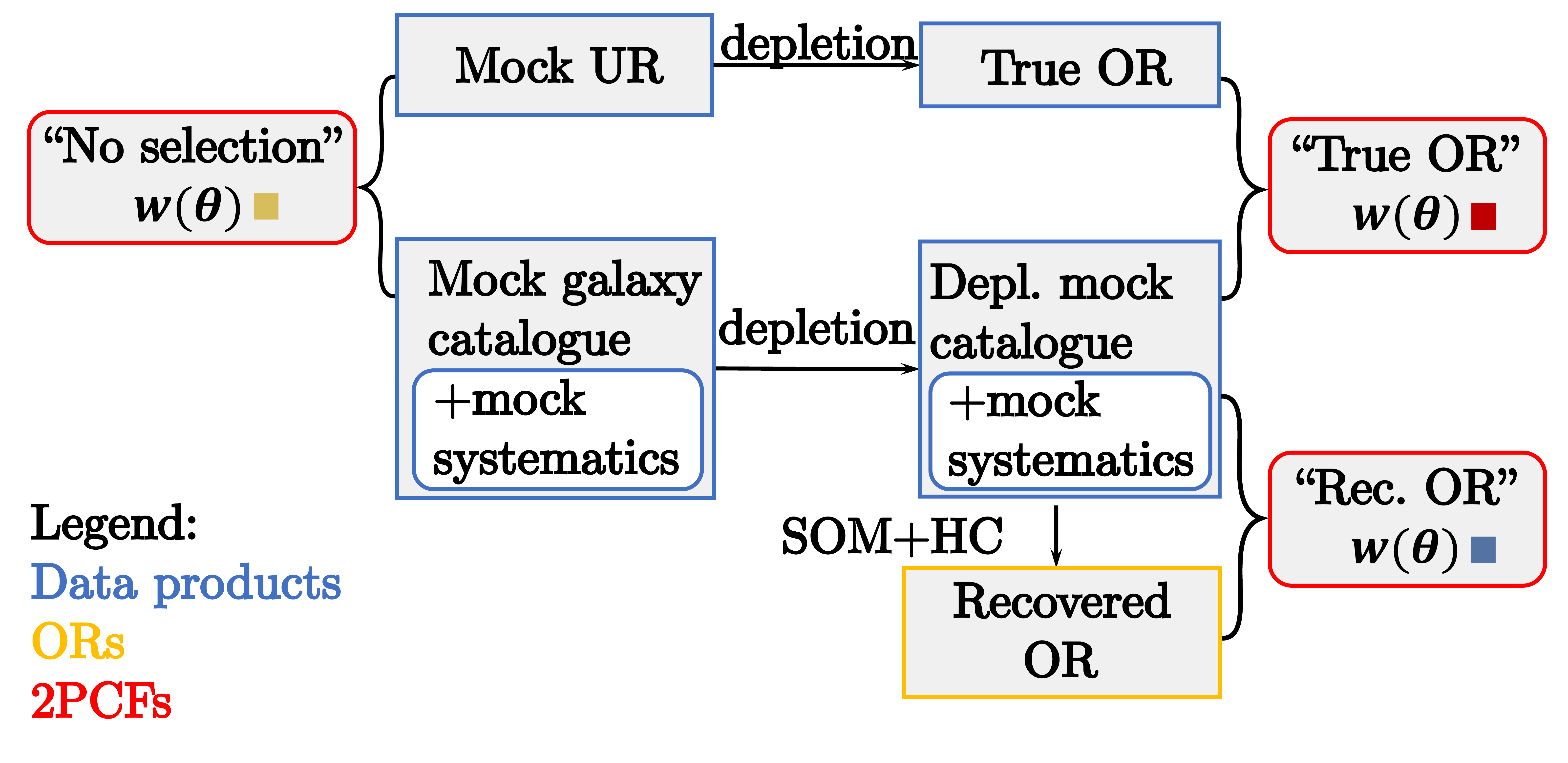}
    \caption{Flowchart of the toy model validation of the SOM+HC method. We note that the UR case ($w(\theta)$ calculated with depleted mock catalogue and mock UR) is not shown in this figure.}
    \label{fig:flowchart_validation}
\end{figure}

In this subsection, we present a validation test of the SOM+HC method on mock galaxy samples with variable depth induced by simple depletion functions of anisotropic systematics. Figure~\ref{fig:flowchart_validation} shows a flowchart of this test. The mock galaxy sample is generated by the \textsc{GLASS} package within a rectangular sky footprint with $0^{\circ}<\mathrm{RA}<100^{\circ}$; $-5^{\circ}<\mathrm{Dec}<5^{\circ}$ under the fiducial cosmology and a Gaussian redshift distribution centred at $z_{\mathrm{mean}}=0.3$ with a standard deviation of 0.1. Before assigning the systematics and applying the depletions, a ``point-source'' mask is applied to mimic the mask in the real observation. The point source mask is a high-resolution (\texttt{Nside}=8192, corresponding to a pixel size of 0.4 arcmin) binary mask, on which we generate a mock point source mask by removing circular holes (with value zero) on random positions with angular sizes of 5-15 arcmin. Galaxies that fall in the holes are removed from the mock catalogue. 

We then divide the footprint into 1000 $1^{\circ}\times1^{\circ}$ tiles. For each galaxy, we assign four systematics with different spatial distributions:

\begin{itemize}
    \item \texttt{Systematics A1} varies discretely across tiles but is constant within them, mimicking per-exposure effects such as limiting depth variations (e.g. background level, PSF size) resulting from the use of a step-and-stare observing strategy.
    \item \texttt{Systematics A2} is the same idea as \texttt{Systematics A1}, but with a different realisation.

    \item \texttt{Systematics B} mimics telescope and camera effects such as PSF shape variations over the focal plane, so in each tile it depends roughly on the angular distance to the tile centre. We take 2-dimensional Gaussian functions in each tile as the \texttt{Systematics B} distribution. The centre of the Gaussian is close to each tile centre with a small random jitter; the covariances are close to diagonal (hence the Gaussian has small ellipticities).
    \item \texttt{Systematics C} varies smoothly over large angles. We model it as a Gaussian function of Galactic latitude. It mimics large-scale variations such as the Galactic foreground.
\end{itemize}

These four systematics are corresponding to type A (\texttt{Systematics A1} and \texttt{Systematics A2}), type B (\texttt{Systematics B}), and type C (\texttt{Systematics C}) systematics introduced in Sect. \ref{sect:data}. Their values are normalised to be between 0 and 1. Since SOM+HC does not change the topology of the galaxy distributions in systematics space, this normalisation does not affect the performance of the method.

We define simple selection functions for \texttt{Systematics A1} (a linear function), \texttt{Systematics B} (a quadratic function), and \texttt{Systematics C} (a trigonometric function modulated by a linear function). \texttt{Systematics A2} has no selection effect and thus acts as a ``distractor'' for the SOM. We show the systematics map and the associated selection function in Fig.~\ref{fig:toy_sys}. We also assume that the selections between the systematics are independent so that the overall selection function is a multiplication of the selection functions of the individual systematics. To make the selection, we generate a uniform random number between 0 and 1 for each galaxy, and if the random number is smaller than the selection function $P_{\mathrm{keep}}(\boldsymbol{\theta})$ at its position, the galaxy is kept in the catalogue. After selection, we have a galaxy catalogue with an angular number density of $\sim$1 arcmin$^{-2}$.

\begin{figure}
    \centering
    \includegraphics[width=\linewidth]{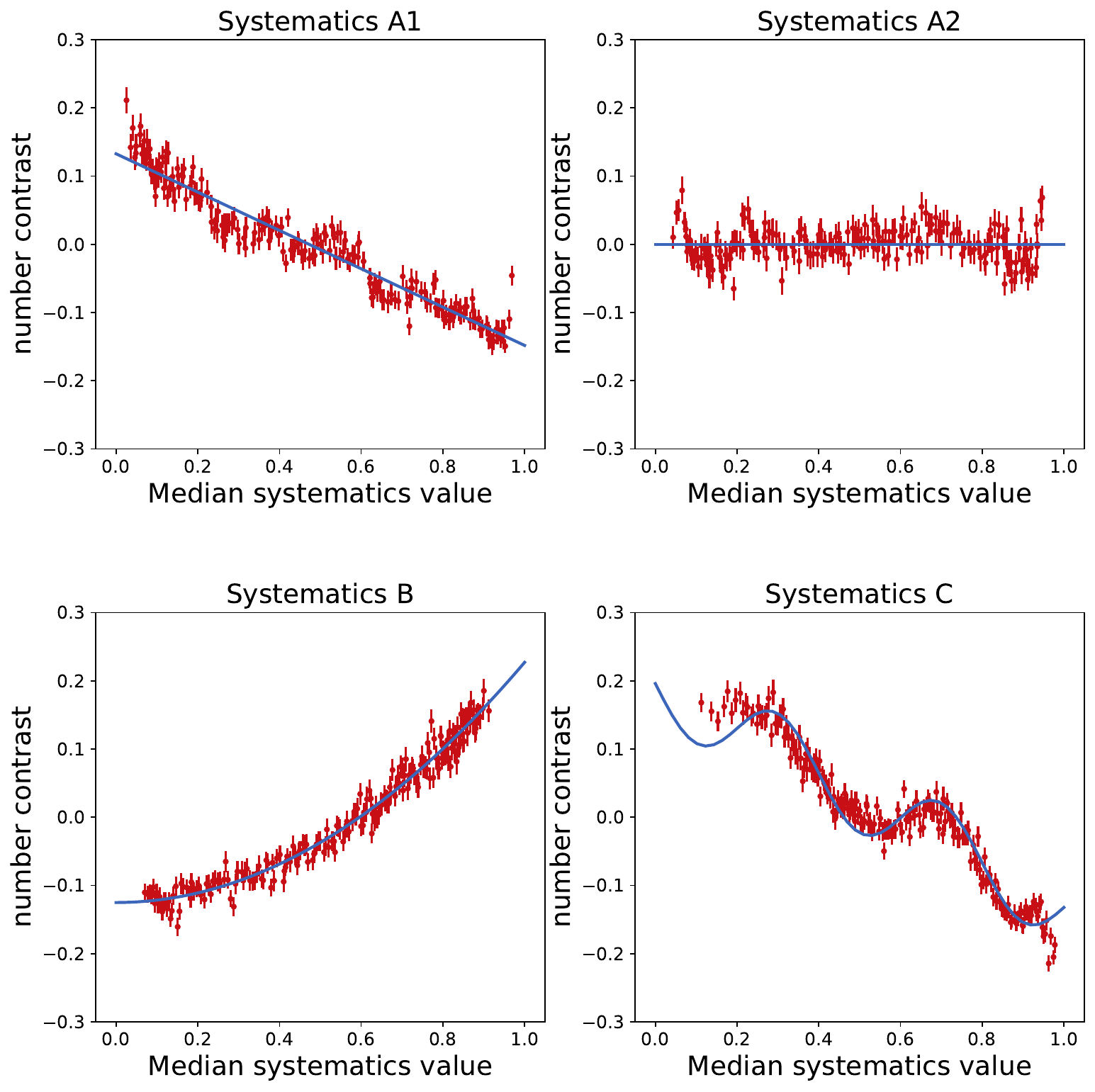}
    \caption{Relationship between the galaxy number contrast and the mean systematic value of each hierarchical cluster. The blue curves are the number contrast derived from the input selection rates (black curves in the colour bars of Fig. \ref{fig:toy_sys}). The average median systematics and number contrast are calculated by averaging the sorted values in each realisation across all the realisations. The standard errors are also calculated and presented as error bars. The errors of the median systematics are too small to be visible. }
    \label{fig:rec_selection}
\end{figure}

We also generate a ``true'' OR catalogue by applying the same selection functions to a UR catalogue in the same footprint. If we use it as the $R$ terms for the 2PCF measurement with the depleted mock catalogue, we should get an unbiased $w(\theta)$. We note that in reality, we do not have access to the true OR. In this validation test, we recover the OR with a 30$\times$30 SOM with hexagonal cells and a toroidal topology (so that the horizontal and vertical edges are adjacent), grouped into 200 hierarchical clusters. To check whether the SOM+HC recovers the input selection functions, we plot the relationship between galaxy number contrast (relative difference between galaxy number in each cluster and the mean galaxy number across clusters) and median systematics values of each cluster in Fig. \ref{fig:rec_selection}. The blue curves are the number contrast derived from the input selection rates (the black curves in the colour bars of Fig. \ref{fig:toy_sys}). The median systematics and number contrast for each point are calculated by first sorting the median systematics for all clusters in a realisation, then averaging the median systematics and number contrasts across realisations. Standard errors are also calculated and presented as error bars. The errors of the median systematics are too small to be visible. The galaxy number contrasts of the HCs generally follow the input selection rate, indicating that SOM+HC is able to recover the input selection functions of each systematics. Notably, SOM+HC will not introduce additional correlation for \texttt{Systematics A2}, which does not select galaxies. We also note that the number contrast slightly mismatches the selection functions at the edges. This is probably because the SOM is less effective closer to the boundary in the systematics space, where there are fewer neighbouring galaxies.

\begin{figure}
    \centering
    \includegraphics[width=\columnwidth]{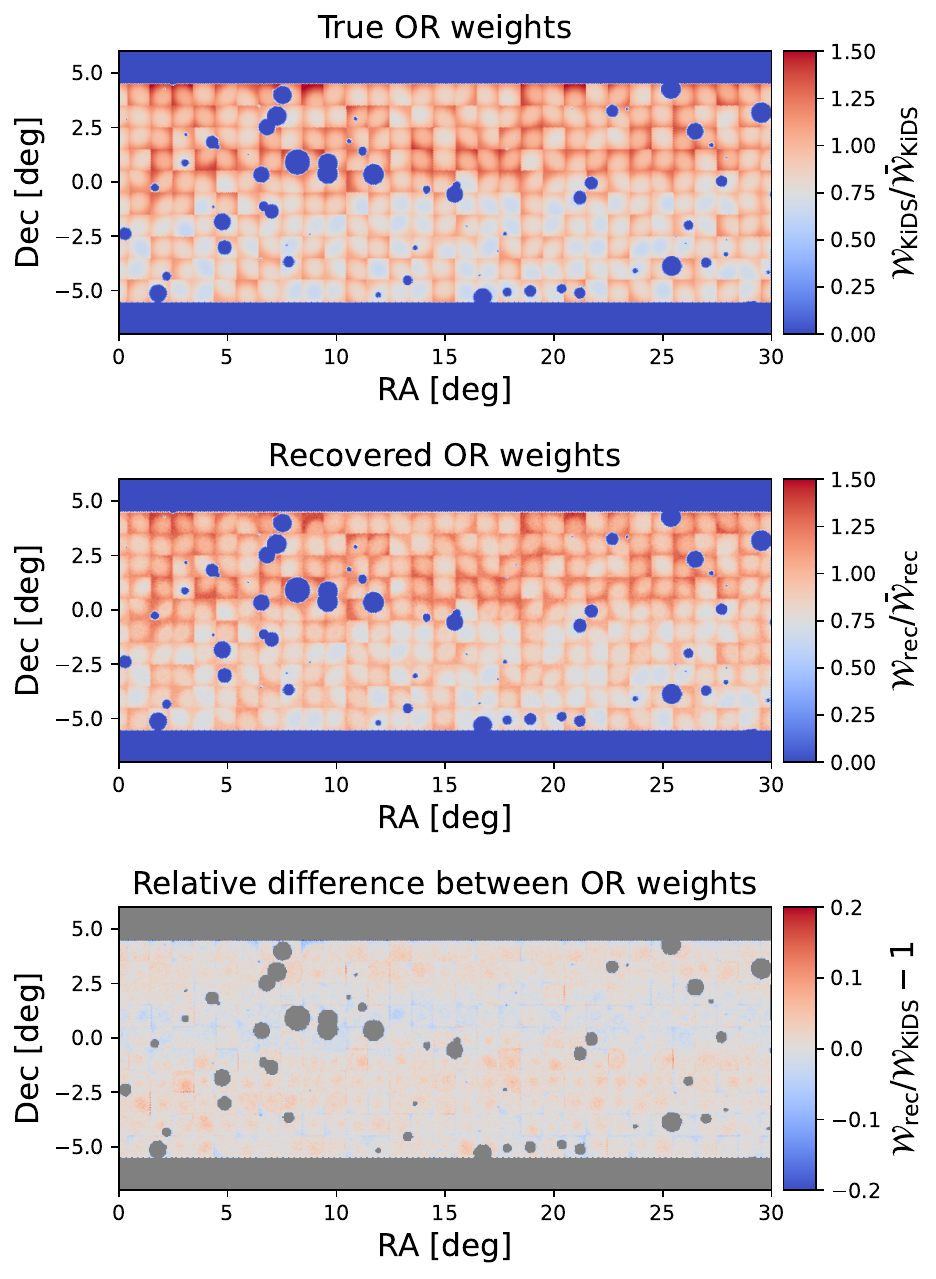}
    \caption{\textit{Top panel}: True OR weights (normalised by their mean) calculated from the total selection function of the toy systematics; \textit{middle panel}: Recovered OR weights generated by the SOM+HC method. Both panels show only part of the footprint. We note that the holes in the maps are masked regions around point sources; \textit{bottom panel}: Relative difference between the recovered OR weights and the true OR weights.}
    \label{fig:toy_or_map}
\end{figure}

The recovered OR weight map is pixelised into a \textsc{HEALPix} map with \texttt{Nside}=1024 \footnote{\ziangtxt{The \texttt{Nside} used here is lower than the optimal \texttt{Nside} for the real data because we generate fewer mock galaxies in this test, and \texttt{Nside}=1024 ensures that each pixel has $\sim$10 galaxies.}}. The true OR weight map, the recovered OR weight map, and the relative difference are shown in Fig.~\ref{fig:toy_or_map}. Visually, the recovered OR shows the same spatial pattern as the true OR, but it appears more discrete. This is due to a finite number of clusters in the systematics space (in other words, if we had an infinite number of galaxies grouped into an infinite number of clusters, we would recover the smooth OR weights). The relative difference (the bottom panel of Fig.~\ref{fig:toy_or_map}) is well within $\sim\pm20\%$. 

In summary, we have an unselected mock galaxy catalogue and a selected catalogue, plus a uniform random, a true organised random and a recovered organised random. To evaluate the SOM+HC method quantitatively, we measure four $w(\theta)$'s from them. They are summarised in Table~\ref{table:w_theta_validate}. In summary, we validate the method by comparing the ``Recovered OR'' and the ``No selection'' 2PCFs. The 2PCF is measured in 20 angular bins between 2.5 and 250 arcmin \citep[following][]{DeRose_2022} with the \textsc{TreeCorr}\footnote{\href{https://rmjarvis.github.io/TreeCorr/_build/html/index.html}{https://rmjarvis.github.io/TreeCorr/\_build/html/index.html}} package \citep{2015ascl.soft08007J}. To evaluate the covariance matrix and reduce the sample variance, we perform the above validation on 40 \textsc{GLASS} realisations and evaluate the covariance as

\begin{equation}
    \tens{C}_{ij} = \frac{1}{40-1} \sum_{r=1}^{40} {\Big( w_r(\theta_i)-\bar{w}(\theta_i) \Big) }{\Big( w_r(\theta_j)-\bar{w}(\theta_j) \Big)}\,,
    \label{eq:covariance}
\end{equation}
where the subscript $r$ denotes the realisation number, and $\bar{w}(\theta)$ is the average 2PCF over realisations. We also compute a theoretical covariance matrix using the \textsc{OneCovariance} code \citep[][]{reischke2024kidslegacycovariancevalidationunified}\footnote{\href{https://github.com/rreischke/OneCovariance}{https://github.com/rreischke/OneCovariance}} and find that it is consistent with the ``No selection'' covariance matrix (see Appendix \ref{sect:app_covmat} for a comparison). We evaluate the bias of the 2PCFs by calculating the $\chi^2$ between each $w(\theta)$ and the ``No selection'' $w(\theta)$

\begin{equation}
\begin{aligned}
       \chi_{\mathrm{d}}^2 &=\Delta\bar{\boldsymbol{w}}^{T}\tens{C}^{-1}\Delta\bar{\boldsymbol{w}}\,,
\end{aligned}
\label{eq:chi2_d}
\end{equation}
where $\Delta\bar{\boldsymbol{w}}$ is the difference between the 2PCF and the ``No selection'' 2PCF. Assuming $\chi_{\mathrm{d}}^2$ follows a $\chi^2$ distribution with degrees-of-freedom equal to the number of angular bins taken into account, we can calculate the corresponding probability-to-exceed (PTE) value. The 2PCF is less biased if the PTE value is closer to 1.

\begin{figure}
    \centering
    \includegraphics[width=\columnwidth]{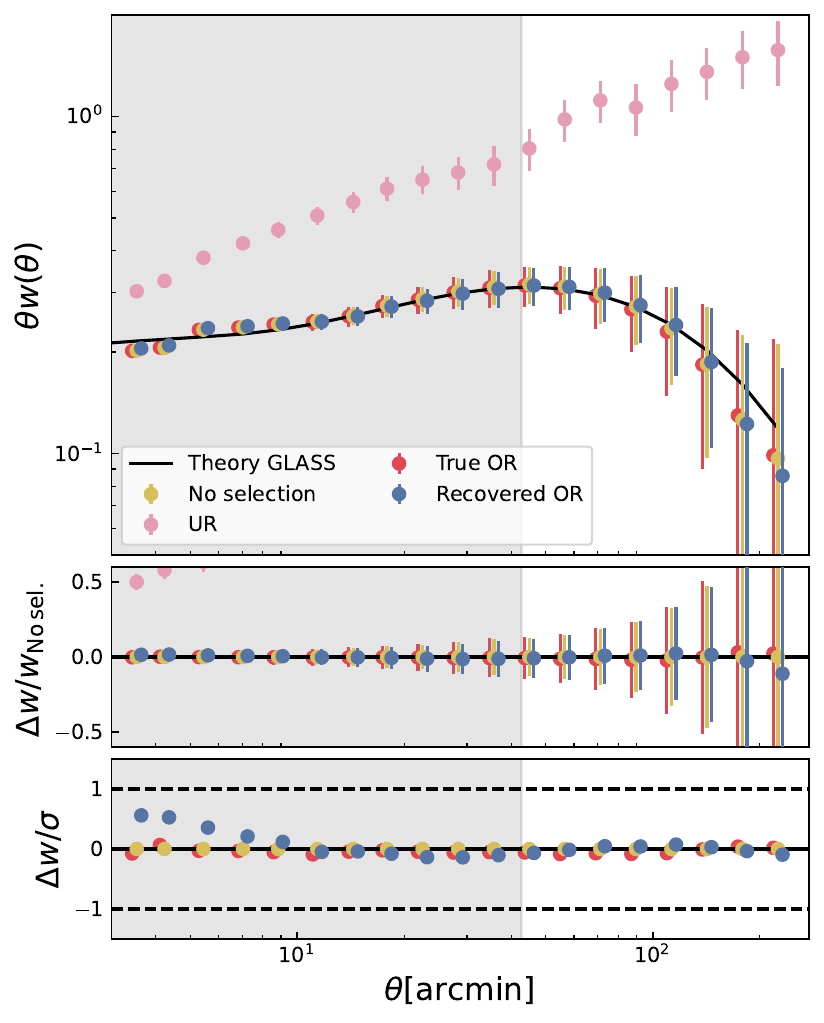}
    \caption{\textit{Top panel}: Measured $w(\theta)$ in the toy-systematics test. The definitions of the four $w(\theta)$ are presented in Table~\ref{table:w_theta_validate}. The data points are the mean $w(\theta)$ from 40 realisations and the error bars are the diagonal elements of the covariance matrices evaluated from the realisations. The black curve is the theoretical $w(\theta)$ computed with \textsc{PYCCL} \citep[][]{Chisari_2019} using the same cosmology and redshift distribution. The shaded region shows the angular scale smaller than 8$h^{-1}\mathrm{Mpc}$ evaluated at the mean redshift. \ziangtxt{The middle panel is the relative bias of each $w(\theta)$ with respect to the No selection case, and the bottom panel is the $w(\theta)$ bias related to the error. Most points of the UR case are drastically biased and are outside the range of the middle and bottom panels.}}
    \label{fig:wtheta_validation}
\end{figure}

The measured 2PCFs and their relative difference to the ``No selection'' case are shown in Fig.~\ref{fig:wtheta_validation}. In this work, we focus on linear scales, so we calculate $\chi_\mathrm{d}$ only on the angular scale $\theta<\theta_{\mathrm{cut}}$ where the cutting scale $\theta_{\mathrm{cut}}$ is the angular scale corresponding to a physical scale of $8h^{-1}\mathrm{Mpc}$ at the mean redshift of the mock sample ($\bar{z}=0.3$ in our case), which has a value of 42.74 arcmin. For the UR case, $\chi_\mathrm{d}=1616$, meaning a huge bias if we use uniform random for the depleted catalogue. For the true OR case, $\chi_\mathrm{d}=0.08$ (corresponding to a PTE=0.99999987), meaning that true organised random can give an unbiased $w(\theta)$ as expected. For the ``Recovered OR'' case, $\chi_\mathrm{d}=0.24$ (corresponding to a PTE=0.999992). This means that the OR recovered by the SOM+HC method also gives unbiased $w(\theta)$. \cite{harryOR} have shown that SOM+HC can correct a slight variable depth bias in the  KiDS-Bright sample. Notably, in this test, SOM+HC can correct $w(\theta)$ that is biased by orders of magnitude. It should also be noted that the wiggle at small scales is due to pixelisation, which changes the angular distance between galaxies as they are effectively moved to the centre of the pixel. This effect particularly affects scales close to the pixel size (3.4 arcmin in our case).

\begin{table*}[!htb]
 \renewcommand{\arraystretch}{1.25}
 \caption{Information for the four $w(\theta)$ measured from the mock sample. }
    \centering
    \begin{tabular}{llllc}
    \toprule
    Galaxy catalogue & Random catalogue & Label & Biased? & Colour \\
    \hline
    Unselected catalogue & UR & ``No selection'' & unbiased & \textcolor{noselection}{$\blacksquare$} \\
    Selected catalogue & UR & ``UR'' & biased & \textcolor{ur}{$\blacksquare$} \\
    Selected catalogue & True OR & ``True OR'' & unbiased & \textcolor{trueor}{$\blacksquare$} \\
    Selected catalogue & Recovered OR & ``Rec. OR'' & unbiased & \textcolor{recor}{$\blacksquare$} \\
    \bottomrule
    \end{tabular}
    \tablefoot{The `Label' column indicates the aliases of the $w(\theta)$ values in the following discussion and figures; the `Biased?' column indicates whether the 2PCF is biased; the `Colour' column indicates the colours of the data points shown in the figures. The `No selection' (first row) and `True OR' (third row) cases are unbiased benchmarks and the `UR' (second row) case is a biased measurement. The `Recovered OR' case (fourth row) is the 2PCF that we used to validate our method.}
    \label{table:w_theta_validate}
\end{table*}

\subsection{Validation with data-driven systematics}

%\tabrealmock
\begin{figure}[!htb]
    \centering
    \includegraphics[width=\columnwidth]{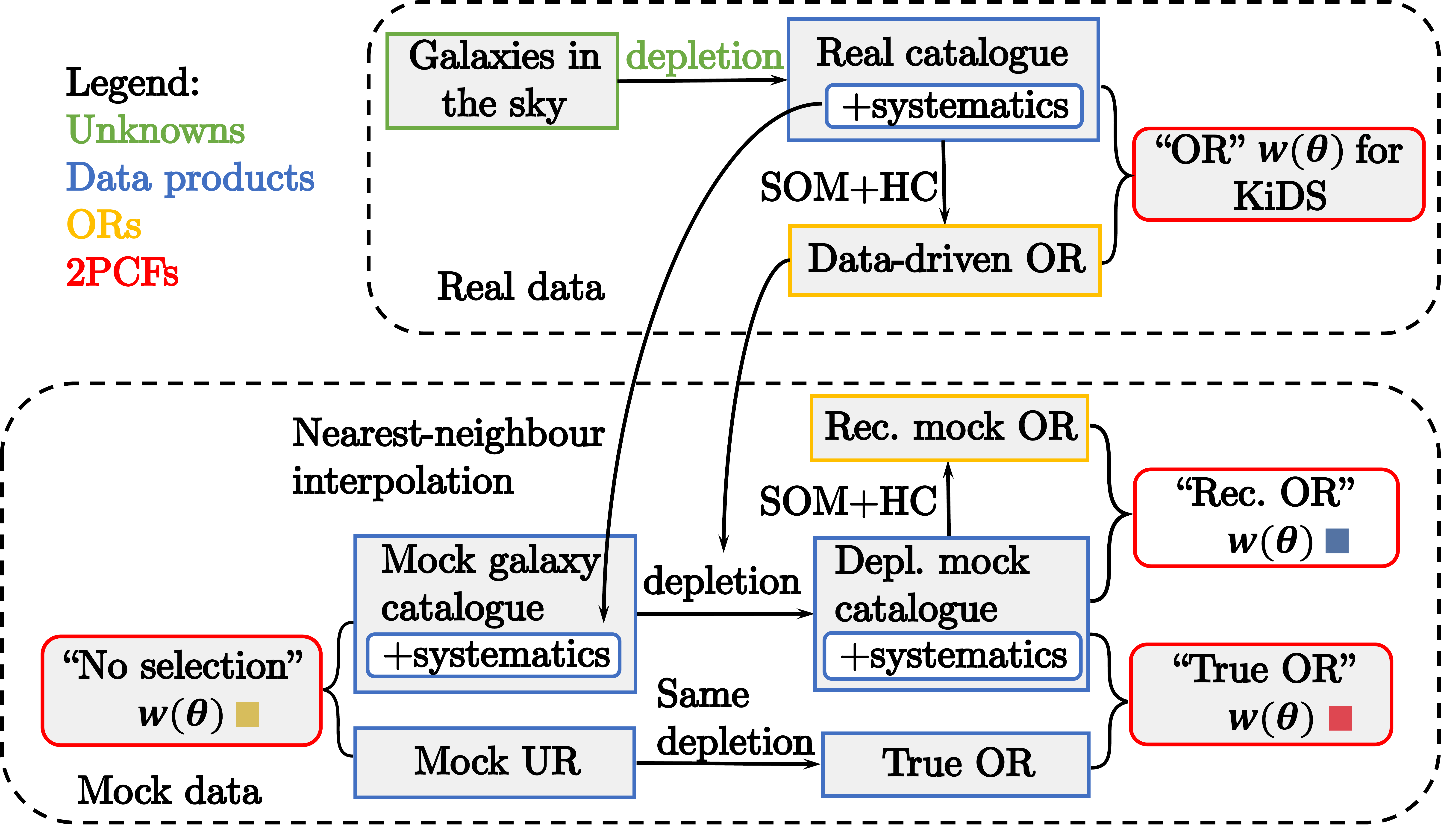}
    \caption{Flowchart of the data-driven test.}
    \label{fig:or_datadriven_flowchart}
\end{figure}

\subsubsection{Introduction to the method}

The previous test with simulated systematics and depletions proves that the SOM+HC can correct for spatially variable depth, but the test is oversimplified and not realistic. In reality, the spatial distribution of systematics can be more complicated, stochastic and correlated. In addition, the selection function can be arbitrary. Therefore, it is difficult to create mocks with realistic variable depth to test the SOM+HC method. In addition, the SOM+HC method might over-correct the 2PCF by obtaining organised randoms ``contaminated'' by the LSS. 

To tackle this problem, we apply ``data-driven systematics'' to the mock catalogue following \citet{harryOR}. \ziangtxt{First, we generate a mock galaxy catalogue following the fiducial cosmology given in Sect. \ref{sect:intro} with a galaxy bias $b=1$. The galaxy sample follows the same calibrated redshift distribution of the KiDS-Legacy sample \footnote{If the variable depth does not encode cosmological information, it is not necessary to use a realistic cosmology and redshift distribution to generate the mock.} and is generated within the KiDS-Legacy footprint}; then we generate the ``data-driven mock systematics'' by assigning systematics values to each mock galaxy with a nearest neighbour interpolation from the real KiDS systematics. Thus, we get a mock galaxy sample with the same spatial distributions of systematics as the real data. We note that even if the spatial distributions of the systematics are correlated with the LSS, this step removes any such correlation because the mock galaxy catalogue has a different spatial distribution. Next, we train a SOM+HC on the real galaxy catalogue (we call it ``SOM+HC+KiDS'') and get an OR weight map $\mathcal{W}_{\mathrm{KiDS}}$ as a proxy of the realistic variable depth (we call it the ``data-driven OR weight''); then we select mock galaxies according to it. We define the ``OR contrast'' as

\begin{equation}
    \delta_{\mathrm{OR,KiDS}}(\boldsymbol{\theta})\coloneqq \frac{\mathcal{W}_{\mathrm{KiDS}}(\boldsymbol{\theta})-\bar{\mathcal{W}}_{\mathrm{KiDS}}}{\bar{\mathcal{W}}_{\mathrm{KiDS}}}\,,
\end{equation}
where $\bar{\mathcal{W}}_{\mathrm{KiDS}}$ is the average OR weight across the footprint. In practice, we generate a large mock catalogue and select galaxies according to the selection function given by $\delta_{\mathrm{OR,KiDS}}$:

\begin{equation}
    P_{\mathrm{keep}}(\boldsymbol{\theta}) = \frac{N_{\mathrm{out}}}{N_{\mathrm{in}}}\left[ 1+\delta_{\mathrm{OR,KiDS}}(\boldsymbol{\theta})\right]\, .
\end{equation}
Here $N_{\mathrm{out}}$ is the desired galaxy number of the depleted mock sample. In this work, we choose $N_{\mathrm{out}}=49,875,861$, the galaxy number in the KiDS-Legacy catalogue. The galaxy number of the input, unselected mock sample $N_{\mathrm{in}}$ is chosen to ensure $P_{\mathrm{keep}}\leq1$. The selection procedure is the same as described in Sect.~\ref{sect:toy_sys}. For each galaxy we generate a uniform random number between 0 and 1 and compare it with $P_{\mathrm{keep}}$ associated with the pixel containing the galaxy. If the random number is less than $P_{\mathrm{keep}}$, we keep the galaxy in the sample; otherwise, we discard it.

The assignment of mock systematics with nearest-neighbour interpolation ensures that the spatial distribution of the mock systematics is realistic. If SOM+HC recovers an unbiased OR weight on the real data, the data-driven OR weight should represent the realistic variable depth. When applied as a selection function to the mock data, it should produce the variable depth caused by the realistic selection function of the mock systematics. We run SOM+HC on the post-selected mock catalogue with the mock systematics to test this consistency. Ideally, the resulting OR weight should match the imported selections (i.e. the data-driven OR weight), and we expect to measure an unbiased 2PCF from the selected mock catalogue corrected with the recovered mock OR. Figure~\ref{fig:or_datadriven_flowchart} shows the flowchart of the data-driven systematics test.

With this test we can also check whether the recovered OR weight contains the LSS. If so, the data-driven OR weight would have an imprint of the LSS from our real Universe, while the mock OR weight should have the imprint of the mock LSS. These two OR weights would not match, and the mock 2PCF corrected by the mock OR weight should be over-corrected.

Several configuration parameters affect the accuracy of the recovered OR weight. In particular, the SOM+HC procedure depends on $NC$, the number of hierarchical clusters, and \texttt{Nside}, the OR weight resolution. If $NC$ and/or \texttt{Nside} is too low, SOM+HC will fail to resolve systematic clustering in systematic/real space, resulting in uncorrected variable depth. If $NC$ is too high, then SOM+HC will start to resolve LSS-induced clustering in systematics space. If \texttt{Nside} is so high that significant amounts of pixels get no galaxy, then the OR weight will follow the LSS. In this section, we aim to perform data-driven systematics tests to find the optimal \{$NC$, \texttt{Nside}\}. We note that the data-driven systematics test depends on these parameters for both data-driven and recovered mock OR. {In the following discussion, we use \{$NC_{\mathrm{KiDS}}$, $\texttt{Nside}_{\mathrm{KiDS}}$\} to denote the set-up for the data-driven OR, and \{$NC_{\mathrm{rec}}$, $\texttt{Nside}_{\mathrm{rec}}$\} for the recovered mock OR.} To avoid endless tests for different parameter value combinations, we assume that the same SOM+HC set-ups on real and mock data, with the same systematics choices, will yield equally good or bad performance on both real and mock data. With this assumption in mind, we can evaluate the performance of a $\{NC_{\mathrm{KiDS}}, \texttt{Nside}_{\mathrm{KiDS}}\}$ combination based on the recovered OR 2PCF {with $NC_{\mathrm{KiDS}}=NC_{\mathrm{rec}}, \texttt{Nside}_{\mathrm{KiDS}} = \texttt{Nside}_{\mathrm{rec}}$}:

\begin{enumerate}
    \item If the recovered OR 2PCF is generally higher than the ``No selection'' case, the $\{NC_{\mathrm{KiDS}},\texttt{Nside}_{\mathrm{KiDS}}\}$ choice is likely to under-estimate the variable depth;
    \item If the recovered OR 2PCF is significantly lower than the ``No selection'' case, the $\{NC_{\mathrm{KiDS}},\texttt{Nside}_{\mathrm{KiDS}}\}$ option is likely to over-correct the variable depth by removing the LSS;
    \item If the recovered OR 2PCF agrees with the ``No selection'' case, it is possible that the choice is optimal. It is also possible that the data-driven OR under-estimates the variable depth and induces too soft selections, which can be corrected by the same $\{NC_{\mathrm{rec}}, \texttt{Nside}_{\mathrm{rec}}\}$ combination, while a realistic variable depth actually requires higher $\{NC_{\mathrm{KiDS}},\texttt{Nside}_{\mathrm{KiDS}}\}$. For the second case, one can think of an extreme case: $NC_{\mathrm{KiDS}}=NC_{\mathrm{rec}}=1$. This case is the same as ``No selection'' and will give an unbiased 2PCF, but the number of clusters is clearly too low for the real data with complex selection effects.
\end{enumerate}

Another consideration is that instead of calculating 2PCF with varying \texttt{Nside}, we calculate the fraction of unmasked empty pixels (i.e. the pixels in the footprint that do not contain a galaxy). If the fraction is large, the OR weight will show the LSS pattern (one can imagine an extreme case: using a pixel size so small that each pixel contains either one or zero galaxies. In this case, the OR weight map is exactly the galaxy map and no 2PCF signal is measured with it). We find that when \texttt{Nside}$\leq$2048, the unmasked empty pixels are less than 0.5\% of the footprint; when $\texttt{Nside}=4096$, the fraction increases to 2\%. Therefore, in this test we fix $\texttt{Nside}_{\mathrm{KiDS}}=\texttt{Nside}_{\mathrm{rec}}=2048$ and only vary $NC$. 

With these assumptions and simplifications in mind, we find the optimal $NC_{\mathrm{KiDS}}$ by first choosing the same $NC_{\mathrm{KiDS}}$ and $NC_{\mathrm{rec}}$ values for data-driven OR and recovered mock OR, respectively. We vary the cluster number from low to high until it is high enough to pick up the LSS and the recovered 2PCF starts to be systematically lower than the unbiased case. On the other hand, if a certain choice of $NC$ gives an unbiased mock 2PCF, it is still possible that $NC_{\mathrm{KiDS}}$ is too low to recover the selection function, which results in a soft data-driven OR map that can be recovered with the same $NC_{\mathrm{rec}}=NC_{\mathrm{{KiDS}}}$. To test if this is the case, we can manually amplify the variance of the data-driven selection function that applies to the mock by

\begin{equation}
    \mathcal{\delta}^{(m)}_{\mathrm{OR, KiDS}}(\boldsymbol{\theta}) \equiv  \frac{\mathcal{W}^{m}_{\mathrm{KiDS}}(\boldsymbol{\theta})}{\overline{\mathcal{W}^{m}_{\mathrm{KiDS}}}}-1\,.
    \label{eq:enhanced_wtheta}
\end{equation}
We note that $\mathcal{W}^{m}_{\mathrm{KiDS}}$ is $\mathcal{W}_{\mathrm{KiDS}}$ to the power of $m$. When the adjustment parameter $m>1$, the variance in $\delta_{\mathrm{OR, KiDS}}$ is enhanced. If the mock OR can still recover an unbiased 2PCF, then we can conclude that this set-up is powerful enough to recover the realistic variable depth. In this test, we choose $m=1.5$ so that the bias in the 2PCF with URs is almost doubled.

If we find the $NC_{\mathrm{KiDS}}$ so that the ``Recovered OR'' 2PCF is also unbiased for the enhanced data-driven selection, we can fix it for the data-driven OR weight. Based on our previous discussion, we can treat the data-driven OR weight with this $NC_{\mathrm{KiDS}}$ as a realistic selection function. With this data-driven selection, we then vary $NC_{\mathrm{rec}}$ for the mock OR and evaluate the bias of the resulting 2PCF. 

\subsubsection{Evaluating the performance of SOM+HC with data-driven systematics}

\label{sect:eva_datadriven}

\begin{figure}
    \centering
    \includegraphics[width=\linewidth]{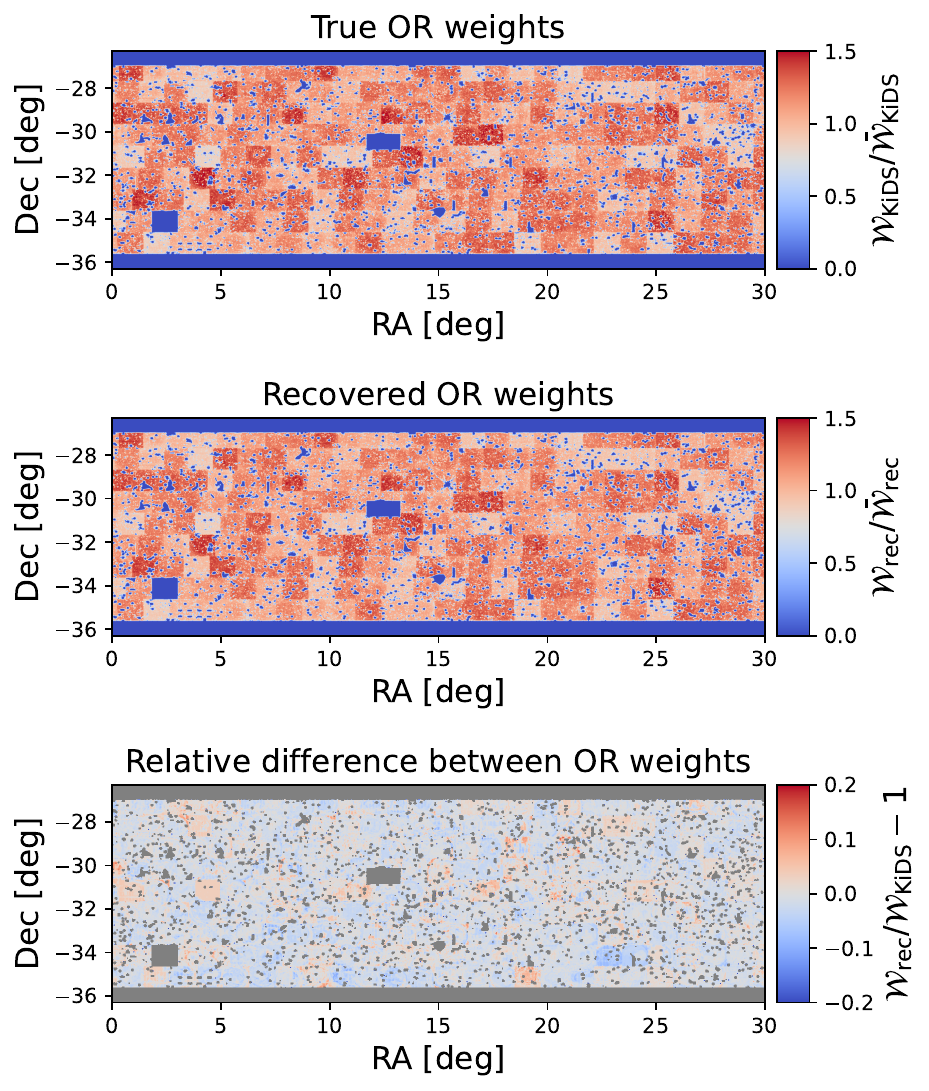}
    \caption{{{\textit{Upper panel}}: Part of the data-driven OR weight generated from the KiDS-Legacy map; {\textit{middle panel}}: Mock OR weight recovered from \textsc{GLASS} mock galaxy sample selected according to the data-driven OR; \textit{bottom panel}: Relative difference between recovered OR weights and true OR weights. Both OR weights are generated with 600 hierarchical clusters and pixelised on a \textsc{HEALPix} map with \texttt{Nside}=2048.}}
    \label{fig:wmap_datadriven}
\end{figure}

Figure \ref{fig:wmap_datadriven} shows an example of the true data-driven OR (the top panel), associated recovered OR (the bottom panel), and the relative difference between them. Both OR weights are generated with 600 hierarchical clusters and pixelised on a \textsc{HEALPix} map with \texttt{Nside}=2048. Visually they look very similar to each other, and the relative difference is well within $\sim\pm20\%$.

Following Sec.~\ref{sect:toy_sys}, we quantitatively evaluate the effectiveness of SOM+HC by comparing the 2PCF from the depleted mock catalogue corrected by the recovered mock OR (the ``Rec. OR $w(\theta)$'') and the ``No selection'' 2PCF (measured from a complete catalogue with the same number of galaxies, corrected by a uniform random). Since the ``No selection'' 2PCF is unbiased, it serves as the reference 2PCF. The consistency between these two 2PCFs indicates the consistency between the recovered mock OR and the data-driven OR (not necessarily between the recovered mock OR and the realistic selections). The angular separation is binned into 20 logarithmic bins from 2.5 arcmin to 250 arcmin. The covariance matrices are again calculated by the \textsc{OneCovariance} code with the same cosmology, redshift distribution, sky coverage and galaxy number density as the \textsc{GLASS} mock (see Fig. \ref{fig:wtheta_datadriven}).

To achieve satisfactory statistical power and to suppress sample variance, we perform the data-driven systematics test on 40 \textsc{GLASS} realisation samples and calculate the average $w(\theta)$. At the data vector level, we compute the $\chi_\mathrm{d}^2$ value between the recovered OR $w(\theta)$ and the no-selection $w(\theta)$ following Eq.~\eqref{eq:chi2_d}. We also use the PTE value to evaluate the goodness of fit. If the PTE is close to 1, then it is almost certain that $\chi^2$ will be larger than a completely random variation, and the difference between ``Recovered OR'' 2PCF and ``No selection'' 2PCF is significantly smaller than the sample variance.

We can also evaluate the consistency at the level of parameter fits. For example, we set the matter density parameter $\Omega_{\mathrm{m}}$ and the galaxy bias $b$ as free parameters. First, we define a Gaussian likelihood,

\begin{equation}
\begin{aligned}
     &\log \mathcal{L}(w(\theta)| \Omega_{\mathrm{m}}, b) = -\frac{\chi_{\mathrm{dm}}^2}{2} \\
     &= -\frac{1}{2}\sum_{i,j} \Big( \hat{w}(\theta_i)- w(\theta_i; \Omega_{\mathrm{m}}, b)\Big) \,\tens{C}^{-1}_{ij}  \left( \hat{w}(\theta_j)- w(\theta_j; \Omega_{\mathrm{m}}, b)\right)\,,
\end{aligned}
\end{equation}
where $\chi_{\mathrm{dm}}^2$ denotes the $\chi^2$ between data and model; $\hat{w}$ is the measured 2PCF averaged across realisations and $w(\theta_i; \Omega_{\mathrm{m}}, b)$ is the theoretical 2PCF. We run an MCMC using the \textsc{emcee} package \citep{2013PASP..125..306F} to sample the posterior distribution. We then compare the consistency between the MCMC chains from the ``No selection'' and ``Recovered OR'' cases. The inference bias of each parameter is described by $\Delta \bar{\Omega}_{\mathrm{m}}$ and $\Delta \bar{b}$, the difference of the mean values of the parameters across the converged chains, with the units of their standard deviations. We also compare the entire posterior with the distance of the mean values of $\{\Omega_{\mathrm{m}}, b\}$ on the parameter plane parametrised with the $\chi^{2}$ value on the ${\Omega}_{\mathrm{m}}-b$ plane:

\begin{equation}
    \chi^{2}_{{\Omega}_{\mathrm{m}},b} =
    \begin{pmatrix}
    \Delta \bar{\Omega}_{\mathrm{m}}\\
    \Delta \bar{b}
    \end{pmatrix}^{\rm{T}}\,
    \tens{C}^{-1}_{\Omega_{\mathrm{m}}, b}\,
    \begin{pmatrix}
    \Delta \bar{\Omega}_{\mathrm{m}}\\
    \Delta \bar{b}
    \end{pmatrix}\,.
    \label{eq:cosmological_parameter_bias}
\end{equation}
Here $\tens{C}$ is the covariance matrix of the inferred parameters derived from the MCMC chains. The PTE value is then derived from the $\chi^2$ value given the degrees of freedom of 2.

\subsection{Results}
\begin{figure*}
    \centering
    \includegraphics[width=\linewidth]{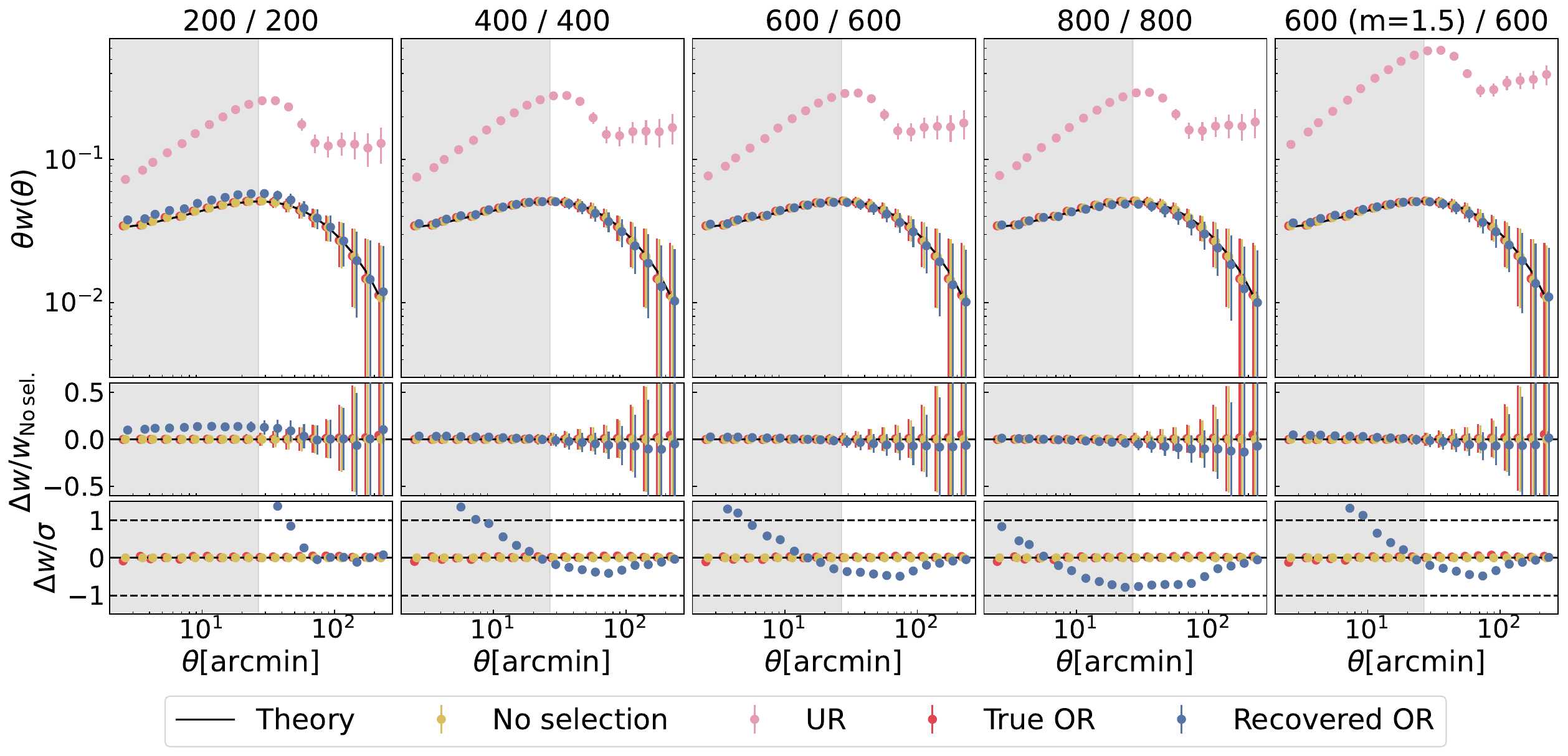}
    \caption{2PCFs measured for the data-driven systematics test with the same choices of $NC_{\mathrm{KiDS}}$ (the number of hierarchical clusters for the data-driven OR) and $NC_{\mathrm{rec}}$ (for the recovered mock OR) Each column of panels corresponds to one set-up. \texttt{Nside} is fixed at 2048 for both the data-driven OR and the recovered OR. The right-most column is an enhanced data-driven OR with $m=1.5$ according to Eq.~\eqref{eq:enhanced_wtheta}. The top panels show the 2PCF data points calculated as the mean values from 40 \textsc{GLASS} realisations, and the error bars are calculated as the square root of the diagonal terms of the theoretical covariance matrix. \ziangtxt{The middle panels show the biases in 2PCFs with respect to the No selection 2PCFs and the bottom panels show the biases relative to the errors in the 2PCFs (The UR case is well beyond the range).} The shaded regions are angular scales corresponding to a physical scale $r<8h^{-1}\mathrm{Mpc}$ at the mean redshift of the galaxy sample.}
    \label{fig:wtheta_datadriven}
\end{figure*}

\begin{table*}[!htb]
\caption{Summary statistics of data-driven systematics test.}
\centering
\begin{tabular}{ll|l|l|ll|l}
\toprule
\multicolumn{2}{c|}{Configuration} &  \multirow{2}{*}{$\chi^2_{\mathrm{d}}$} & \multirow{2}{*}{$\mathrm{PTE}_{\mathrm{d}}$}  & \multirow{2}{*}{$\Delta \Omega_{\mathrm{m}}[\sigma]$}  & \multirow{2}{*}{$\Delta b [\sigma]$} & \multirow{2}{*}{$\chi^2_{\Delta \Omega_{\mathrm{m}}, \Delta b}$} \\ $NC_{\mathrm{KiDS}}$  &  $NC_{\mathrm{rec}}$ & & & &  \\ 
\hline
200 & 200 & 11.66 & 0.3088 & 0.36 & 1.69 & 4.26 \\ 
400 & 400 & 0.32 & 1.0 & 0.31 & 0.33 & 0.12 \\ 
600 & 600 & 0.29 & 1.0 & 0.23 & 0.16 & 0.05 \\ 
800 & 800 & 0.58 & 1.0 & 0.44 & 0.09 & 0.3 \\ 
600 ($m=1.5$) & 600 & 0.45 & 1.0 & 0.35 & 0.32 & 0.14 \\ 
\hline
600 & 200 & 11.19 & 0.3431 & 0.27 & 1.54 & 3.35 \\ 
600 & 400 & 0.4 & 1.0 & 0.38 & 0.37 & 0.17 \\ 
600 & 800 & 0.57 & 1.0 & 0.44 & 0.09 & 0.33 \\ 
\hline
\multicolumn{2}{c|}{True OR} & 0.08 & 1.0 & 0.09 & 0.08 & 0.01 \\ 
\multicolumn{2}{c|}{UR} & 8475.94 & 0.0 & -0.72 & 27.39 & 1759.8 \\ 

\bottomrule
\end{tabular}
\tablefoot{Four statistics are used to evaluate the consistency between 2PCFs and ``No selection'' 2PCFs. They are 1) $\chi^2_{\mathrm{d}}$ (defined in Eq.~\eqref{eq:chi2_d}) describes the difference between 2PCF data vectors; 2) probability-to-exceed (PTE) indicates the probability for a random data vector having higher $\chi^2_{\mathrm{d}}$ value; 3) $\Delta \Omega_{\mathrm{m}}$ and $\Delta b$ describes the individual parameter bias in the unit of constraining error; 4) $\chi^2_{\Delta \Omega_{\mathrm{m}},\,\Delta b}$ describes the overall constraining bias on the parameter space. The first five rows evaluate the difference between ``Recovered OR'' and ``No selection'' with the same $NC$ set-up for the data-driven OR and recovered mock OR, and the following three rows evaluate the difference for fixed $NC_{\mathrm{KiDS}}$ test, and the last two rows evaluate the difference between ``True OR'', ``UR'' cases, and ``No selection'' cases, respectively; }
\label{table:datadriven_summary}
\end{table*}

\begin{figure}[!htb]
    \centering
    \includegraphics[width=\linewidth]{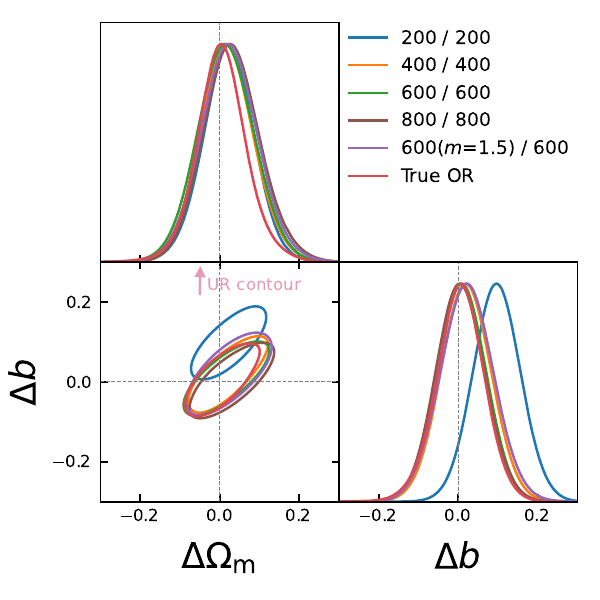}
    \caption{$1-\sigma$ confidence contours of the parameter posterior shift with respect to the best-fit values constrained from the No selection 2PCF. The pink arrow indicates the direction of the UR contour with the best-fit values of $\Delta\Omega_{\rm m}=-0.03$, $\Delta b=1.33$, which is well outside the dynamic range of the plot.}
    \label{fig:mcmc_datadriven}
\end{figure}

In the upper panels of Fig. \ref{fig:wtheta_datadriven}, we present the 2PCFs measured from the data-driven systematics test. Each column of panels shows a set-up specified by the number of clusters for data-driven OR and recovered mock OR. The right-most column corresponds to an enhanced data-driven OR by setting $m=1.5$ in Eq~\eqref{eq:enhanced_wtheta}. The colour scheme and meaning of data points are the same as introduced in Table \ref{table:w_theta_validate}. The data points are the mean 2PCF computed from 40 \textsc{GLASS} realisations and the error bars are the square root of the diagonal terms of the theoretical covariance matrix given by the \textsc{OneCovariance} code. \ziangtxt{The middle panels show the fractional bias of the 2PCFs with respect to the ``No selection'' 2PCF and the bottom panels show the bias of the 2PCFs with respect to the error.} The $1\sigma$ confidence contours of the parameter posterior shift with respect to the best-fit values constrained from the ``No selection'' 2PCF are shown in Fig. \ref{fig:mcmc_datadriven}. We note that the ``UR'' 2PCF gives such biased constraints that the contour is well outside the plot area.

Table \ref{table:w_theta_validate} summarises the statistics defined in Sect. \ref{sect:eva_datadriven}. The uniform random gives extremely biased 2PCF, highlighting the necessity to mitigate the variable depth. The ``True OR'' case agrees well with the ``No selection'' case as expected, {and this means that using the true OR in Eq.~\ref{eq:wtheta_def} can indeed correct the variable depth for the 2PCF}. From $NC=200$ to $NC=800$ the SOM+HC changes from under-correcting the 2PCF to slightly over-correcting it. The measured 2PCF and constrained cosmological parameters are in best agreement with the ``No selection'' case at $NC=600$. When we use an enhanced data-driven selection on the mock catalogue by setting $m=1.5$, $\chi_{\mathrm{d}}$ for UR increases from $8\,475$ (when $m=1$) to $20\,578$, {meaning that the bias in UR 2PCF is more than doubled when $m=1.5$}. The SOM+HC method with $NC=600$ can still correct the bias in the 2PCF. Therefore, we conclude that SOM+HC with $NC\sim600$ is the optimal choice for the KiDS-Legacy galaxy clustering 2PCF measurement. 

Now we fix $NC_{\mathrm{KiDS}}=600$ so that the data-driven selection reflects the realistic variable depth, then we vary $NC_{\mathrm{rec}}$ from 200 to 800. 
The evaluation statistics are summarised from the 6th row to the 8th row of Table \ref{table:datadriven_summary}, plus the third row with $NC_{\mathrm{rec}}=600$. Again we see that the recovered 2PCF goes from biased high (when $NC_{\mathrm{OR}}=200$) to well corrected. We also note that the recovery biases are generally low with $NC_{\mathrm{rec}}$ varying from 400 to 800, for which $\chi^2_{\mathrm{d}}$ values are at a level of $\sim0.5\sigma$, with a minimum value of 0.39 when $NC_{\mathrm{rec}}=600$. The goodness of recovery is also robust to the choice of $NC_{\mathrm{rec}}$ from 400 to 800. In conclusion, we  choose $NC_{\mathrm{rec}}=600$ for the galaxy clustering measurement with KiDS-Legacy data.

\section{2PCF measurement with KiDS-Legacy sample}
\label{sect:kidslegacy}

\subsection{Application of SOM+HC on the real data}
\label{sect:kidslegacy-somhc}

\begin{figure*}
    \centering
    \includegraphics[width=\linewidth]{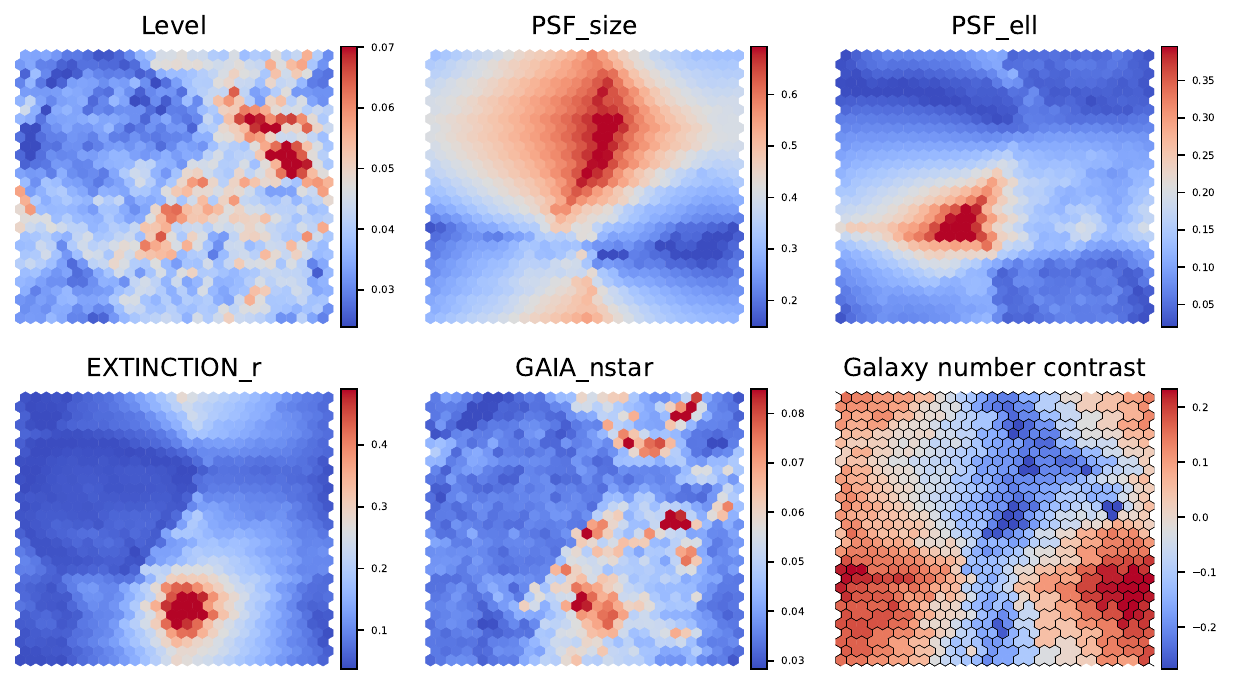}
    \caption{Self-organising maps trained on the KiDS-Legacy catalogue, with the dimension $30\times30$. The first five panels are SOMs coloured by the average systematics values in each cell. The last panel (bottom right) is the SOM coloured by the galaxy number contrast of each hierarchical cluster. The black lines are the boundaries of each HC. We note that we use a toroidal topology for the SOM, so the left and right edges and the top and bottom edges are adjacent.}
    \label{fig:som_plot}
\end{figure*}

\begin{figure}
    \centering
    \includegraphics[width=\linewidth]{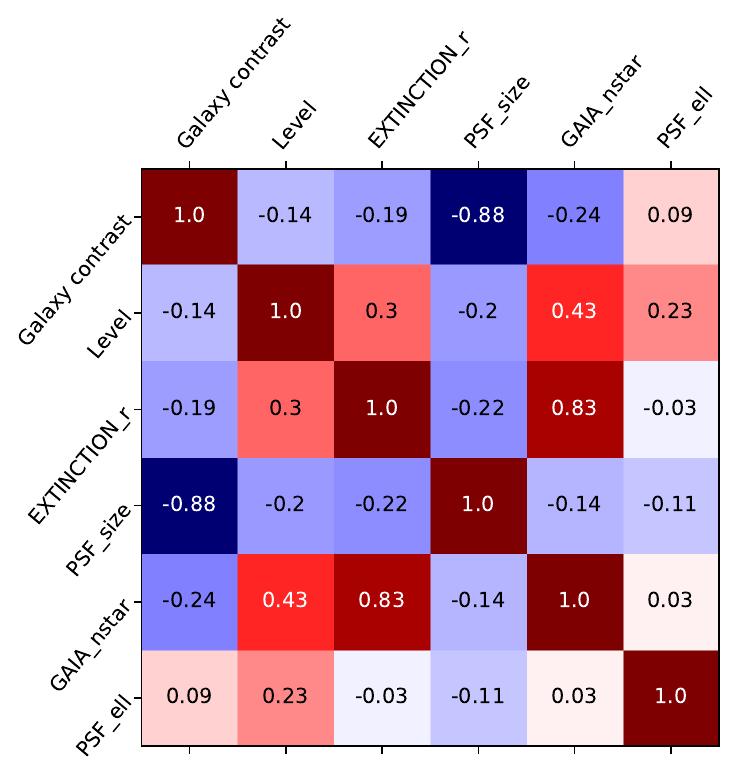}
    \caption{Spearman correlation coefficient matrix. The numbers in each grid are the correlation coefficient between the median systematics and the galaxy contrast in each hierarchical cluster.}
    \label{fig:sys_corr}
\end{figure}

In this section we apply the SOM+HC method to the real KiDS-Legacy galaxy catalogue. The systematics used to train the SOM are described in Sec.~\ref{sect:data}. The SOM+HC set-up is the same as the data-driven systematics test: the SOM dimension is 30, with hexagonal cells (so each cell has six adjacent neighbourhoods), toroidal topology (so the left and right edges are adjacent, and the top and bottom edges are adjacent). Training lasts for 10 epochs, with an initial width of the Gaussian neighbourhood function $\sigma=30/2=15$, decreasing linearly to $\sigma=1$ in the last epoch. The initial learning rate is 0.1 and decreases linearly to 0.01. The trained SOM cells are grouped into 600 clusters, from which the OR weight is generated on a HEALPix map with \texttt{Nside}=2048. %We refer the reader to Section \ref{sect:method} for an introduction to the SOM+HC method.

Figure \ref{fig:som_plot} shows the results of the post-trained SOM+HC. The first five panels show the SOM colour-coded by the average systematics values of each cell. We note that the SOM inherits the grained patterns of the mean \texttt{Level} and \texttt{GAIA\_nstar} spatial distributions (see Fig. \ref{fig:sys_maps}). The last panel shows the galaxy number contrast of the 600 hierarchical clusters (denoted by cells with black borders). Due to the toroidal topology, clusters near the edge may actually cross the edge and continue on the other side. One can see the correlation between galaxy number contrast and each systematics from the SOM maps. For example, the galaxy number contrast is strongly anticorrelated with \texttt{PSF\_size}. This is expected, as a larger PSF size indicates poorer atmospheric seeing, making galaxies harder to detect under these conditions. A quantitative correlation can be evaluated using Spearman correlation coefficients. Figure \ref{fig:sys_corr} shows the correlation coefficient matrix. The bottom row shows the correlation between systematics and galaxy number contrast, and it is clear that PSF size has a strong anti-correlation with galaxy number contrast, \ziangtxt{in agreement with the contrast-systematics relation shown as black curves in Fig.~\ref{fig:sys_maps}}. On the other hand, the PSF shape only has a weak correlation with the contrast in galaxy numbers. We also find that \texttt{Level} and \texttt{GAIA\_nstar} are correlated with extinction. 

\subsection{Blinded measurement of 2PCF}

Since the KiDS-Legacy cosmological analysis is not yet unblinded, in this paper we only make blinded measurements of the 2PCF. We leave more sophisticated measurements, including tomographic galaxy clustering, to later work. This section serves as a showcase of the SOM+HC method on real data.

The 2PCF is measured the same way as those from Sect.~\ref{sect:validation}. The ``UR'' 2PCF is measured by using the coverage map as the random term $N^{a}_{R,p}$ in Eq~\eqref{eq:ddpix_def} (for the similar definition of the $RR$ and $DR$ terms), and the ``OR'' 2PCF uses the OR weight from SOM+HC. Both 2PCFs are then blinded using the method given in \cite{Muir_2020}. The blinding method includes the following steps:

\begin{enumerate}
    \item Select a reference cosmology and calculate the associated theoretical 2PCF. In our case, we use the fiducial cosmology defined in Sect. \ref{sect:intro}.
    \item Shift the cosmological parameters to be blinded with a Gaussian random number. The standard deviation of the Gaussian random defines how much one wants to blind the parameters. In this paper we shift \{$\Omega_{\mathrm{m}}$, $b$\} with standard deviation \{0.1, 0.1\} respectively.
    \item Calculate the shifted 2PCF given the shifted parameters and obtain the data vector shift by subtracting it from the reference 2PCF.
    \item Add the data vector shift to the measured 2PCF to obtain the blinded 2PCF.
\end{enumerate}

Both the UR and OR 2PCFs are shifted by the same amount to make a meaningful comparison. We then run an MCMC on the blinded 2PCFs to sample the posterior of \{$\Omega_{\mathrm{m}}$, $b$\}. To prevent accidental unblinding, we never save the blinded data vector, but directly pass it to the MCMC code. The covariance matrix is computed from the \textsc{OneCovariance} code. Since we do not know the true parameter values {\textit{a priori}}, we use an iterative fitting procedure: first, we compute the covariance matrix with the fiducial cosmology and run MCMCs with it, then we compute the best-fit parameters from the posterior and update the covariance matrices for OR and UR respectively. The posterior is then sampled with the updated covariance matrices. The MCMCs are sampled with the same modelling code when we blind the data vector. We note that we only constrain the linear model with data points at $\theta>8h^{-1}\mathrm{Mpc}/D_A(\bar{z})$, where $D_A(\bar{z})$ is the angular diameter distance at the mean KiDS-Legacy redshift $\bar{z}=0.7$. \ziangtxt{The redshift distribution is calibrated with a combination of SOM and the clustering-redshift method, which will also be presented in a companion paper.}

Data-driven systematics tests (see Sect. \ref{sect:validation}) suggest that $NC=600$ is the optimal choice. In this section, we use this as the fiducial setting, but also try $NC=\{200, 400, 800\}$ to test the robustness of the $NC$ choice. All 2PCFs are shifted with the same blinding shift. The posteriors are also fitted from these measurements.

\begin{figure}
    \centering
    \includegraphics[width=\linewidth]{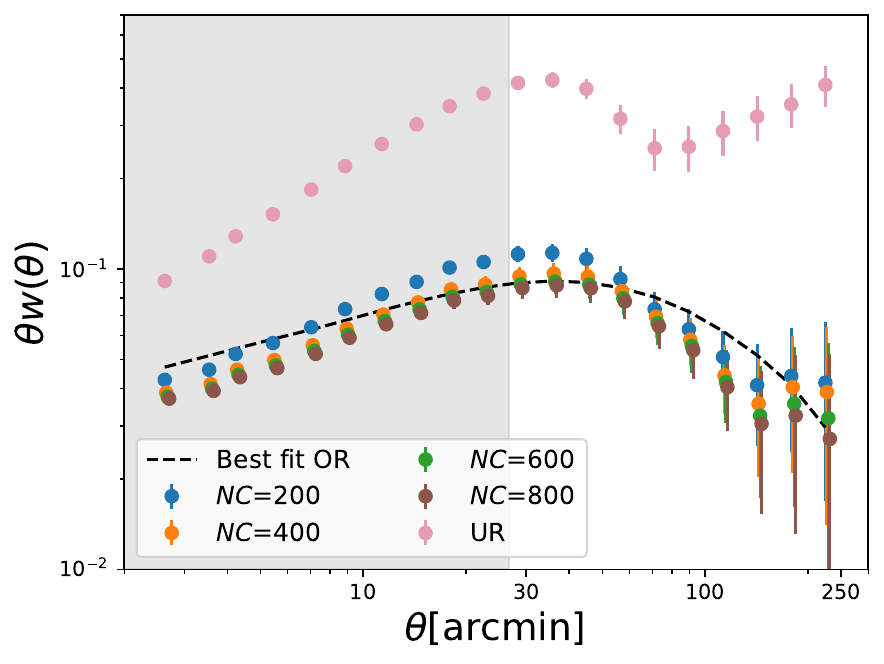}
    \caption{Blinded 2PCFs measured from the KiDS-Legacy sample. The blue and pink dots are the measurements corrected by the recovered organised random and by the uniform random, respectively. The shaded regions are angular scales corresponding to physical scales smaller than $8h^{-1}\mathrm{Mpc}$ estimated at the mean redshift. The error bars are the standard deviations derived from the covariance matrix computed by the \textsc{OneCovariance} code. The black dashed curve shows the best-fit 2PCF from the MCMC.}
    \label{fig:wtheta_legacy}
\end{figure}

\begin{figure*}

        \centering
    \subfloat{\includegraphics[width=0.5\linewidth]{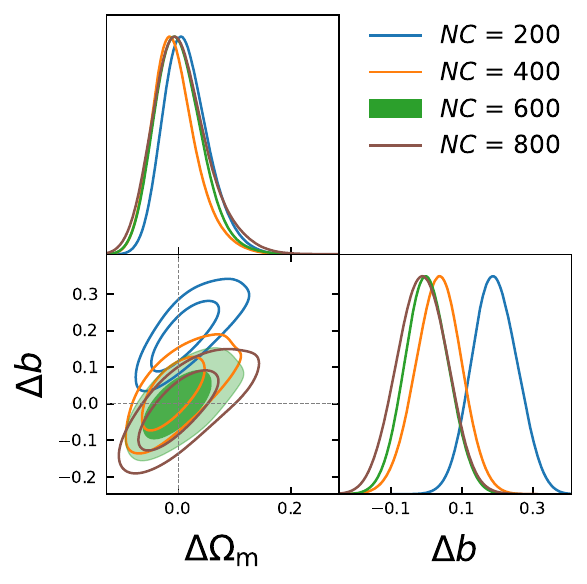}
    }
    \subfloat{\includegraphics[width=0.5\linewidth]{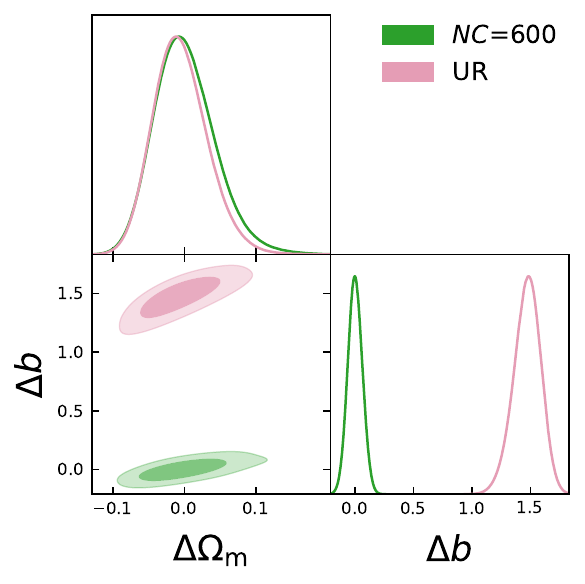}
    }
    \caption{Contours of 68.3\% and 95.4\% credible levels of the parameter posterior shift with respect to the best-fit values constrained from $NC=600$ case. {\textit{Left panel}}: Contours from the OR 2PCFs with different $NC$ choices. The fiducial choice $NC=600$ is shown as green filled contours; {\textit{Right panel}}: Contours of OR 2PCF ($NC=600$, green) and UR 2PCF (pink).}
     \label{fig:wtheta_legacy_mcmc}
\end{figure*}

The blinded 2PCFs are shown in Fig. \ref{fig:wtheta_legacy} with error bars derived from the updated covariance matrices. The black dashed line shows the theoretical 2PCF calculated from the best-fit parameters from the blinded 2PCF measurement with $NC=600$. The reduced $\chi^2$ value between the OR ($NC=600$) 2PCF and the best-fit 2PCF is 1.28, corresponding to a PTE of 0.25, indicating a good fit between the model and the blinded data. Figure \ref{fig:wtheta_legacy_mcmc} shows the posteriors sampled from the MCMC, shifted with respect to the best-fit value of the $NC=600$ case. The left panel shows the contours of shifted posteriors constrained from 2PCF corrected by OR with different $NC$, while the right panel shows contours corresponding to 2PCF corrected by OR ($NC=600$) and UR, respectively. On linear scales, the OR and UR 2PCFs differ at a level of $70\sigma$ and the galaxy bias parameter differs at a level of $40\sigma$. Interestingly, the matter density $\Omega_{\mathrm{m}}$ does not change significantly with the correction. This is because the selection does not cause significant bias in $\Omega_{\mathrm{m}}$ ($\Omega_{\mathrm{m}}$ is not biased even for the UR 2PCF), but this might be not the case for other surveys.

The 2PCF measurements with different $NC$ suggest that the OR 2PCFs decrease with $NC$ (i.e. the corrections increase with $NC$). If we treat $NC=600$ as the fiducial choice, the corresponding $\chi_{\mathrm{d}}$ values are 5.75, 0.27, 0.07 with $NC=200, 400, 800$, so the change is significant from $NC=200$ to $NC=400$, while moderate from $NC=400$ to $NC=800$. This is consistent with the data-driven validation test. Therefore, we conclude that the OR correction is quite robust around $NC=600$.

\section{Discussions and conclusions}
\label{sect:discussions}

The aim of this work was to correct the bias in the galaxy 2PCF from deep surveys due to complicated selection effects. We showed that for deep surveys such as KiDS, the selection effect is very pronounced and can lead to a bias in the 2PCF about ten times the signal, especially on large scales. Therefore, a critical step is to correct for the variable depth caused by complex selection effects. We introduce the SOM+HC method proposed by \citet{harryOR} on the KiDS-Legacy sample and show that it can correct the bias for faint galaxy samples.

SOM+HC is a machine-learning method for recovering organised randoms that have the same selection pattern as the galaxy sample. The method uses a combination of self-organising maps and hierarchical clusters that group galaxies in systematics space and redistributes them across the survey footprint. Compared to other methods used to mitigate selection bias in the 2PCF, SOM+HC has the advantage of being model-independent. As an unsupervised machine learning algorithm, it does not need to parametrise selection functions or biases in statistics, but directly captures arbitrary selection-induced patterns in the systematics space. Therefore, it can recover complex selection functions and their correlations without making \textit{a priori} assumptions about them. In addition, the SOM+HC method is purely data-driven, meaning that it can be trained on the real data itself without relying on mock data or any external information. 

\citet{harryOR} proved that SOM+HC can correct the slight bias in the bright sample, while this work further validates the method on a faint galaxy sample with more complex selection effects and a more significant bias in the 2PCF. The validations are performed on mock galaxies from \textsc{GLASS} with both toy selections and data-driven selections. The toy test is a direct demonstration of how SOM+HC can recover the selection functions and the unbiased 2PCF. The data-driven test demonstrates the performance of SOM+HC on deep surveys such as KiDS and determines the optimal set-up for realistic 2PCF measurements. From Fig. \ref{fig:wtheta_datadriven}, we note that the bias in the UR 2PCF is scale-dependent. This scale dependency may arise because different scales are dominated by selection effects of different systematics, but it is hard to quantitatively isolate them in different scales. OR is able to correct such complicated selection effects. From this test, we see that the selection effects in the real KiDS-Legacy galaxy sample bias the 2PCF by $\chi^2\sim 8500$ and that the SOM+HC with $NC=600$ can correct this bias down to a level of $\chi^2\sim 0.3$ given a degree of freedom of 10, resulting in the reduction of bias on the $(\Omega_{\rm m}, b)$ from a level of $27\sigma$ to $0.3\sigma$. Notably, this set-up can recover an unbiased 2PCF even with an enhanced selection, correcting a 2PCF bias of $\chi^2\sim 20000$ to $\chi^2\sim 0.4$.

Several set-ups of the SOM+HC algorithm can affect its performance, so the optimal set-up varies from survey to survey. For example, the number of galaxies (data volume) is important for the SOM+HC set-up. A catalogue with a larger number of galaxies is less sensitive to Poisson noise and has a more obvious variable depth, making the SOM+HC more efficient in recovering an accurate OR. On the other hand, smaller galaxy catalogues require lower $NC$ and $\texttt{Nside}$ to reduce cosmological contamination in the OR, but this will limit the spatial resolution of the OR. \ziangtxt{Furthermore, data-driven regression methods, including SOM+HC, aim to nullify modes of variance in the training data, which carries the risk of over-correction. A potential solution is to train the model on data from a disjoint sky region and run it on the region of interest. This approach requires ensuring that the training region is both representative and uncorrelated with the testing region (the region of interest), making it practically challenging.} For datasets with weaker selection effects than the LSS fluctuation (such as the KiDS bright sample), SOM+HC is more prone to over-correction. Therefore, SOM+HC should perform better on larger, fainter galaxy samples. For KiDS-Legacy, we show that $NC=600$ gives a parameter constraint accuracy of $\sim0.3\sigma$ and performance is found to be stable around this choice. In this work we also set the pixel size to 1.4 arcmin, equivalent to \texttt{Nside}=2048, to avoid unintentionally masking pixels within the footprint, while maintaining the best performance of the OR. We expect smaller pixels to give a more accurate OR for future surveys with higher galaxy number densities. 

Using the validated SOM+HC method, we performed a preliminary blinded 2PCF measurement from the KiDS-Legacy galaxy catalogue. We find that the UR 2PCF is significantly higher than the OR 2PCF by an order of magnitude. Furthermore, the corrected 2PCF is robust to the choice of $NC$ around 600. We applied the SOM+HC method to the photometric galaxy clustering measurement for the KiDS-Legacy 6$\times$2pt and the bright galaxy clustering measurement for the KiDS-Legacy 3$\times$2pt measurements. More detailed discussions of the methodology (e.g. tomographic galaxy clustering) will be given in the forthcoming papers.

We moreover note that the tests carried out in this work were focused on linear scales. \ziangtxt{The bottom panels in Fig. \ref{fig:wtheta_datadriven}} show that the performance of the method is worse on small scales. This is because the variance is very small on small scales, so the bias tends to be significant compared to it. Therefore, we need to take extra care at these scales. For example, future work could combine different algorithms to correct for variable depth at different scales.

Combinations of multiple 2PCFs, namely the $N\times2$pt measurements, will form the main analysis components of the next generation of LSS surveys, including LSST and {\textit{Euclid}}. Galaxy 2PCF will be a critical part of these measurements. For these deep surveys, selection effects will also lead to variable depths. For example, the rolling cadence survey strategy of  LSST \citep{Bianco_2021} will introduce stripe-like non-uniformity. Based on \citet{hang2024impactsurveyspatialvariability}, this will cause a significant bias up to an order of magnitude for the LSST Y3 galaxy clustering measurement. The \textit{Euclid} survey, on the other hand, will combine ground-based multi-band photometry to estimate photo-$z$ \citep{2020A&A...644A..31E}, so any photo-$z$-based selection will introduce variable depth. With large sky coverage and depth, the measurement precision of these surveys will be greatly improved, requiring cleaner and more reliable bias correction. As discussed above, SOM+HC is more effective with larger galaxy samples. Therefore, we expect that as the data volumes of the next generation of surveys increase, SOM+HC will become more powerful in recovering selection-induced clustering. In addition, one can use smaller pixels (or higher \texttt{Nside}) for such galaxy catalogues, and obtain the OR weights at higher angular resolutions. To this end, we published the code used in this work as the \textsc{tiaogeng} package\footnote{ \href{https://github.com/yanzastro/tiaogeng/}{https://github.com/yanzastro/tiaogeng/}} for future implementation in pipelines for next-generation surveys (e.g.~\textsc{TXPipe}\footnote{\hyperlink{https://txpipe.readthedocs.io/en/latest/}{https://txpipe.readthedocs.io/en/latest/}}). A combination with other methods, such as template-based correction methods, could be even more effective in mitigating the complex selection effects for future deep surveys.

\begin{acknowledgements}

We appreciate fruitful discussions with Harry Johnston, Andrina Nicola, and Anna Porredon.

ZY acknowledges support from the Max Planck Society and the Alexander von Humboldt Foundation in the framework of the Max Planck-Humboldt Research Award endowed by the Federal Ministry of Education and Research (Germany). 

AHW is supported by the Deutsches Zentrum für Luft- und Raumfahrt (DLR), made possible by the Bundesministerium für Wirtschaft und Klimaschutz, and acknowledges funding from the German Science Foundation DFG, via the Collaborative Research Center SFB1491 "Cosmic Interacting Matters - From Source to Signal".

NEC and CG acknowledge support from the project ``A rising tide: Galaxy intrinsic alignments as a new probe of cosmology and galaxy evolution'' (with project number VI.Vidi.203.011) of the Talent programme Vidi which is (partly) financed by the Dutch Research Council (NWO). 

SJ acknowledges the Dennis Sciama Fellowship at the University of Portsmouth and the Ramón y Cajal Fellowship from the Spanish Ministry of Science.

AL acknowledges support from the research project grant `Understanding the Dynamic Universe' funded by the Knut and Alice Wallenberg Foundation under Dnr KAW 2018.0067.

RR is supported by an ERC Consolidator Grant (No. 770935).

MA acknowledges the UK Science and Technology Facilities Council (STFC) under grant number ST/Y002652/1 and the Royal Society under grant numbers RGSR2222268 and ICAR1231094

MB is supported by the Polish National Science Center through grants no. 2020/38/E/ST9/00395, 2018/30/E/ST9/00698, 2018/31/G/ST9/03388 and 2020/39/B/ST9/03494.

AD acknowledges support from the ERC Consolidator Grant (No. 770935)

CH acknowledges support from the Max Planck Society and the Alexander von Humboldt Foundation in the framework of the Max Planck-Humboldt Research Award endowed by the Federal Ministry of Education and Research, and the UK Science and Technology Facilities Council (STFC) under grant ST/V000594/1

H. Hildebrandt is supported by a DFG Heisenberg grant (Hi 1495/5-1), the DFG Collaborative Research Center SFB1491, an ERC Consolidator Grant (No. 770935), and the DLR project 50QE2305.

PJ is supported by the Polish National Science Center through grant no. 2020/38/E/ST9/00395.

BJ acknowledges support by the ERC-selected UKRI Frontier Research Grant EP/Y03015X/1 and by STFC Consolidated Grant ST/V000780/1.

LL is supported by the Austrian Science Fund (FWF) [ESP 357-N].

CM acknowledges support from the Beecroft Trust, the Spanish Ministry of Science under the grant number PID2021-128338NB-I00, and from the European Research Council under grant number 770935.

LM acknowledges the financial contribution from the grant PRIN-MUR 2022 20227RNLY3 “The concordance cosmological model: stress-tests with galaxy clusters” supported by Next Generation EU and from the grant ASI n. 2024-10-HH.0 “Attività scientifiche per la missione Euclid – fase E”

NRN acknowledges financial support from the National Science Foundation of China, Research Fund for Excellent International Scholars (grant n. 12150710511), and from the research grant from China Manned Space Project n. CMS-CSST-2021-A01.

BS acknowledges support from the Max Planck Society and the Alexander von Humboldt Foundation in the framework of the Max Planck-Humboldt Research Award endowed by the Federal Ministry of Education and Research.

MvWK acknowledges the support by the UK Space Agency.

MY acknowledges funding from the European Research Council (ERC) under the European Union’s Horizon 2020 research and innovation program (Grant agreement No. 101053992)

\par
\\
The data in this paper is analysed with open-source python packages \textsc{numpy} \citep{harris2020array}, \textsc{scipy} \citep{2020SciPy-NMeth}, \textsc{astropy} \citep{astropy:2018}, \textsc{matplotlib} \citep{Hunter:2007}, \textsc{GLASS}\citep{glass}, \textsc{somoclu}\citep{Wittek_2017}, \textsc{healpy} \citep{Zonca2019}, \textsc{NaMaster} \citep{2019namaster}, \textsc{CCL} \citep{Chisari_2019}, \textsc{emcee} \citep{Foreman_Mackey_2013}, and \textsc{GetDist} \citep{Lewis:2019xzd}.
We also use \textsc{WebPlotDigitizer} \citep{Rohatgi2020} to digitise some external data from plots in the literature.
\\
{{\it Author contributions:} All authors contributed to the development and writing of this paper. The authorship list is given in three groups: the lead authors (ZY \& AHW) followed by two alphabetical groups. The first alphabetical group includes those who are key contributors to both the scientific analysis and the data products. The second group covers those who have either made a significant contribution to the data products or the scientific analysis.
 }\\
{{\it Data availability:} Our software is \href{https://github.com/yanzastro/tiaogeng}{open-source} for future usage.}
\end{acknowledgements}

\bibliographystyle{aa}
\bibliography{main}

\begin{appendix}

\section{Covariance matrix comparison}
\label{sect:app_covmat}

In this Appendix we compare the covariance matrix from mock realisations and the theoretical covariance computed by the \textsc{OneCovariance} code. The mock covariance is computed based on Eq. \eqref{eq:covariance}. To ensure consistency, we configure the input file of the \textsc{OneCovariance} code so that the cosmology, footprint, redshift distribution and galaxy number density match those of the \textsc{GLASS} mock. In addition, since the \textsc{GLASS} package only generates a Gaussian field, we assume that the theoretical covariance matrix contains only the Gaussian covariance plus the super-sample covariance \citep{Takada_2013} to account for the covariance of modes larger than the observed field.

\begin{figure}
    \centering
    \includegraphics[width=\linewidth]{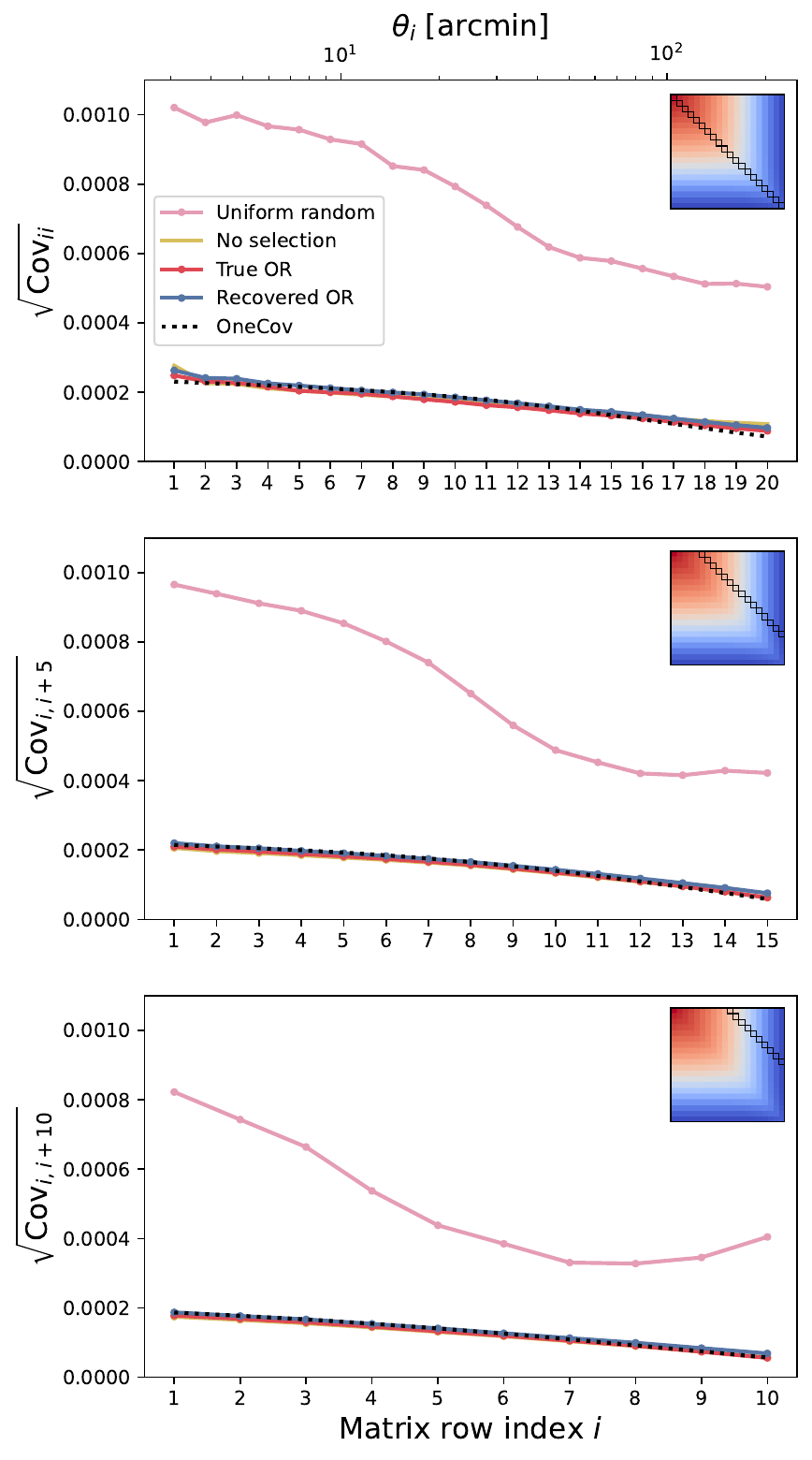}
    \caption{Comparison of the covariance matrices. Panels from top to bottom show the square root of the main covariance diagonal, the 5th and 10th diagonals above the main diagonal, as shown by the black grids in the top right theoretical covariance matrices in each panel. The coloured curves show the covariance terms of the four 2PCFs defined in Table \ref{table:w_theta_validate} from the data-driven test with $\{NC_{\mathrm{KiDS}}=NC_{\mathrm{rec}}=600\}$; the dotted lines are calculated from the \textsc{OneCovariance} code. The lower x-ticks are the row indices, while the upper x-ticks on the top panel are the corresponding angular scale for the diagonal term.}
    \label{fig:covmat_compare}
\end{figure}

In this section we compare only the covariance of the 2PCFs from the data-driven test with $\{NC_{\mathrm{KiDS}}=600, NC_{\mathrm{rec}}=600\}$. The three panels in Fig. \ref{fig:covmat_compare} show the square root of the main diagonal, the 5th and the 10th diagonal above the mean diagonal. We note that the covariance of the uniform random case is significantly higher than that of the ``no selection'' case, implying that variable depth contamination also introduces additional covariance into the data. The covariance for the ``No selection'' and ``True OR'' cases matches the theoretical covariance given by \textsc{OneCovariance} as expected. The covariance of ``Recovered OR'' agrees with that of the unbiased cases, so we conclude that the SOM+HC method can also recover accurate covariance in the correlation function in the linear regime.

\section{Correcting selection effects for angular power spectra with SOM+HC}
\label{sect:pcl_or}

\begin{figure}
    \centering
    \includegraphics[width=1\linewidth]{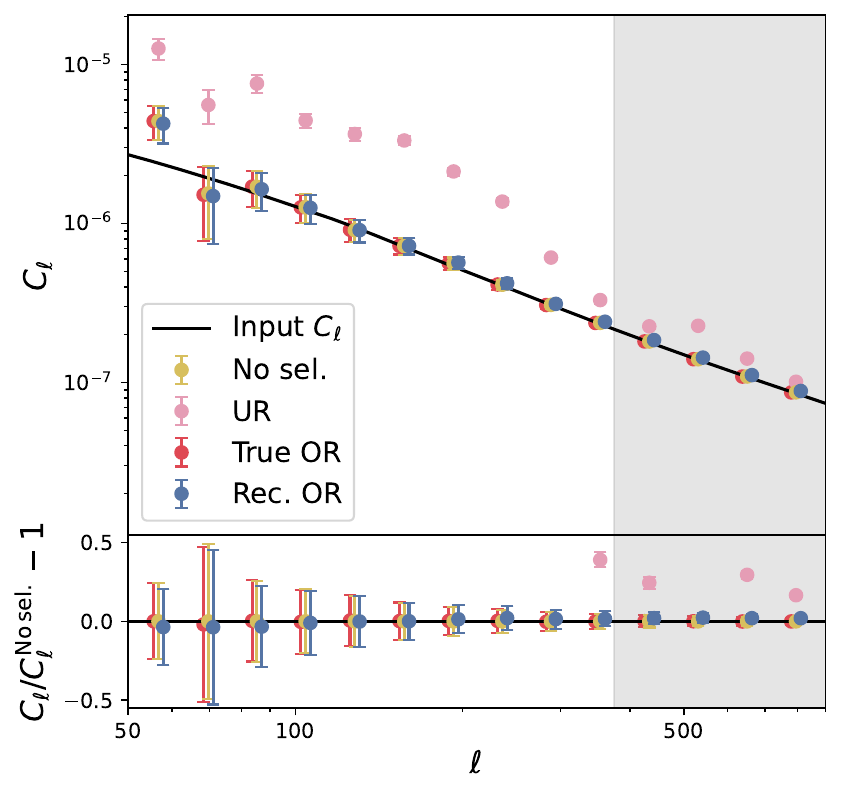}
    \caption{{\textit{Upper panel:}} Pseudo-$C_{\ell}$ measured from \textsc{GLASS} mock samples. The data points of each series are the average PCL from 40 realisations; the error bars are the square root of the covariance diagonal given by \textsc{NaMaster}. The $\chi^2_{\mathrm{d}}$ values, which describe the difference between each dataset and the No selection PCL, are calculated similarly to Eq.~\eqref{eq:chi2_d}.  {\textit{Lower panel}}: Relative difference of each case with respect to the no selection case. The shaded regions are the $\ell$ modes corresponding to a physical scale smaller than 8 $h^{-1}$Mpc estimated at the mean redshift.}
    \label{fig:pcl}
\end{figure}

\begin{figure}
    \centering
    \includegraphics[width=1\linewidth]{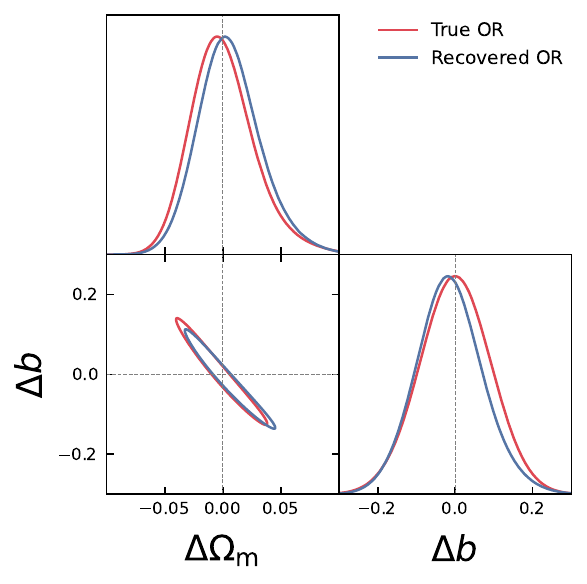}
    \caption{1-$\sigma$ credible contours of parameter posterior shift with respect to the best-fit values constrained from the ``No selection'' PCL}
    \label{fig:post_pcl}
\end{figure}

Another widely used two-point statistic is the angular power spectrum, which is defined as the correlation of galaxy over-density in harmonic space. In general, angular power spectra have weak correlations between $\ell$ modes, allowing us to study different angular scales independently. 

In practice, there are two estimators of the power spectra. One is the ``band power'', where we measure the correlation functions first, then invert Eq.~\eqref{eq:wtheta_def} by performing the integration $C_\ell = \int_0^\infty w(\theta) J_0(\ell \theta) \theta \, d\theta$ at the central $\ell$ in each band. The accuracy of the band power depends on the integral limit of $\theta$ and the discrete $\theta$'s when calculating the correlation function \citep[][]{Schneider_2002}.

The other estimator is the pseudo-$C_{\ell}$ \citep[PCL,][]{Wandelt_2001, Alonso_2018} , where one calculates the coupled $C_{\ell}$ from the weighted sky map,   then decouples it with the mode coupling matrix of the weight, and bins it. The weight can be a binary mask specifying the footprint of the survey (so that a source gets a weight of 1 if it is in the footprint, otherwise 0); or a weight for each source to suppress errors (like the lensing weight) or to correct for selection effects (like the organised random weight in this paper). In this appendix we briefly discuss the usage and performance of SOM+HC on the measurement of PCL $C_{\ell}$ from the same \textsc{GLASS} mock samples. Detailed discussions will be given in a companion paper presenting the methodology of the KiDS-Legacy 6$\times$2pt cosmology.

The galaxy PCL is based on pixelised galaxy over-density maps. On a weighted sky, the galaxy over-density in the $p$-th pixel is given by \citep{Nicola_2020}:

\begin{equation}
    \delta(p) = \frac{N(p)}{w(p)\overline{N}}-1\,,
    \label{eq:delta_g}
\end{equation}
where $N(p)$ is the number of galaxies in the pixel, $w(p)$ is the weight value (the organised random weight in our case) in the pixel. By dividing the galaxy number by $w(p)$, we have effectively corrected for the variable depth. The average galaxy number is given by

\begin{equation}
    \bar{N} \equiv \sum_{p\in w}\frac{N_p}{w_p}\,,
\end{equation}
where the sum is taken within the footprint.   

We can measure the PCL directly between the maps of two galaxy samples $a$ and $b$ as

\begin{equation}
    \tilde{C}^{ab}_{\ell} = \frac{1}{2\ell+1}\sum_{m}\tilde{a}^a_{\ell m}\tilde{a}^{b *}_{\ell m}\,,
    \label{eq:cell_def}
\end{equation}
where $\tilde{a}^a_{\ell m}$ is the harmonic coefficient of sample $a$ on weighted sky and $\tilde{a}^{a *}_{\ell m}$ is its complex conjugate. For weighted galaxy maps, it is linked to the underlying power spectra $C_{\ell}$ via

\begin{equation}
    \tilde{C}^{ab}_{\ell} = \sum_{\ell^{\prime}} M_{\ell\ell^{\prime}}(w^a, w^b) C^{ab}_{\ell}\,,
\end{equation}
where $M_{\ell\ell^{\prime}}(w^a, w^b)$ is the mode-mixing matrix determined by the weight maps $(w^a, w^b)$ of the two fields.  Therefore, the directly measured $\tilde{C}^{ab}_{\ell}$ is called ``coupled PCL''. An unbiased angular power spectrum is estimated by decoupling it via the inverse of the mode-coupling matrix. Assuming Poisson-distributed objects, the OR weight is proportional to the inverse variance of the field, we use OR weight to calculate the mode-coupling matrix. In practice, one also wants to bin the PCL into $\ell$ bins. For technical details, we refer  to \citet{2019namaster}. The calculations are implemented in the \textsc{NaMaster} package \citep{2019namaster} which we  use in this section to measure the PCL.

The measured galaxy auto-power spectrum contains a shot noise which needs to be subtracted. We assume a Poissonian shot noise, for which the coupled noise spectrum (the ``noise bias'' termed in \citealt{2019namaster}) is given by:

\begin{equation}
    \tilde{N}_{\ell} = \Omega_{\mathrm{pix}}\frac{\left\langle w\right\rangle}{\overline{N}}\,,
\end{equation}
where $\Omega_{\mathrm{pix}}$ is the pixel area in the units of steradians; $\langle w \rangle$ is the mean weight value per pixel across the whole sky.

The Gaussian covariance matrix depends on unbiased estimations of the angular power spectra. In this section we only measure angular power spectra from \textsc{GLASS} mock catalogue, so we take the input angular power spectra for \textsc{GLASS} to estimate the theoretical Gaussian covariance matrix.\footnote{In real measurement when the true angular power spectra are unknown, an iterative estimation is usually used to evaluate the covariance \citep{Eifler_2009}. That is, one uses the theoretical angular power spectra calculated from reasonable cosmological parameters to calculate the covariance and constrain the parameters with the corresponding likelihood. The covariance matrix is updated with the best-fit parameter. This process is performed iteratively until the best-fit parameter converges.} The mode-coupling induced by the OR weights is also taken care of by the \textsc{NaMaster} package. The non-Gaussian term includes a connected covariance matrix which is dominated by the galaxy trispectra \citep{Krause_2017} which only affects the small scales, so we neglect it in this section. Another non-Gaussian term is the super-sample covariance \citep[SSC;][]{Takada_2013}{}{} which accounts for the correlated modes that are larger than the survey footprint. We calculate this term using the \textsc{OneCovariance} package, but we note that this term is negligible for KiDS-Legacy.

Therefore, for PCL, the variable depth affects the mapping process, the mode-mixing matrix, the shot noise and the covariance matrix. In this section we perform the data-driven test for PCL with the same 40 \textsc{GLASS} mock samples with interpolated systematics. Here we show the case $NC_{\mathrm{KiDS}}=NC_{\mathrm{rec}}=600$. That is, the mock sample is selected with the data-driven OR weight with $NC_{\mathrm{KiDS}}=600$ and the recovered OR weight is from 600 HCs trained on the selected mock sample. We first generate the galaxy fluctuation map according to Eq.~\eqref{eq:delta_g} and measure the four PCLs defined similarly to Table \ref{table:w_theta_validate}: the UR case is a depleted galaxy sample for which the mode-mixing matrix is calculated from the footprint; the `No selection' case is the unselected galaxy sample for which the mode-mixing matrix is calculated from the footprint. 

The PCL measure is shown in Fig. \ref{fig:pcl}. Data points in the top panel show the averaged PCL across realisations. Error bars are the standard deviation of each $C_{\ell}$ calculated from the covariance matrix given by \textsc{NaMaster}. The bottom panel shows the relative difference between each $C_{\ell}$ measurement and the ``no selection'' case. The shaded region corresponds to physical scales smaller than $8h^{-1}\mathrm{Mpc}$ estimated at the mean redshift. We ignore these scales in this analysis. To assess the consistency between the OR-corrected PCL and the ``no selection'' PCL, we calculate the $\chi_{\mathrm{d}}$ (defined similarly to Eq. \eqref{eq:chi2_d} but using $C_{\ell}$ instead of $w(\theta)$) between them. The values are quoted in the upper panel of Fig. \ref{fig:pcl}. The uniform random again gives a very large bias ($\chi_{\mathrm{d}}=3230$) in the PCL and the true OR completely corrects the bias as expected. The OR recovered by SOM+HC gives a residual of $\chi_{\mathrm{d}}=0.42$. 

In the same way as for the 2PCF, we run an MCMC on the measured PCL. The Gaussian likelihood is defined with the Gaussian covariance given by \textsc{NaMaster} plus the SSC given by \textsc{OneCovariance}. We calculate the shifted posterior with respect to the best-fit values constrained from the ``No selection'' case. The contours of the 68\% credible level are shown in Fig. \ref{fig:post_pcl}. We calculate the parameter constraint bias following the 2PCF test and get $\Delta\Omega_{\mathrm{m}}[\sigma]=0.21$, $\Delta b[\sigma]=-0.17$, $\chi^2_{\Delta\Omega_{\mathrm{m}}, \Delta b}=0.07$ and for the recovered OR weight case. For the true OR case, the constraining bias is at the $0.001\sigma$ level. We note that both posteriors have the same shape, indicating a close degeneracy between the galaxy bias and $\Omega_{\mathrm{m}}$.

From this exercise, we can conclude that the OR weight recovered by the SOM+HC method can also correct for variable depth in 2-point statistics in harmonic space. Furthermore, the optimal choice of $NC$ that we found with data-driven systematics also provides accurate PCL measurements and parameter constraints. We leave further tests, including tomographic galaxy clustering PCL, to a future KiDS-Legacy $6\times2$pt methodology paper.

\section{SOM+HC with all the systematics from KiDS-Legacy}
\label{sect:all_sys}

\begin{figure}
    \centering
    \includegraphics[width=\columnwidth]{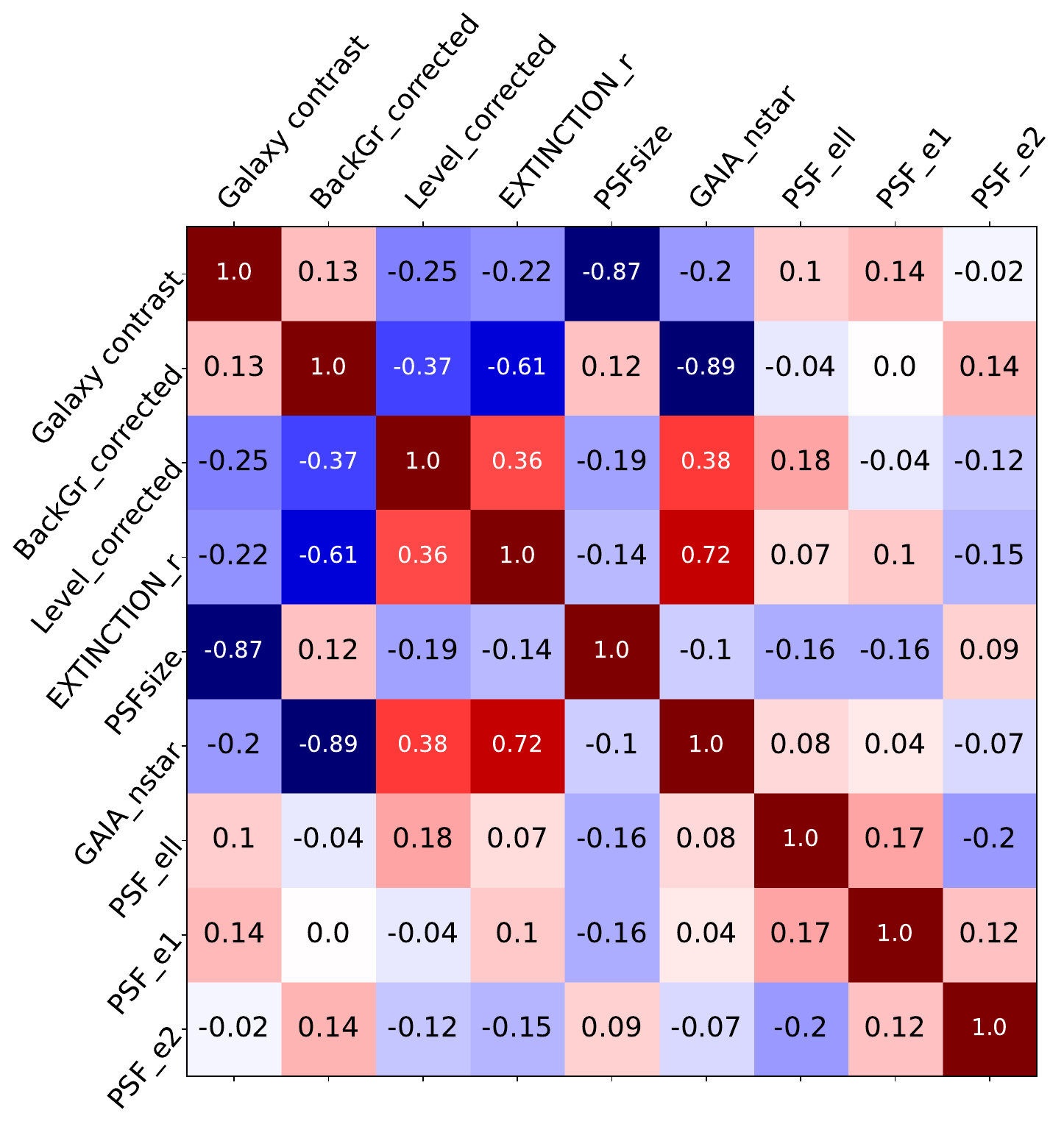}
    \caption{Spearman correlation coefficient matrix. The numbers in each grid are the correlation coefficient between the median systematics and the galaxy contrast in each hierarchical cluster.}
    \label{fig:syscorr_full}
\end{figure}

\begin{figure}
    \centering
    \includegraphics[width=\linewidth]{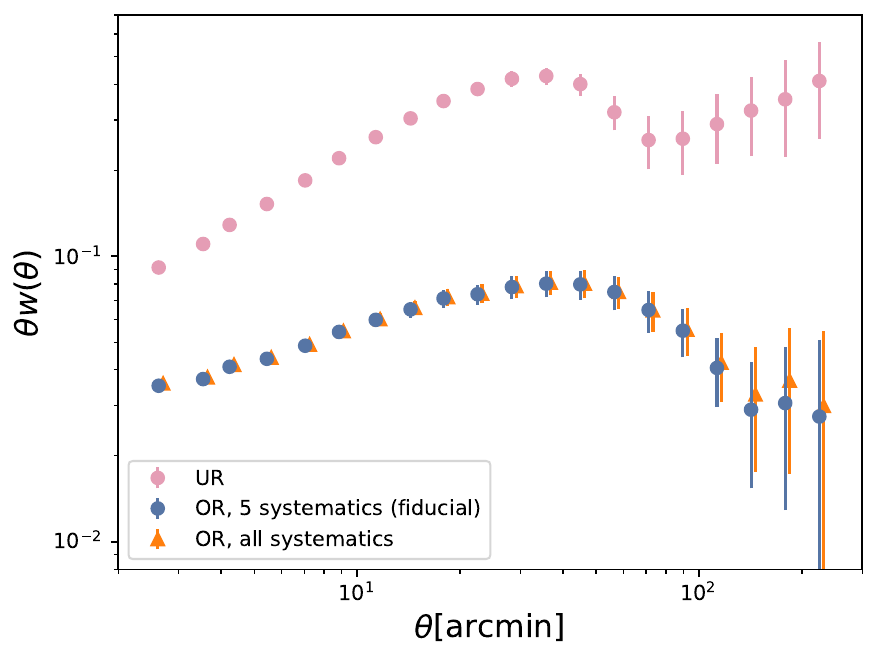}
    \caption{Blinded 2PCFs measured from the KiDS-Legacy catalogue. The pink dots are UR measurements; the blue dots are our fiducial OR measurements (OR recovered from 5 systematics) and the orange dots are measured with OR recovered from 16 systematics. The error bars are derived from the theoretical covariance matrices with best-fit UR and OR parameters.}
    \label{fig:wtheta_legacy_fullsys}
\end{figure}

Our fiducial choice of five systematics is based on the consideration that some systematics are strongly correlated with each other (such as extinction in different bands), so introducing them will not add information to the SOM. In addition, some systematics have no selection effect in the galaxy field. Therefore, including them in the training vector will not improve performance, but will increase the time and computational resources required (especially in the validation procedure). In this section we validate our fiducial choice of systematics described in Section \ref{sect:data} by training the SOM with all the available systematics in the KiDS-Legacy catalogue. This increases the number of systematics from 5 to 16 and also increases the number of training epochs required for the SOM to converge.

We group the galaxy into 600 clusters as the fiducial set-up and calculate Spearman's correlation coefficient between median systematics and galaxy contrast in each hierarchical cluster. Figure \ref{fig:syscorr_full} shows the correlation coefficient matrix. The first row of the correlation matrix shows the correlation between galaxy number contrast and each systematics, indicating the selection effect captured by SOM+HC. The extinction in different bands is fully correlated because it is calculated by scaling the reddening template given by \cite{Schlafly_2011} according to the band, so we only include $r$-band here as the fiducial case. PSF size has the highest negative correlation, which is in agreement with the fiducial run. Level and GAIA star number density are both slightly correlated with galaxy number contrast. We do not include \texttt{Background} in our fiducial run because it correlates weakly with galaxy number contrast, while it correlates strongly with GAIA star number density, so including it does not add much information.

To further justify our choice of systematics, we measure the 2PCF corrected by the OR weight generated from all the systematics. Figure \ref{fig:wtheta_legacy_fullsys} shows the UR (pink), fiducial OR (blue) and full systematics OR (orange) 2PCFs. The $\chi_{\mathrm{d}}^2$ value between the fiducial 2PCF and all differences are insignificant over the entire angular scale considered. Therefore, we conclude that the 5 systematics that we choose are representative of the whole set of systematics to recover the organised random.

\section{Angular galaxy clustering for the KiDS-1000 bright sample}
\label{sect:app_k1000b}

\begin{figure}
    \centering
    \includegraphics[width=\columnwidth]{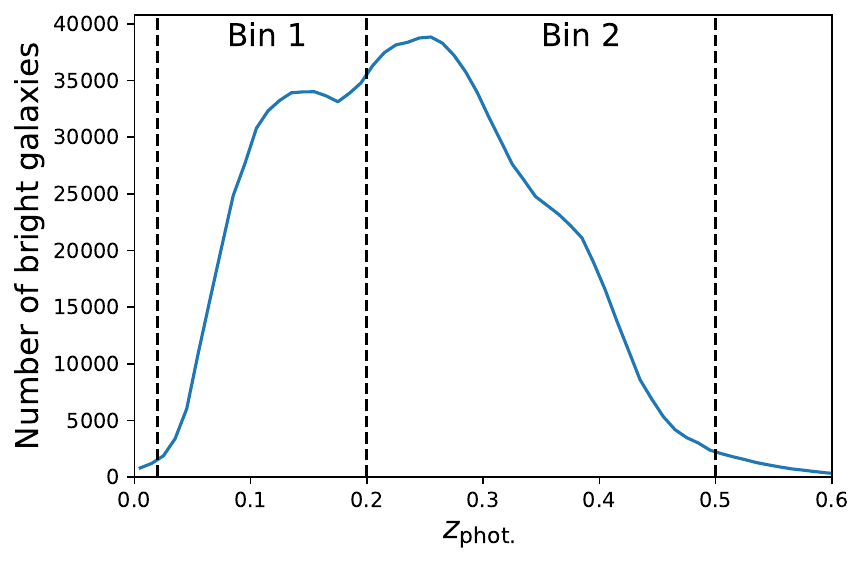}
    \caption{Photometric redshift distribution of the KiDS-1000 bright sample, with the $z_{\mathrm{phot.}}$ = \{0.02, 0.2, 0.5\} redshift bins (dashed lines) employed in our \wtheta measurements}
    \label{fig:k1000b_dndz}

\end{figure}

\citet{harryOR} measured the 2PCF from the KiDS-1000 Bright sample \citep{Bilicki_2021} selected with a magnitude cut $r<20$ from the 4th KiDS data release. Its redshift distribution is reliably calibrated from the overlap with Galaxy And Mass Assembly (GAMA) spectroscopy using the neural network algorithm implemented in the ANNz2 software (see Fig.~\ref{fig:k1000b_dndz} for the redshift distribution). The sample covers a sky area of 789 $\mathrm{deg}^2$ and has a number density of 0.36 $\mathrm{arcmin}^{-2}$. 

In this section we re-measure the 2PCF of the KiDS-Bright sample with the new SOM+HC implementation to check the consistency between our code and the pipeline used in \citet{harryOR}. Following the fiducial set-up (the ``\texttt{100A}'' set-up) in \citet{harryOR}, we recover the OR with a 100$\times$100 SOM trained on the same systematics ($r$-band detection threshold, PSF size and PSF shape) grouped into 100 hierarchical clusters. The OR weight map has a \texttt{Nside}=1024, which corresponds to an angular resolution of 3.4 arcmin, slightly larger than the fiducial set-up in \citet{harryOR}. The correlation function is then measured in 30 angular bins between 3 and 300 arcmin.

\begin{figure*}
    \centering
    \includegraphics[width=\linewidth]{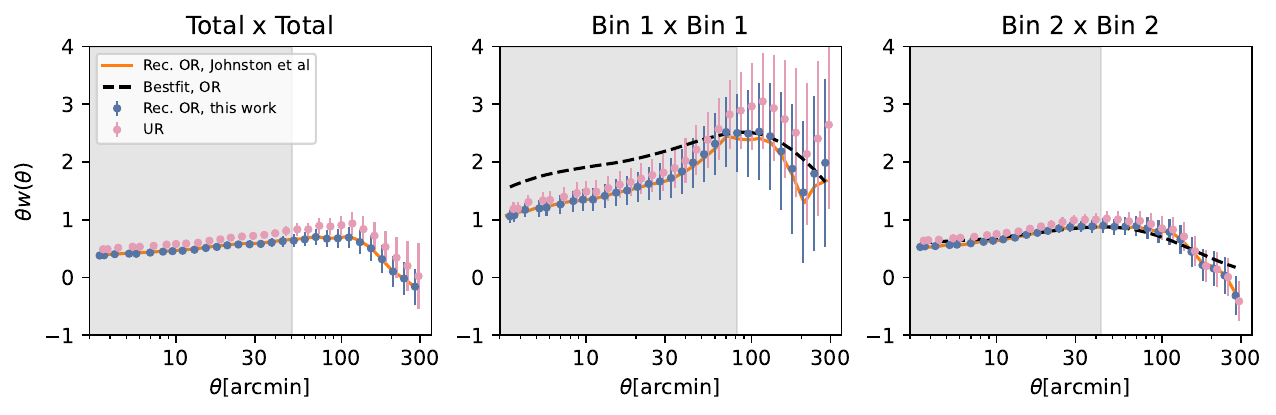}
    \caption{2PCFs measured from the KiDS-1000 bright sample. The three panels are the auto-correlation \wtheta of the whole sample, and the first and the second tomographic bins, respectively. The blue and pink dots are the measurements corrected by recovered organised random and uniform random. The shaded regions are angular scales corresponding to physical scales smaller than $8h^{-1}\mathrm{Mpc}$ estimated at the mean redshift. The error bars are the standard deviations derived from the covariance matrix provided by the \textsc{OneCovariance} package. The orange curve is the OR-corrected 2PCF data points measured by \citet{harryOR}. The black dashed curves in the second and third panels are the best-fit 2PCF from MCMC.}
    \label{fig:wtheta_k1000b}
\end{figure*}

After recovering the OR weight, we measure the correlation functions of the two tomographic bins of the bright sample, defined by selecting galaxies with ANNz-calibrated photo-$z$ with the cut $\{0.02, 0.2, 0.5\}$. The 2PCFs corrected by the uniform random and the recovered organised random are presented in Fig.~\ref{fig:wtheta_k1000b} with pink points and blue points, respectively. We notice that our measurements are fairly consistent with that from \citet{harryOR}, shown as orange curves. The error bars are the standard deviation derived from the covariance matrix given by the \textsc{OneCovariance} code with the same redshift distribution and the best-fit $\Omega_{\mathrm{m}}$ and galaxy biases to be determined below. The shaded regions are angular scales corresponding to physical scales smaller than $8h^{-1}\mathrm{Mpc}$ estimated at the mean redshift. We then fit $\{\Omega_{\mathrm{m}}, b_1, b_2\}$ in the linear model with the 2PCFs in the linear scale. We note that we don't know the parameters {\textit{a priori}} to compute the covariance matrix, so we do an iterative parameter fit. We choose the initial value of \omm=0.33 and $b_1=1.1$, $b_2=1.25$ to compute the covariance matrix. The pilot galaxy bias values are taken from \citet{van_Uitert_2018}, which apply to a GAMA-like subsample of KV-450 with a similar redshift distribution to the bright sample used here. We define a Gaussian likelihood and run an MCMC to obtain the best-fit parameters, and then update the covariance matrix with them. We then run the MCMC again with the updated covariance matrix to obtain the posterior.

\begin{figure}
    \centering
    \includegraphics[width=\columnwidth]{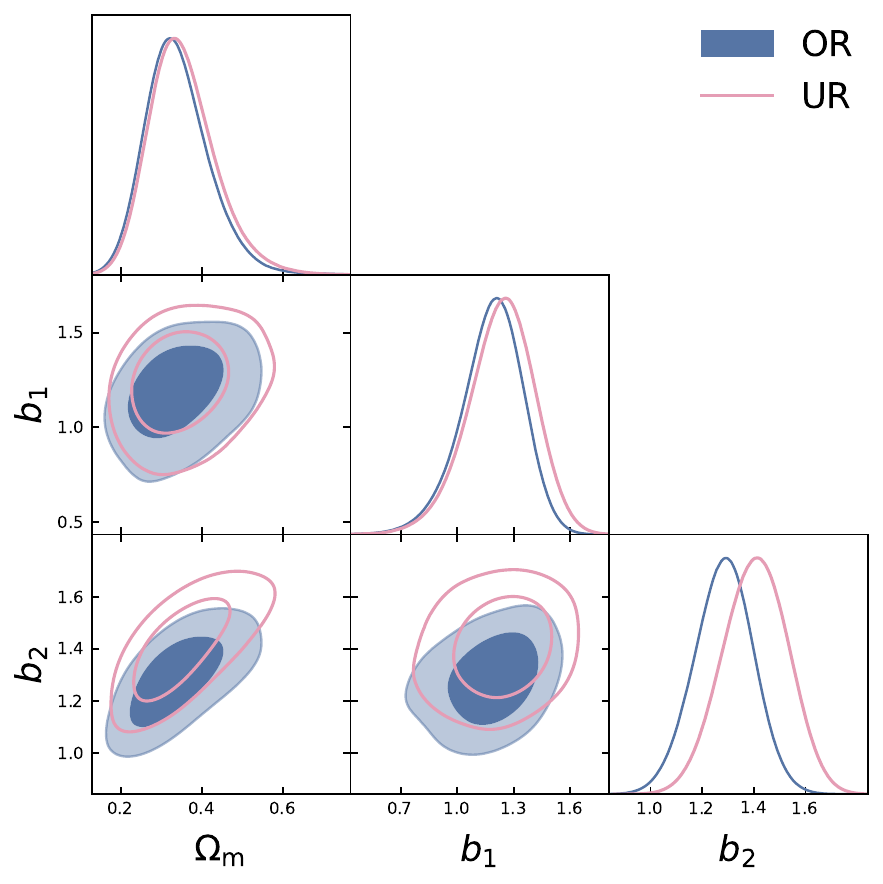}
    \caption{Posterior of $\Omega_{\mathrm{m}}, b_1, b_2$ fit by the 2PCF from KiDS-1000 bright samples. The deep shaded and lightly shaded contour is the 68\% and 95\% credible levels, respectively. The blue and pink contours correspond to the posterior from 2PCF corrected with OR and UR respectively.}
    \label{fig:post_k1000b}
\end{figure}

The posteriors of the UR and OR cases are shown as pink and blue contours in Fig.~\ref{fig:post_k1000b}. The mean value of the parameters in the converged MCMC chains and the $1-\sigma$ levels are summarized in Table~\ref{tab:k1000b}. The theoretical $w(\theta)$ calculated from the best-fit parameters of the OR case is shown as dashed black curves in Fig.~\ref{fig:wtheta_k1000b}. The reduced $\chi^2$ value between the OR 2PCF and the best-fit 2PCF is 1.06, corresponding to a PTE of 0.38, indicating a good fit between the model and the data. All three parameters are constrained. Notably, the matter density is constrained in agreement with previous cosmological probes. The galaxy biases are constrained to be close to 1 with a slightly increasing trend. However, we note  that the bias parameters are highly degenerate with $\sigma_8$, which is fixed in our case. To break this degeneracy we need to introduce matter field tracers such as cosmic shear or use the halo model like in \citet{2023A&A...675A.189D}. This is the content of the ongoing KiDS-Legacy $3\times2$pt and $6\times2$pt projects. 

From the data and the posterior, we found that the difference between UR and OR is at the $\sim 1\sigma$ level, suggesting that the variable depth in the bright sample is much less pronounced than in the faint sample. This is expected, as the detectability of bright galaxies should be less affected by the Galactic or atmospheric foreground.

\begin{table}
\caption{Parameter fit from KiDS-1000 bright galaxy 2PCF.}
 \renewcommand{\arraystretch}{1.25}

    \centering
    \begin{tabular}{llll}
\toprule
Parameter & Prior range & Best-fit, OR & Best-fit, UR \\ 
\hline
$\Omega_{\mathrm{m}}$ & [0.01, 1] & $0.338^{+0.066}_{-0.085}$ & $0.350^{+0.068}_{-0.089}$ \\ 
 $b_1$ & [0, 5] & $1.19^{+0.18}_{-0.14}$ & $1.24^{+0.19}_{-0.16}$ \\ 
 $b_2$ & [0, 5] & $1.28\pm 0.11$ & $1.41\pm 0.12$ \\ 
 
\bottomrule
    \end{tabular}
    \label{tab:k1000b}
     \tablefoot{The best-fit values are those with maximum posterior and the errors are the 68.3\% credible level.}
\end{table}

\end{appendix}

\end{document}